%% file: main-techrep.tex
\documentclass[12pt,fullpage]{article}

\pdfoutput=1

\usepackage{amssymb}
\usepackage[fleqn]{amsmath}

\usepackage[usenames]{xcolor}
\usepackage{proof}
\usepackage{multirow}
\usepackage{xspace}
\usepackage{url}
\usepackage{tikz} 
\usetikzlibrary{matrix,arrows}
\usepackage{subcaption}
\usepackage{todonotes}
\usepackage{enumerate}
\usepackage{times}
\usepackage{graphicx}
\usepackage{mdframed}
\usepackage{wrapfig}
\usepackage{mathtools}
\usepackage{amsthm}

\newcommand\onlyintechrep[1]{#1}
\newcommand\onlyinpaper[1]{}

\newcommand\rtext[1]{{#1}}
\newcommand\ftext[1]{{#1}}

\newtheorem{thm}{Theorem}
\newtheorem{lem}{Lemma}
\newtheorem{exmp}{Example} 
\newtheorem{defn}{Definition} 
\newtheorem{prop}{Proposition} 
\newtheorem*{pf}{Proof}

\input{./macros}

\graphicspath{{./}}

\title{Distributed First Order Logic\thanks{This paper is a substantially revised and extended version of a paper with the same title presented at the 1998 International Workshop on Frontiers of Combining Systems (FroCoS'98)}}

\author{Chiara Ghidini \quad Luciano Serafini \\
Fondazione Bruno Kessler, Trento, 38100, Italy\\
\tt{\{ghidini,serafini\}@fbk.eu}}

\begin{document}

\maketitle

\begin{abstract}
Distributed First Order Logic (DFOL) has been introduced more than \ftext{ten} years ago with the purpose of formalising distributed knowledge-based systems, where knowledge about heterogeneous domains is scattered into a set of \ftext{interconnected modules}. \ftext{DFOL formalises the knowledge contained in each module by means of first\ftext{-}order theories, and the interconnections between modules by means of special inference rules called \emph{bridge rules}.}
Despite their \ftext{restricted form in the original DFOL formulation, bridge rules have} influenced several works in the areas of heterogeneous knowledge integration, \ftext{modular knowledge representation,} and schema/ontology matching. This, in turn, has fostered extensions and modifications of the original DFOL \ftext{that} have never been systematically described and published.
This paper \ftext{tackles the lack of a comprehensive description of DFOL} by \ftext{providing} a systematic account of a completely revised and extended version of \ftext{the logic, together with a sound and complete axiomatisation of a general form of bridge rules based on Natural Deduction}. The resulting DFOL framework is then proposed as a clear formal tool for the representation \ftext{of} and reasoning \ftext{about distributed knowledge and bridge rules.}
\end{abstract}

%


\input{introduction.tex}
\input{examples.tex}

\input{semantics.tex}

\input{interpretationConstraints.tex}

\input{calculus.tex}

\input{soundAndcomplete.tex}

\input{relatedwork.tex}
\input{conclusions.tex}

\appendix
\onlyintechrep{\input{appendixB.tex}}
\input{soundness}
\input{completeness}

\bibliographystyle{plain}
\bibliography{biblio}

\end{document}

%% file: macros.tex
\newcommand\Mc{{\M^c}}
\newcommand\GGamma{{\mathbf \Gamma}}
\newcommand\SSigma{{\mathbf \Sigma}}

\newcommand{\doublebar}[1]{\bar{\bar{#1}}}

\newcommand\fun{{\sf F}}
\newcommand\tot{{\sf T}}
\newcommand\inj{{\sf J}}
\newcommand\sur{{\sf S}}
\newcommand\inv{{\sf I}}
\newcommand\com{{\sf C}}
\newcommand\euc{{\sf E}}
\newcommand\congr{{\sf G}}
\newcommand\incp{{\sf IP}}

\newcommand{\MC}[1]{{\sf ML}(#1)}
\newcommand{\shortreason}[1]{\parbox[t]{29ex}{#1}}
\newcommand{\longformula}[1]{
 \begin{minipage}[t]{.33\textwidth}
	{#1}
 \end{minipage}
}
\newcommand{\reason}[1]{\parbox[t]{31ex}{#1}}

\newenvironment{ndproof}
   {\def\arraystretch{1.2}
\begin{array}{lll@{\hspace{2pt}}ll} \hline 
       {\text{\it Label}} &
       {\text{\it Formula}} &
       {\text{\it L.A.}} &
       {\text{\it G.A.}} &
       {\text{\it Inference rule}} \\ \hline}
{\end{array}}

\newcommand{\gap}{\vspace*{0.2cm}}

\newenvironment{pot}[1]{\ \\[\parskip] \noindent{\textsc{Proof of Theorem
	    #1.}}}{\mbox{\hspace{0cm}}\hfill $\Box$ \vspace*{.2cm}}

\DeclareMathSymbol{\NAT}{\mathord}{AMSb}{"4E}

\newcommand{\exi}[1]{\exists{\rm I}_{#1}}
\newcommand{\exe}[1]{\exists{\rm E}_{#1}}
\newcommand{\alle}[1]{\forall{\rm E}_{#1}}
\newcommand{\alli}[1]{\forall{\rm I}_{#1}}
\newcommand{\ori}[1]{\dis{\rm I}_{#1}}
\newcommand{\ore}[1]{\dis{\rm E}_{#1}}
\newcommand{\andi}[1]{\con{\rm I}_{#1}}
\newcommand{\ande}[1]{\con{\rm E}_{#1}}
\newcommand{\impi}[1]{\imp{\rm I}_{#1}}
\newcommand{\impe}[1]{\imp{\rm E}_{#1}}

\newcommand{\eqi}[1]{=\!{\rm I}_{#1}}
\newcommand{\eqe}[1]{=\!{\rm E}_{#1}}
\newcommand{\cut}[1]{{\sf Cut}_{#1}\xspace}

\newcommand{\der}[1]{\vdash_{#1}}

\newlength{\satindexwidth}
\newlength{\satindexskip}
\newcommand{\sat}[1]{%
 \settowidth{\satindexwidth}{{\tiny #1}}%
 \ifdim\satindexwidth>.5ex%
  \setlength{\satindexskip}{-.5ex}%
 \else\setlength{\satindexskip}{\satindexwidth}\fi%
 \models_{_{\mbox{\hspace{\satindexskip}\tiny #1}}}}

\newcommand{\imp}{\supset}
\newcommand{\dis}{\vee}
\newcommand{\con}{\wedge}

\newcommand{\npla}[1]{\langle#1\rangle}

\newcommand\bdom{{\bf dom}}
\newcommand\bx{{\bf x}}

\newcommand\co[2]{#2\!:\!#1}

\newcommand\IC{{\sf BR}\xspace}

\newcommand\dr{{r}}

\newcommand\fromvar[2]{\overset{#2\rightarrow}{#1}}
\newcommand\tovar[2]{\overset{\rightarrow #2}{\negthickspace #1}}
\newcommand\bothvar[1]{\overset{\rightarrow}{#1}}
\newcommand\fromI[1]{\overset{#1\rightarrow}{}}
\newcommand\toI[1]{\overset{\rightarrow #1}{}}

\newcommand\M{{\cal{M}}\xspace}

\newcommand\dom{\bdom}
\newcommand{\brmodels}{\models_{\IC}}
\def\ext#1#2{|\!|#1|\!|_{#2}}

\def\price{\ensuremath{price}\xspace}

\def\mediator{\ensuremath{m}\xspace}
\def\selleruno {\ensuremath{\it Sel.1}\xspace}

\def\sellerdue {\ensuremath{Sel.2}\xspace}
\def\available{\ensuremath{Available}\xspace}
\def\offer{\ensuremath{\it Offer}\xspace}
\def\price{\ensuremath{Price}\xspace}

\def\obsuno{\ensuremath{Mr.1}\xspace}
\def\obsdue{\ensuremath{Mr.2}\xspace}
\def\ball{\ensuremath{b}}
\def\inbox{\ensuremath{inbox}}
\def\positions{\ensuremath{p}}
\def\bianco{\ensuremath{white}}
\def\nero{\ensuremath{black}}

\def\bempty{\ensuremath{empty}}

\def\fls{\phi}
\def\frs{\psi}

\def\DI{{\frak I}}
\def\I{{\mathcal I}}
\def\Ii{{\I_i}}
\def\Ij{{\I_j}}

\def\D{\Delta}

\def\isa{\sqsubseteq}
\def\into{\stackrel{\mbox{\tiny $\sqsubseteq$}}{\longrightarrow}}
\def\onto{\stackrel{\mbox{\tiny $\sqsupseteq$}}{\longrightarrow}}

\def\I{{\mathcal I}}
\def\BR{{\sf BR}}

\def\relation#1{\textsf{\small #1}}

\newcommand\rtovar[3]{\overset{\overset{#3}{\rightarrow} #2}{\negthickspace #1}}
\newcommand\rfromvar[3]{\overset{#2\overset{#3}{\rightarrow}}{\negthickspace #1}}

\newcommand{\shoiq}{
  \ensuremath{\mathcal{SHOIQ}}\xspace}
\newcommand\pdl[1]{\stackrel{\mbox{\small $#1$}}{\rightarrow}}

%% file: introduction.tex
\section{Introduction}

The method of structuring complex knowledge-based systems in a set of largely autonomous modules has become common practice in several areas such as Semantic Web, Database, Linked Data, \ftext{Ontologies,} and Peer-to-Peer systems. \ftext{In these practices,} knowledge is \ftext{often} structured in \ftext{multiple} interacting sources and systems, hereafter indicated as local knowledge bases or simply knowledge bases (KBs).
%
Several efforts have been devoted to provide a well\ftext{-}founded theoretical background able to represent and reason about distributed knowledge. \ftext{Several examples} can be found in well established areas of Database and Knowledge Representation such as
federated and multi-databases \cite{sheth1,lakshmanan1-schemasql,grant1-query-lang-for-multidb}, 
database and information integration \cite{hull1,subrahmanian1,catarci1,calvanese2,genesereth1-infomaster,levy1-information-manifold},
database schema matching \cite{rahm1-matching-survey}, and
contextual reasoning~\cite{mccarthy26,buvac2,ghidini10}. 
Further \ftext{examples} can also be found in more recent areas of the Semantic Web, such as 
ontology matching~\cite{SerafiniStuckenschmidtWache-IJCAI05,ontology-matching-EuzenatShvaiko07}, 
ontology integration~\cite{maedche1-kaon,e-connections,SerafiniBorgidaTamilin-IJCAI05}, 
ontology modularisation~\cite{Mossakowski:2012uq,KoLuWaWo-OntMod-08,Bao:2009:PDL:1560559.1560578}, 
linked data~\cite{Bouquetetal:identity:aswc2009, Hartig:2012yq}, and in  
Peer-to-Peer systems~\cite{bernstein1,franconi1,halevy1,calvanese-pods-2004}. 

%


\ftext{The formalisms mentioned above share several aspects: they all focus on static and boolean knowledge\footnote{\ftext{It is important to mention here that in this paper we discard aspects tied to the non-monotonic evolution of knowledge and to its many valued/probabilistic/fuzzy nature.}};} local knowledge is expressed using \ftext{a} (restricted form of) first\ftext{-}order \ftext{language}; \ftext{each module is associated with a specific (first\ftext{-}order) language, called local language; the domains of interpretation of the different local languages can be heterogeneous;
the same symbol in different local languages can have different interpretations;
knowledge within the different modules is related through some form of cross-language axioms.}
Despite their commonalities, these formalisms are mainly tailored to the characterisation of specific phenomena of distributed knowledge\ftext{.} \ftext{L}ittle work exists on the definition of a general logic, comprehensive of a sound and complete calculus and of a rigorous investigation of its properties, \ftext{as well as} able to represent generic semantically heterogeneous distributed systems, based on first\ftext{-}order logic and comprised of heterogeneous domains.

%
%

As a step towards the definition of such a logic, \emph{Distributed First Order Logic (DFOL)} was introduced in~\cite{ghidini5}.
\ftext{As explained in detail in Section~\ref{sec:originalDFOL},
the original DFOL was able to capture only limited interconnections between local KBs. Nonetheless}, the idea presented in~\cite{ghidini5} of connecting different domains of interpretation by means of directional \textbf{domain relations}, and \ftext{a number of} unpublished efforts to substantially extend DFOL to increase its flexibility and expressiveness, have strongly influenced several frameworks which include \emph{Package-Based Description Logics (P-DL)}~\cite{Bao:2009:PDL:1560559.1560578}, \emph{Distributed Description Logic (DDL)}~\cite{SerafiniBorgidaTamilin-IJCAI05}, and \emph{C-OWL}~\cite{Bouquet-CWOL-ISWC03}.

In this paper we overcome the limitations of the original formulation of DFOL and present a systematic account of a completely revised and extended version of the formalism, which was elaborated in conjunction with most of the efforts listed above.
The unpublished elements described in this paper include: (i) a general version of \emph{bridge rules} based on the introduction of \emph{arrow variables} as a way to express general semantic relations between local KBs (Section~\ref{sec:syntax-semantics}); (ii) a notion of \emph{logical consequence between bridge rules} (Section~\ref{subsec:derived-bridge-rules}); (iii) a thorough investigation of the \emph{properties} of DFOL (Section~\ref{sec:syntax-semantics}) and of how to use it to represent important types of relations between local KBs (Section~\ref{sec:interpretation-cosntraints}); and (iv) a \emph{general sound and complete calculus} able to capture the semantic relations enforced by arrow variables, to infer new bridge rules and to discover unsatisfiable distributed knowledge-based systems (Section~\ref{sec:calculus}).

To make the presentation clearer, but also to show the generality of the approach, we informally describe, and then formalise using DFOL, two examples of distributed knowledge, namely reasoning with viewpoints, and information integration. This material is covered in Section~\ref{sec:examples} (informal presentation) and Examples \ref{ex:formalization-magic-box}, \ref{ex:formalization-mediator}, and \ref{ex:calculus-magic-box} (formalisation using DFOL).

The extended version of DFOL presented in this paper is also used, in Section~\ref{sec:rel-work}, as a framework for the encoding of different static and \ftext{boolean} knowledge representation formalisms grounded in first\ftext{-}order logic. \ftext{In line with the work presented in~\cite{SerafiniStuckenschmidtWache-IJCAI05} these formalisms are} tailored to the representation of semantically heterogeneous distributed knowledge-base systems (e.g., ontologies, databases, and contexts) \ftext{with} heterogeneous domains. 


%% file: examples.tex
\section{Two explanatory examples} 
\label{sec:examples} 

The examples introduced in this section are used throughout the paper to discuss and illustrate the ideas and the formalisation of DFOL we propose.

\subsection{Reasoning with viewpoints}
\label{sub:reasoning_with_viewpoints}

\begin{figure}[t]
\centering
\begin{subfigure}[b]{.4\linewidth}
\includegraphics[width=\linewidth]{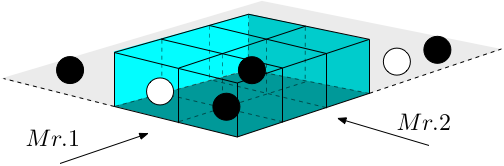}
\caption{The global view\ftext{.}}
\label{fig:panorama}
\end{subfigure}
\qquad
\begin{subfigure}[b]{.45\textwidth}
\includegraphics[width=\linewidth]{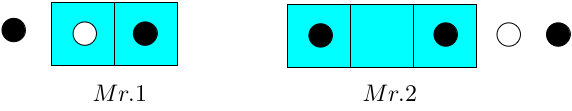}
\caption{$\obsuno$ and $\obsdue$'s points of view.}
\label{fig:visioni-parziali}
\end{subfigure}
\caption{The magic box.}
\label{fig:magic-box}
\end{figure}

\begin{exmp}[The magic box]
Consider the scenario in Figure~\ref{fig:panorama}: there are two observers, $\obsuno$ and $\obsdue$, each having a partial viewpoint of a box and of an indefinite number of balls. The balls can be black or white and the box is composed of six sectors, each possibly containing a ball. Balls can be inside the box or in the grey area outside the box. From their perspectives observers cannot distinguish the depth inside the box. Moreover they cannot see balls hidden behind other balls and balls located behind the box.
Figure~\ref{fig:visioni-parziali} shows what $\obsuno$ and $\obsdue$ actually see in the scenario depicted in Figure~\ref{fig:panorama}.\footnote{The example is an extension of the ``magic box'' example originally proposed in
\cite{ghidini10}.}
\end{exmp}

The magic box, together with the balls, represents a ``complex'' environment corresponding to the domain of the agents' local knowledge bases. The agents' points of view correspond to their local knowledge. The local knowledge of the agents is
constrained one another by the fact that they describe views over the same environment.
Assuming that we have a complete description of the box we can build the agents' local knowledge (bases) as views over this complete description. However\ftext{,} such \ftext{a} complete description is often not available. What we often have are only the partial views, and a set of constraints between these views, with no representation of the external world (in our example case, the entire box).
In cases like this we need a logical formalism able to describe the point of view of the different agents ($\obsuno$, and $\obsdue$, in our example) and the constraints among these views, without having to represent the entire box
as we see it in Figure~\ref{fig:panorama}. The formalism should be able to represent
and reason about statements such as:
\begin{enumerate}
	\itemsep=-.2\parsep
	\item\label{item:wp1} \emph{``the domain of \obsuno contains 3 balls and a box with 2 sectors''};
	\item\label{item:wp2} \emph{``\obsdue sees a black ball in the right sector''};
	\item\label{item:wp3} \emph{``\obsuno and \obsdue  agree on the colour of the balls they both see''};
	\item\label{item:wp4} \emph{``if \obsuno sees an empty box, then \obsdue sees an empty box too''};
	\item\label{item:wp5} \emph{``if \obsdue sees 3 balls in the box, then the leftmost is also seen by \obsuno''}\ftext{.}
\end{enumerate}

This example involves, in a very simple form, a number of crucial aspects of distributed knowledge representation: first, it deals with \textbf{heterogeneous local domains} which correspond to the different sets of balls in the different viewpoints. Second, it has to do with \textbf{cross\ftext{-}domain identity}. In fact, we need to represent the connections between the perceptions of the balls by each agent, without having a\ftext{n} objective model that completely and correctly describes all the objects (balls) present in the box. An example is statement 5 above. Third, we have \textbf{heterogeneous local properties}. In our example \obsuno sees a box composed of two sectors, while for \obsdue the box is composed of three sectors. Thus \obsdue has a a notion of ``a ball being in the central sector'' which \obsuno does not have. Fourth, it deals with \textbf{constrained viewpoints}. The viewpoints of the agents are, in fact, not independent, since they are the result of they observing the the environment. Thus, if \obsuno sees an empty box, then \obsdue is constrained to see an empty box too, as described in statement 4 above.

\subsection{Mediator-based Information Integration}
\label{sub:information_integration}

Information integration is often based on architectures that make use of a \emph{mediator}~\cite{ullman1}, as in the following example.

\begin{figure}[ht]
    \centering
      \includegraphics[width=\textwidth]{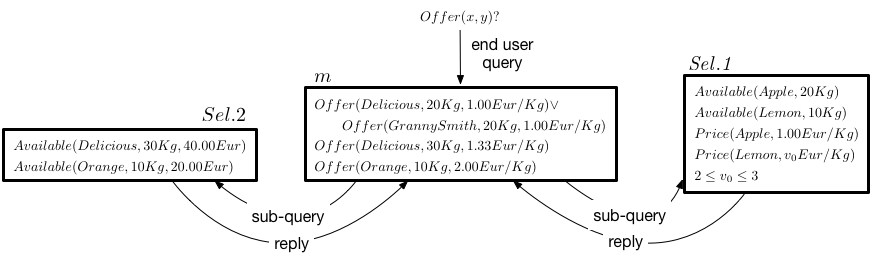}
    \caption{An example of mediator system\ftext{.}}
\label{fig:mediator}
\end{figure}

\begin{exmp}
\label{ex:mediator}
Consider the databases of two fruit sellers, \selleruno\ and \sellerdue, depicted in Figure \ref{fig:mediator}. The information about fruits sold by \selleruno\ is contained in two relations $\available(fruit,qty)$, and $\price(fruit,x)$ with the intuitive meaning that a quantity $\ensuremath{qty}$ of $\ensuremath{fruit}$ is available for selling and that its price is fixed to
$\ensuremath{x}$ Euros per kilogram (Eur/Kg for short). The value of $\ensuremath{x}$ could be a number or an interval $[x_1,x_2]$, expressing the fact that a specific price has not been fixed yet but it is contained within $x_1$ and $x_2$.
\sellerdue, instead, stores information about fruit prices in a single relation $\available(fruit,qty,x)$, where $x$ indicates the total price of quantity $\ensuremath{qty}$ of $\ensuremath{fruit}$, and not its price per kilo.
A mediator $\mediator$ collects the data of \selleruno and \sellerdue and integrates them into a single relation $\offer(fruit,qty,x)$, meaning that a quantity $qty$ of $fruit$ is available at price $x$ Euros per kilo from (at least)
one of the two sellers. Customers looking for information about fruit prices can submit a query to the mediator, instead of asking the two sellers separately as shown in Figure~\ref{fig:mediator}.
\end{exmp}

Even if we discard details on how the information is integrated, and the process of query-answering is performed, we can observe that a logic for the representation of such a scenario must \ftext{be} able to represent the heterogeneous schemata and domains of the three subsystems $m$, \selleruno, and \sellerdue. In particular the formalism should be able to represent the following facts:
\begin{enumerate}
	\itemsep=-.2\parsep
	\item \selleruno sells ``apples'', \ftext{whereas} \sellerdue and \mediator represent the domain of apples at a greater granularity, and are able to offer specific varieties of apples (ranging among Delicious and Granny Smith in our example).
	Moreover\ftext{,} for the sake of the example, the ``apples'' of \selleruno correspond to both ``Delicious'' and ``GrannySmith'' in the mediator. This justifies the disjunctive statement retrieved by the mediator as a ``translation'' of the statements about apples contained in the database of \selleruno;
	\item total prices of \sellerdue are transformed in prices per kilo in \mediator to be homogeneous with price format of \selleruno;
	\item \mediator is not interested in retrieving information about fruits whose price is not yet defined (lemons in our case);
	\item the information goes from \selleruno (resp.~\sellerdue) to \mediator and not from the mediator to the sellers.
\end{enumerate}
Again, this example involves \textbf{heterogeneous local domains} and \textbf{cross\ftext{-}domain identity}, as described in statement 1 above. Moreover\ftext{,} it involves \textbf{heterogeneous local properties} represented by the different relations, which are nonetheless constrained by the fact that they all represent the availability of fruit at a certain price. Thus, for example, if 10 Kg of oranges cost 20 Euros in the database of \sellerdue, then oranges cost 2 Euros per Kilo in the database of the mediator. In addition, information is required to be \textbf{directional}: in our example it flows from the sellers to the mediator and not vice-versa, since the sellers must be prevented to retrieve knowledge about potential competitors that could be stored in the mediator.

%% file: semantics.tex
\section{Syntax and Semantics of DFOL}
\label{sec:syntax-semantics}

In this section we provide the syntax and semantics of Distributed First Order Logic (DFOL). They are based on the syntax and semantics of \ftext{f}irst\ftext{-}\ftext{o}rder \ftext{l}ogic and provide an extension of the Local Models Semantics presented in~\cite{ghidini10} to the case where  each local KB is described by means of a first\ftext{-}order language. 

\subsection{DFOL Syntax}
\label{sec:syntax}
Let $\{L_i\}_{i \in I}$ (hereafter $\{L_i\}$) be a family of
first\ftext{-}order languages defined over a non empty set $I$ of indexes. For the sake of simplicity we assume, without loss of generality, that all the languages $L_i$ contain the same set $X$ of infinitely many variables. Each language $L_i$ is the language used by the $i$-th local knowledge base to partially describe the world from its own perspective. 
For instance, in the magic box example $I=\{1,2\}$. 

In DFOL, each $L_i$ is a first\ftext{-}order language with equality, extended with a new set of symbols, called \emph{arrow variables}, which are of the same syntactic type as constants and standard individual variables (hereafter often called non-arrow variables).  
Formally, for each variable $x \in X$, and each index $i,j\in I$, with $i\neq j$, the signature of $L_i$ is extended to contain the two arrow variables\index{arrow variable} $\tovar{x}{j}$ and $\fromvar{x}{j}$.

The arrow variables $\tovar{x}{j}$ and $\fromvar{x}{j}$ in $L_i$ intuitively denote an object in the domain of interpretation of $L_i$ that corresponds to the object $x$ in the domain of $L_j$. The difference between $\tovar{x}{j}$ and $\fromvar{x}{j}$ will become clearer later in the paper. We often use $\fromvar{x}{}$ to denote a generic arrow variable (that is, either of the form $\tovar{x}{j}$ or $\fromvar{x}{j}$). 

Terms of $L_i$, also called \emph{$i$-terms}, are recursively defined as in first\ftext{-}order logic starting from the set of constants, variables, and arrow variables, and by recursively applying function symbols. Formally: 
\begin{enumerate}
	\item Any constant, variable, and arrow variable of $L_i$ is a $i$-term\ftext{.}
	\item If $f$ is a function symbol of arity $n$ in $L_i$ and $t_1,\ldots, t_n$ are $i$-terms, then $f(t_1,...,t_n)$ is a $i$-term.
\end{enumerate}
Formulas of $L_i$, called {\em $i$-formulas}, \index{$i$-formula} are defined as in first\ftext{-}order logic, with the discriminant that we only quantify over non-arrow variables. Formally:
\begin{enumerate}
	\item If $P$ is a n-ary predicate symbol in $L_i$ and $t_1,\ldots, t_n$ are $i$-terms, then $P(t_1,...,t_n)$ is a $i$-formula.
	\item If $t_1$ and $t_2$ are $i$-terms, then $t_1 = t_2$ is a $i$-formula.
	\item If $\phi$ and $\psi$ are $i$-formulas, then $\lnot \phi$, $\phi \imp \psi$, $\phi \wedge \psi$, $\phi \vee \psi$, are $i$-formulas. 
	\item If $\phi$ is a formula and $x$ is a non-arrow variable, then $\forall x \phi$ and $\exists x \phi$ are $i$-formulas.
\end{enumerate}

Examples of $i$-terms are $x$, $c$, $\tovar{x}{j}$, $f(c,d)$, and $f(\fromvar{x}{j},f(g(d)))$. Examples of $i$-formulas are $P(x,y,z)$, $P(\tovar{x}{j},w,a)$, $\bot\imp P(f(c),d)$, $\forall x.P(x,y)$, $\forall x.x=\fromvar{x}{j}$, $\exists 
y.P(y,\tovar{x}{j})$. Instead $\forall \fromvar{x}{j}.P(\fromvar{x}{j})$ is not an $i$-formula as we do not allow quantification on arrow variables.  

A $i$-formula $\phi$ is \emph{closed}\index{closed formula} if it does not contain  arrow variables and all the occurrences of the variable $x$ in $\phi$ are in the scope of a quantifier $\forall x$ or $\exists x$. $\phi$ is {\em open}\index{open formula} if it is not closed.  A variable $x$ {\em occurs free} \index{free occurrence of a variable} in a formula if $x$ occurs in $\phi$ not in the scope of a quantifier $\forall x$ or $\exists x$. Notice that $x$, $\tovar{x}{i}$ and $\fromvar{x}{i}$ are different variables, and therefore $x$ does not occur free in an expression of type $p(\tovar{x}{i})$. The notation $\phi(\bx)$ \index{$\phi(\bx)$} is used to denote the formula $\phi$ and the fact that the free variables of $\phi$ are $\bx = \{x_1,\ldots, x_n\}$.

Languages $L_i$ and $L_j$ are not necessarily disjoint and the same formula $\phi$ can occur in different languages with different meanings. A \emph{labeled formula} is a pair $\co{\phi}{i}$%
\footnote{Similar notations are introduced in \protect{\cite{mccarthy26,gabbay11,subrahmanian1,dinsmore1,masini1}}.}
and is used to denote that $\phi$ is a formula in $L_i$. Given a set of $i$-formulas $\Gamma$, we use $\co{\Gamma}{i}$ as a shorthand for the set of labelled formulas $\{\co{\gamma}{i} | \gamma \in \Gamma\}$. 
Note that we do not admit formulas which are composed of symbols coming from different alphabets. Thus $\co{P(x)}{1}\wedge\co{a=b}{2}$ and $\forall x \co{P(x)}{1}$ are not well\ftext{-}formed labeled formulas in DFOL.

\begin{exmp}[Languages for the magic box]
	\label{ex:languages-example}
The DFOL languages $L_1$ and $L_2$ that describe the knowledge of $\obsuno$ and
$\obsdue$ in the magic box example are defined as follows.
\begin{itemize}
\item 
$L_1$ contains an infinite set of constants $\ball_1$, $\ball_2$,
$\ldots$ used to denote balls, two constants $l$ and $r$ used to indicate the left-hand side and right-hand
side positions in the box, the binary predicate $\inbox(x,y)$ which stands for ``the ball $x$ is
in the position $y$ of the box'', and the unary predicates $\bianco(x)$ and
$\nero(x)$ for ``the ball $x$ is white'' (resp. black). 
\item 
$L_2$ is obtained by extending $L_1$ with a new constant
$c$ for the centre position in the box. 
\end{itemize}
Examples of labeled formulas describing the knowledge of $\obsuno$ and
$\obsdue$ are:
\begin{itemize}
	\item ``According to $\obsuno$, ball $b_3$ is in the left slot of the box and ball $b_1$ is the same as ball $b_3$''
	$$\co{\inbox(b_3,l) \con b_1=b_3}{1}$$
\item \ftext{``According to $\obsdue$ all the balls inside the box are black''}
	$$\ftext{\co{\forall x(\forall y \  \inbox(x,y)\imp\nero(x))}{2}}$$ 
\end{itemize}
\end{exmp}

\subsection{Denoting cross-domain objects}
DFOL associates different domains of interpretation to the local knowledge bases; therefore it needs a mechanism to denote cross-domain identity.
Arrow variables provide such a mechanism, and are used to refer to counterpart objects which belong to other domains. In particular, arrow variables of the form $\fromvar{x}{j}$ and $\tovar{x}{j}$ occurring in a $i$-formula are used to denote an object in the domain of interpretation of $L_i$, which corresponds to the object denoted by $x$ in the domain of $L_j$. 

\begin{figure}[b]
  \centering
    \includegraphics[width=.65\textwidth]{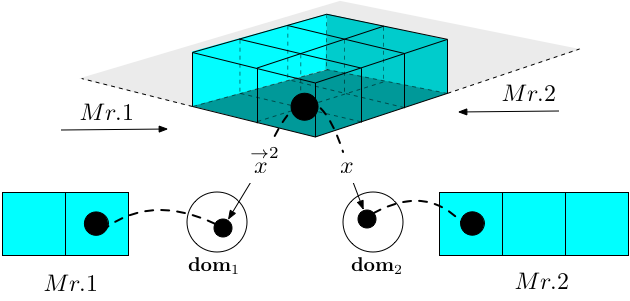}
  \caption{Denoting cross-domain objects in the magic box example.}
  \label{fig:figure_arrowVariablesSample}
\end{figure}
Consider, for instance, statement 3 at page \pageref{item:wp3}. The formalisation of this statement requires the ability to represent a ball that is seen by both observers. 
Since DFOL represents the partial viewpoints of \obsuno and \obsdue, each one with its own domain of interpretation, there is no object that directly represents a ball seen by both. Indeed, consider the black ball in the corner of the magic box represented at the top of Figure~\ref{fig:figure_arrowVariablesSample}. \obsuno and \obsdue have their own representation of this ball in their different domains, as graphically depicted at the bottom of Figure~\ref{fig:figure_arrowVariablesSample}. The way we represent the connection between these two different objects is by using an arrow variable, say $\tovar{x}{2}$, interpreted in the domain of \obsuno which \emph{corresponds} to the ball denoted by $x$ seen by \obsdue. We can then predicate that both $\tovar{x}{2}$ and $x$ are black using the formulas $\co{\textit{black}(\tovar{x}{2})}{1}$ and $\co{\textit{black}(x)}{2}$. The precise way in which DFOL binds the interpretation of $\tovar{x}{2}$ and $x$ in the different domains will become clear with the definition of Assignment (Definition~\ref{def:assignment}). 

The notion of arrow variable introduced here is connected to the notion of counterparts introduced by Lewis in~\cite{lewis12}. Roughly speaking, the language of Lewis' Counterpart Theory contains a binary predicate $C(x,y)$ meaning that $x$ is the counterpart of $y$, where $x$ and $y$ are supposed to denote two objects in two different possible worlds. 
In DFOL, we have local knowledge bases with different local languages instead of possible worlds. Therefore\ftext{,} we cannot
explicitly state that $x$ is counterpart of $y$, when $x$ and $y$ belong to two different languages\ftext{,} but only state it implicitly by means of arrow variables. That is, we can name in the language $L_i$ a counterpart of $x$ in $L_j$ by using the arrow variables $\fromvar{x}{j}$ and $\tovar{x}{j}$. 

\subsection{DFOL Semantics}
\label{sec:semantics}

\begin{figure}[htbp]
	\centering
	\includegraphics[width=.5\textwidth]{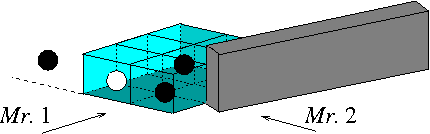}
	\caption{Partial knowledge in the magic box.}
	\label{fig:figure_wall}
\end{figure}

The semantics of a family of DFOL languages $\{L_i\}$ is defined by associating a set of interpretations, called \emph{local models}, 
to each $L_i$ in  $\{L_i\}$ and by relating objects in different domains via, so-called, \emph{domain relations}. 
This semantics is an extension of Local Models Semantics as defined in~\cite{ghidini10}.
If we look at the knowledge contained in a knowledge base $i$ we can distinguish three cases. First, $i$ can be complete, that is, for each formula $\phi \in L_i$ either $\phi$ or $\neg \phi$ belongs to the (deductive closure of the) knowledge base; second, it can be incomplete, if there exist at least a formula $\phi$ such that neither $\phi$ or $\neg \phi$ belongs to it; third, it can be inconsistent, that is, both $\phi$ and $\neg \phi$ belong to it. 
To represent these three possible statuses, each $i$ is associated with a (possibly empty) \emph{set} of local models. That is, each $i$ is associated with an epistemic state. A singleton corresponds to a complete KB, the empty set corresponds to an inconsistent KB, \ftext{whereas} all the other sets correspond to an incomplete KB. While completeness w.r.t. the entire language $L_i$ may be unrealistic, and even undesirable, it may be a good property to require for certain types of formulas, as we will see in the following paragraphs.      
To characterise the portion of knowledge upon which $i$ has complete knowledge we introduce the notion of complete sub-language $L_i^c$ and we restrict the definition of complete knowledge to the formulas of $L_i^c$. 
Let $L_i^c$ be a sub-language of $L_i$ built from a subset of constants, functional symbols, and predicate symbols of $L_i$, including equality, plus the set of arrow and not-arrow variables of $L_i$. We call $L_i^c$ the \emph{complete sub-language} of $L_i$. \emph{Complete terms} and \emph{complete formulas} are terms and formulas of $L_i^c$. Otherwise they are called \emph{non complete}. Note that in DFOL $L_i^c$ must contain the equality predicate as we impose that each $i$-th knowledge base is able to evaluate whether two objects are equal or not. Additional constants, functional symbols, or predicates can be added to $L_i^c$ to represent domain\ftext{-}specific complete knowledge. For instance, in the magic box example we may assume that $\obsuno$ and $\obsdue$ have complete knowledge about the position of the balls. That is, they know if a ball is in a slot or not. On the contrary, assume that $\obsdue$'s view over the box is partially concealed by a big wall, as depicted in Figure \ref{fig:figure_wall}. In this scenario $\obsdue$ is able to see one box sector and knows that there are two sectors behind the wall with balls inside and outside the box. In this case $\obsdue$ has complete knowledge about the left hand side position of the box but is uncommitted to whether there are balls in the sectors behind the wall. This is formalised by including the formulas $inbox(b,l)$ into $L_2^c$ for all the balls $b$ in the language of $\obsdue$, and by letting, e.g., sentences of the form $inbox(b,c)$ to be non complete, that is, true in some local model of $\obsdue$ and false in others. 

\begin{defn}[Set of Local Models]
	\label{def:set-of-models}
A \emph{set of local models} of $L_i$ is a set of first\ftext{-}order interpretations of $L_i$ on a (non empty) domain $\dom_i$, which agree on the interpretation of $L_i^c$, the complete sub-language of $L_i$. 
\end{defn}

The \emph{semantic overlap} between different knowledge bases is explicitly represented in DFOL by means of \emph{domain relations}.

\begin{defn}[Domain relation]
\label{def:domain-relation}
A \emph{domain relation} $r_{ij}$ from $\dom_i$ to $\dom_j$ is a binary relation contained in $\dom_i\times\dom_j$.
\end{defn}

We often use the simpler expression \emph{domain relation from $i$ to $j$} to denote a domain relation $r_{ij}$ from $\dom_i$ to $\dom_j$. We also use the functional notation $\dr_{ij}(d)$ to denote the set $\{d'\in\dom_j | \npla{d,d'}\in\dr_{ij}\}$.

A domain relation from $i$ to $j$ illustrates how the $j$-th knowledge base represents the domain of the $i$-th knowledge base in its own domain.
Therefore\ftext{,} a pair $\npla{d,d'}$ being in $\dr_{ij}$ means that, from the point of view of $j$, $d$ in $\bdom_i$ is the representation of $d'$ in $\bdom_j$. 
Thus, $\dr_{ij}$ formalises $j$'s subjective point of view on the relation between $\bdom_i$ and $\bdom_j$, and not an absolute and objective point of view; this implies that $\npla{d,d'}\in\dr_{ij}$ must not be read as if $d$ and $d'$ were the same object in a domain shared by $i$ and $j$. This latter fact could only be formalised by an external (above, meta) observer to both $i$ and $j$. 

Domain relations are not symmetric by default. This represents the fact that the point of view of $j$ over the domain of $i$ may differ from the point of view of $i$ over the domain of $j$, which may even not exist. For instance, in the mediator system example, it is plausible to impose that \mediator has a representation of the domains of \selleruno and \sellerdue, in its own domain while the opposite is prevented. Domain relations are conceptually analogous to \emph{conversion functions} between semantic objects, as defined in~\cite{sciore1}. 

Specific relations between the domains of different knowledge bases can be modelled by adding constraints \ftext{about} the form of $\dr_{ij}$. For instance, two knowledge bases with different but \emph{isomorphic} representations of the same domain
can be modelled by imposing $\dr_{ij}=\dr^{-1}_{ji}$. Likewise, completely unrelated domains can be represented by
imposing $\dr_{ij}=\dr_{ji}=\emptyset$. Transitive mappings between the domains of three knowledge bases $i$, $j$ and $k$ can be 
represented by imposing $\dr_{ik}=\dr_{ij}\circ\dr_{jk}$.
Moreover, if $\dom_{i}$ and $\dom_{j}$ are ordered according to two ordering relations $<_i$ and $<_j$ respectively, then a domain relation that
satisfies the following property
\begin{equation}
	\begin{split}
		\text{for all\ } & d_1,d_2\in\bdom_{i}, d_1<_id_2  \rightarrow\ \\
		& \text{for all\ } d_1'\in\dr_{ij}(d_1),\
		\text{for all\ } d_2'\in\dr_{ij}(d_2).\  d_1'<_j d_2'
	\end{split}	
\end{equation}
formalises a mapping which preserves the ordering. An example of this last property is a domain relation that captures a currency
exchange function. Further constraints on $\dr_{ij}$ are discussed in Section~\ref{sec:interpretation-cosntraints}.

\begin{defn}[DFOL Model]
\index{DFOL Model}
\label{def:fol-model}
A DFOL \emph{model}, or simply a {\em model} $\M$ (for $\{L_i\}$) is a pair $\M = \npla{\{M_i\},\{\dr_{ij}\}}$ 
where, for each $i, j \in I$, $M_i$ is a set of local models for $L_i$, and $\dr_{ij}$ is a domain relation from $i$ to $j$. 
\end{defn}

\begin{exmp}
\label{ex:dfol-model}
A DFOL model for the scenario shown in Figure~\ref{fig:magic-box} is a 4-tuple $\M=\npla{\{m_1\},\{m_2\},\dr_{12},\dr_{21}}$,
where $m_1=\npla{\{{\it left},{\it right},a,b,c\},\cdot^{\I_1}}$ and $m_2=\npla{\{{\it left}, {\it centre}, \-{\it right},a,b,c,d\},\cdot^{\I_2}}$ with 
$$ 
{\scriptstyle \cdot^{\I_1} = \left\{
\begin{array}{l}
b_1^{\I_1} = a \quad 
b_2^{\I_1} = b \quad
b_3^{\I_1} = c  \\
{\it inbox}^{\I_1} = \{\npla{b,{\it left}},\npla{c,{\it right}}\} \\ 
{\it black}^{\I_1} = \{a,c\} \\
{\it white}^{\I_1} = \{b\}
\end{array}
\right.
\quad \
\cdot^{I_2} = \left\{
\begin{array}{l}
b_1^{I_2} = a \quad
b_2^{I_2} = b \\
b_3^{I_2} = c \quad 
b_4^{I_2} = d  \\
{\it inbox}^{I_2} = \{\npla{a,{\it left}},\npla{b,{\it right}}\} \\ 
{\it black}^{I_2} = \{a,b,d\} \\ 
{\it white}^{I_2} = \{c\}
\end{array}
\right.}
$$
Moreover, $\dr_{12}=\{(c,a)\}$ and  $\dr_{21}=\{(a,c)\}$. 
\end{exmp}

\begin{defn}[Assignment]
	\label{def:assignment}
Let $\M = \npla{\{M_i\},\{\dr_{ij}\}}$ be a model for $\{L_i\}$ and $X_i$ be a set containing all the non-arrow variables plus a subset of the arrow variables of $L_i$. An \emph{assignment} $a$ is a family $\{a_i\}$ of functions $a_i$ from $X_i$ to $\dom_i$ which satisfies the following: 
\begin{enumerate}[(i)]  
\item if $a_i(\tovar{x}{j})$ is defined, then $a_i(\tovar{x}{j}) \in \dr_{ij}^{-1}(a_j(x))$;
\item if $a_i(\fromvar{x}{j})$ is defined, then $a_i(\fromvar{x}{j})\in\dr_{ji}(a_j(x))$. 
\end{enumerate}
\end{defn}
The definition above extends the classical notion of assignment given for first\ftext{-}order logic to deal with extended variables. 
Intuitively, if the non-arrow variable $x$ occurring in the $j$-th knowledge base is a placeholder for the element $d\in\bdom_j$, then the occurrence of the arrow variable $\tovar{x}{j}$ in a formula of the $i$-th knowledge base is a placeholder for an element $d'\in\bdom_i$ which is a pre-image (via $\dr_{ij}^{-1}$) of $d$. Analogously\ftext{,} the arrow variable $\fromvar{x}{j}$ occurring in $\co{\psi}{i}$ is a placeholder for any element $d''\in\dom_i$ which is an image (via $\dr_{ji}$) of $d$.

An assignment $a'$ is an \emph{extension} of $a$, in symbols $a\leq a'$, if $a_i(v)=d$ implies $a_i'(v)=d$ for all the non-arrow and arrow variables $v$. Notationally, given an assignment $a$, a (non-arrow or arrow) variable $x$, and an element $d\in\dom_i$, we denote with $a(x:=d)$ the assignment obtained from $a$ by letting $a_i(x)=d$.

\begin{defn}[Admissible assignment]
An assignment $a$ is \emph{(strictly) admissible for a formula} $\co{\phi}{i}$ if $a_i$ assigns all (and only) the arrow variables
occurring in $\phi$. $a$ is (strictly) admissible for a set of formulas $\Gamma$ if it is (strictly) admissible for all $\co{\phi}{j}$ in $\Gamma$.
\end{defn}

\begin{defn}[Satisfiability]
\label{def:satisfiability-with-arrow-variables}
A formula $\co{\phi}{i}$ is \emph{satisfied} by a DFOL model $\M$ w.r.t.~the assignment $a$, in symbols $\M\models\co{\phi}{i}[a]$, if 
\begin{enumerate}[(i)]
\item\label{item:satisfiability-admissibility} $a$ is admissible for $\co{\phi}{i}$; and 
\item\label{item:satisfiability-sets} for all $m\in M_i$, $m\models\phi[a_i]$ according to the classic definition of 
 first\ftext{-}order satisfiability.
\end{enumerate}
$\M\models\Gamma[a]$ if\ftext{,} for all $\co{\phi}{i}\in\Gamma$, $\M\models\co{\phi}{i}[a]$.
\end{defn}

With an abuse of notation we use the symbol $\models$ of satisfiability to denote both first\ftext{-}order satisfiability and DFOL satisfiability. The context will always make clear the distinction between the two.

If we compare satisfiability of a formula in a DFOL model with the standard notion of satisfiability of a first\ftext{-}order formula in a first\ftext{-}order model we can observe three differences: first, assignments do not force all arrow variables to denote objects in the domain; second\ftext{,} we admit partial knowledge as we evaluate the satisfiability of a formula in a \emph{set} of local models, rather than into a single one; third, we admit \emph{islands of inconsistency}, by allowing some $M_i$ to be empty. In the following we analyse these three aspects one by one. 

\subsubsection{Satisfiability and arrow variables}
\label{sec:satisfiability-and-arrow-variables}
Definition \ref{def:assignment} requires assignments to be defined for all non-arrow variables, but not necessarily for all arrow variables.\footnote{This, in order \ftext{to not} constrain the existence of pairs in the domain relation, if not required by explicit bridge rules which we will introduce in Section \ref{subsec:interpretations-constraints}.}
To avoid many of the ontological issues raised by free logics~\cite{bencivenga1}, where special truth conditions are given for $\phi(t)$ when $t$ does not denote any object in the domain, condition \eqref{item:satisfiability-admissibility} in Definition~\ref{def:satisfiability-with-arrow-variables} guarantees that satisfiability of $\co{\phi}{i}$ is defined over admissible assignments for $\co{\phi}{i}$. This provides the first difference between satisfiability in DFOL and satisfiability in first\ftext{-}order logic, whose consequences are highlighted in the proposition below.

\begin{prop} 
	\label{prop:satisfiability-and-arrow-variables}
	Let $\bothvar{x}$ denote either $\tovar{x}{j}$ or
  $\fromvar{x}{j}$ for some $j\neq i$, and $\M$ be a 
  DFOL model such that $M_i$ contains a single first\ftext{-}order model $m$. Then the following properties hold: 
\begin{enumerate}[(i)]
	\item\label{th:satisfiability-1} if $a$ is admissible for $\phi$, then $\M\models\co{\phi}{i}[a]$ if and only if $m\models\phi[a_i]$;
	\item\label{th:satisfiability-2} if $a$ is not admissible for $\co{\phi}{i}$, then $\M\not\models\co{\phi}{i}[a]$ and $\M\not \models \co{\neg\phi}{i}[a]$; 
	\item \label{th:satisfiability-exist} if $a_i(\bothvar{x})$ is not defined, then $\M\not\models\co{\exists y.y=\bothvar{x}}{i}[a]$; 
	\item \label{th:satisfiability-3} $\M\models\co{\forall x.\phi(x)}{i}[a]$ does not imply that $\M\models\co{\phi(\bothvar{x})}{i}[a]$ for an arbitrary arrow variable $\bothvar{x}$;
	\item \label{th:satisfiability-4} $\M\models\co{\neg\phi}{i}[a]$ (resp., $\M\models\co{\psi}{i}[a]$) does not imply that $\M\models\co{\phi\imp\psi}{i}[a]$; 
	\item\label{th:satisfiability-6} $\M\models\co{\phi}{i}[a]$ does not imply that $\M\models\co{\phi\vee\psi}{i}[a]$;
	\item\label{th:satisfiability-7} if $\M\models\co{\phi\imp\psi}{i}[a]$, then $\M\models\co{\phi}{i}[a]$ implies that $\M\models\co{\psi}{i}[a]$;
	\item\label{th:satisfiability-8} $\M\models\co{\phi(\bothvar{x})}{i}[a]$ implies that $\M\models\co{\exists x.\phi(x)}{i}[a]$.
\end{enumerate}
\end{prop}

Property~\eqref{th:satisfiability-1} shows that DFOL satisfiability and first\ftext{-}order logic satisfiability coincide when $M_i$ is a single first\ftext{-}order model, provided that $a$ is admissible for $\phi$.  
Property~\eqref{th:satisfiability-2} states that $\M$ does not satisfy any formula containing arrow variables $\bothvar{x}$ which are not assigned by $a$, including formulas which have the form of classical tautologies.  Property~\eqref{th:satisfiability-exist} shows that the existence of an individual equal to $\bothvar{x}$ is not always guaranteed in DFOL. Another important difference w.r.t. satisfiability in first\ftext{-}order logic is the fact that a universally quantified variable cannot be instantiated to an arbitrary term that contains arrow variables (property~\eqref{th:satisfiability-3}). The term must contain arrow variables $\bothvar{x}$ that are assigned to some value by $a$. Properties~\eqref{th:satisfiability-4}--\eqref{th:satisfiability-6} state  that the ``introduction'' of classical connectives in a formula cannot be done according to the rules for propositional logic, since extending a formula with new terms may introduce new arrow variables not assigned by $a$.
Finally, properties \eqref{th:satisfiability-7} and \eqref{th:satisfiability-8} provide examples of first\ftext{-}order properties which still hold in DFOL. In particular \eqref{th:satisfiability-7} shows that modus ponens is a sound inference rule for satisfiability in DFOL, while property \eqref{th:satisfiability-8} shows that if $\phi$ holds for a certain arrow variable $\bothvar{x}$, then there is an object of the world (i.e., $\exists x$) such that $\phi$ holds for it. 
All the above properties are consequences of the fact that $\M\models i:\phi[a]$ does not only mean that all the models $m$ in $M_i$ satisfy $\phi$, but also that the arrow variables contained in $\phi$ actually denote elements in $\bdom_i$. 

\subsubsection{Satisfiability in a set of local models}
\label{sec:satisfiability-and-local-models}
Interpreting each $L_i$ into a set of models, rather than into a single model, enables the formalisation of \emph{partial} knowledge about values of terms and about truth values of formulas, as informally described at page \pageref{def:set-of-models}. Proposition~\ref{prop:satisfiability-and-local-models} describes the main effects of partial knowledge on the notion of satisfiability in DFOL.  

\begin{prop} 
	\label{prop:satisfiability-and-local-models}
	Let $t$ be a non-complete term and $\phi$ and $\psi$ be non-complete formulas
	of $L_i$ which do not contain arrow variables. There \ftext{exist} a DFOL model $\M$ and an assignment $a$ such as: 
\begin{enumerate}[(i)]
	\item\label{th:satisfiability-local-models-1} $\M\not\models\co{x=t}{i}[a]$;
	\item\label{th:satisfiability-local-models-2} $\M\models\co{\phi\vee\psi}{i}[a]$ but neither $\M\models\co{\phi}{i}[a]$ nor $\M\models\co{\psi}{i}[a]$;
	\item\label{th:satisfiability-local-models-3} $\M\models\co{\exists x.\phi(x)}{i}[a]$ but there is no $d\in\dom_i$ with $\M\models\co{\phi(x)}{i}[a(x:=d)]$. 
\end{enumerate}
\end{prop}

Properties~\eqref{th:satisfiability-local-models-1} and~\eqref{th:satisfiability-local-models-2} emphasise that the value of non-complete terms and of disjuncts of non\ftext{-}complete formulas can be undetermined. An interesting instance of property~\eqref{th:satisfiability-local-models-2} is when $\psi = \neg \phi$. In this case neither $\M\models\co{\phi}{i}[a]$ nor $\M\models\co{\neg \phi}{i}[a]$, as in property~\eqref{th:satisfiability-2} of  Proposition~\ref{prop:satisfiability-and-arrow-variables}, but for a different reason: Proposition~\ref{prop:satisfiability-and-arrow-variables} states that a model $\M$ does not satisfy a formula and its negation if assignment $a$ is not complete for that formula. Instead, Proposition~\ref{prop:satisfiability-and-local-models} states that $\M$ does not satisfy a formula and its negation because it contains two local models\ftext{,} one satisfying $\phi$ and the other satisfying $\neg \phi$. Finally, property~\eqref{th:satisfiability-local-models-3} states that the value of an existentially quantified variable can be unknown in a given knowledge base. 

Satisfiability of complete formulas w.r.t. a set of local models shares the same properties of satisfiability w.r.t. a single local model. This is a consequence of the fact that complete formulas are interpreted in the same way in all the local models in $M_i$. Thus, Proposition~\ref{prop:satisfiability-and-local-models} does not hold for complete formulas. 

\begin{prop}
\label{prop:complete-equality}
Let $t$ be a complete term and $\phi$ and $\psi$ be complete formulas
of $L_i$ which do not contain arrow variables. For all models $\M$: 
\begin{enumerate}[(i)]
\item\label{th:satisfiability-local-models-1-bis} there is an assignment $a$ such that $\M\models\co{x=t}{i}[a]$;  
\item\label{th:satisfiability-local-models-2-bis} for all assignments $a$, $\M\models\co{\phi\vee\psi}{i}[a]$ iff $\M\models\co{\phi}{i}[a]$ or $\M\models\co{\psi}{i}[a]$;
\item\label{th:satisfiability-local-models-3-bis} for all assignments $a$, $\M\models \co{\exists x \phi(x)}{i}[a]$ iff for some $d\in\dom_i$
   $\M\models\co{\phi(x)}{i}[a(x:=d)]$.  
\end{enumerate}
\end{prop}

\subsubsection{Local inconsistency}
Models $\M$ where $M_i = \emptyset$ and $M_j \neq \emptyset$ formalise the idea of \emph{local inconsistency} of the $i$-th knowledge base. 
That is, of a situation where one (or more) inconsistent knowledge base can coexist with consistent ones. This basic property of local inconsistence is formally described by the following proposition: 

\begin{prop} 
	\label{prop:local-inconsistency}
Let $\{L_i\}$ be a family of first\ftext{-}order languages. There exists a DFOL model $\M$ for $\{L_i\}$ such that $\M\models\co{\bot}{i}$ but $\M\not\models\co{\bot}{j}$. 
\end{prop}
To prove this statement consider a trivial model $\M$ with $M_i=\emptyset$ and $M_j\not=\emptyset$.

\subsection{Denoting cross-KB constraints via bridge rules}
\label{subsec:interpretations-constraints}

The DFOL language described so far is able to represent the different local KBs, but cannot be used to express formulas spanning over different knowledge bases. {We enrich DFOL with this ability by introducing} a class of ``cross language formulas''. These formulas are an extension of the notion of \emph{bridge rule}, first introduced in~\cite{giunchiglia38} in a proof-theoretic setting. 

\begin{defn}[Bridge rule]
	\label{def:bridge-rule}
Given $i, i_1, \ldots, i_n \in I$, a \emph{bridge rule from $i_1,\dots,i_n$ to $i$} is an expression of the form 
$\co{\phi_1}{i_1},\ldots,\co{\phi_n}{i_n}
\rightarrow\co{\phi}{i}$.
\end{defn}
A bridge rule can be seen as an axiom spanning between different logical theories (the local knowledge bases); it restricts
the set of possible DFOL models to those in which $\co{\phi}{i}$ is a logical consequence of $\co{\phi_1}{i_1}, \dotsc, \co{\phi_n}{i_n}$. We call $\co{\phi_1}{i_1},\ldots,\co{\phi_n}{i_n}$ the premises of the rule and $\co{\phi}{i}$ the conclusion. As an example, the bridge rule
$$
\co{\inbox(x,r)}{1} \rightarrow  \co{\exists y \ \inbox(\fromvar{x}{1},y)}{2} 
$$
represents the fact that the rightmost ball seen by $\obsuno$ inside the box is seen also by $\obsdue$. 


\begin{defn}[Satisfiability of bridge rules]
	\label{def:satisfiability-bridge-rules}
A model $\M$ \emph{satisfies} a bridge rule
$\co{\phi_1}{i_1},\ldots,\co{\phi_n}{i_n}
\rightarrow\co{\phi}{i}$ if for all the assignments $a$ strictly admissible for
$\co{\phi_1}{i_1},\ldots,\co{\phi_n}{i_n}$ the following holds:
\begin{equation*}
\begin{split}
	&  \text{if } \M\models i_1:\phi_1[a], \ldots, \M\models i_n:\phi_n[a] \text{ then } \\ 
	& \quad \text{there is an extension } a'\geq a, 
	 \text{ admissible for } \co{\phi}{i}, \text{ such that } \M\models i:\phi[a'].
\end{split}	
\end{equation*}
\end{defn}
Given a set of bridge rules $\IC$ on the family of languages $\{L_i\}$, a \emph{$\IC$-model}
is a DFOL model for $\{L_i\}$ that satisfies all the bridge rules of $\IC$.

Definition~\ref{def:satisfiability-bridge-rules} enables us to illustrate the difference between $\co{\phi \imp \psi}{i}$ and $\co{\phi}{i}\rightarrow\co{\psi}{i}$. Let us disregard here the requirement of the existence of $a'$ extension of $a$. $\M \models \co{\phi \imp \psi}{i}$ is satisfied if all local models $m_i \in \M$ satisfy $\phi \imp \psi$. Instead, $\M \models \co{\phi}{i}\rightarrow\co{\psi}{i}$ is satisfied if, whenever all local models $m_i \in \M$ satisfy $\phi$ it is also the case that all the local models $m_i \in \M$ satisfy $\psi$. This difference is analogous to the one between $\Box(\phi \imp \psi)$ and $\Box \phi \imp \Box \psi$ in modal logic.

Bridge rules, together with arrow variables, are used to relate cross\ftext{-}domain objects and knowledge. We illustrate this with the help of simple bridge rules, together with their intuitive reading: 
\begin{align}
	\label{eq:simple-br1}
\co{P(\tovar{x}{j})}{i}\rightarrow\co{Q(x)}{j} \quad & \parbox[t]{.55\textwidth}{Every object of $\dom_j$, that is a translation of an object of $\dom_i$ that has property $P$, has property $Q$\ftext{.}}\\
	\label{eq:simple-br2}
\co{P(x)}{i}\rightarrow\co{Q(\fromvar{x}{i})}{j} \quad & \parbox[t]{.55\textwidth}{Every object of $\dom_i$ that has property $P$ can be translated into an object of $\dom_j$ that has property $Q$\ftext{.}}\\ 
	\label{eq:simple-br3}
\co{Q(\fromvar{x}{i})}{j}\rightarrow\co{P(x)}{i} \quad & \parbox[t]{.55\textwidth}{Every object of $\dom_i$, that is translated into an object of $\dom_j$ that has property $Q$, has property $P$\ftext{.}}\\
	\label{eq:simple-br4}
\co{Q(x)}{j}\rightarrow\co{P(\tovar{x}{j})}{i} \quad & \parbox[t]{.55\textwidth}{Every object of $\dom_j$ that has property $Q$ is the translation of some object of $\dom_i$ that has property $P$\ftext{.}}
\end{align}

The intuitive (and formal) reading of bridge rules~\eqref{eq:simple-br1}--\eqref{eq:simple-br4} (and of bridge rules in general) can be expressed also in terms of query containment, given the appropriate transformation via domain relation. Let $\ext{P}{i}$ be the answer of query $P(x)$ to a database $i$, then bridge rules~\eqref{eq:simple-br1}--\eqref{eq:simple-br4} can be read as:

\noindent
\centerline{\scalebox{.9}
{\begin{equation*}
	\dr_{ij}(\ext{P}{i})\subseteq\ext{Q}{j}\quad \quad
	\ext{P}{i}\subseteq\dr_{ij}^{-1}(\ext{Q}{j})\quad \quad
	\ext{P}{i}\supseteq\dr_{ij}^{-1}(\ext{Q}{j})\quad \quad
	\dr_{ij}(\ext{P}{i})\supseteq\ext{Q}{j}
\end{equation*}}
}

\vspace{.3cm}
Definition \ref{def:satisfiability-bridge-rules} states that a bridge rule is satisfied if \emph{for all} the assignments $a$ strictly admissible for the premises of the rule,  \emph{there exists} an extension $a'$ of $a$ admissible for the conclusion. This implies that arrow variables occurring in the premise of a bridge rule are intended to be \emph{universally quantified}, while arrow variables occurring in the consequence of a bridge rule are intended to be \emph{existentially quantified}. 
In other words, if we use an arrow variable in the consequence of a bridge rule we impose the existence of certain mappings between domains. This happens in \eqref{eq:simple-br2}, where every element of $P$ must have at least one translation into $Q$ (via $\dr_{ij}$), and in \eqref{eq:simple-br4}, where every element that is $Q$ has at least a pre-image in $P$ (via $\dr_{ij}$). Conversely, if we use arrow variables in the premise of a bridge rule we restrict the way domain relations can map elements of the different domains without imposing the existence of certain mappings. This happens in \eqref{eq:simple-br1}, where the elements of $P$ are not forced to have a translation into some elements of $Q$, and in \eqref{eq:simple-br3}, where the elements of $Q$ are not forced to be the translation of some element of $P$.

%

\begin{defn}[Logical Consequence]
\index{Logical consequence in DFOL}
\label{def:logical-consequence}
$\co{\phi}{i}$ is a \emph{logical consequence} of a set of formulas $\Gamma$ w.r.t. a set of bridge rules $\IC$, in symbols
$\Gamma\models_{\IC}i:\phi$, if for all the $\IC$-models $\M$ and for all the assignments $a$, strictly admissible for $\Gamma$, the following holds: 
\begin{equation*}
\begin{split}
	& \text{if } \M\models\Gamma[a], \text{ then}\\
	& \quad \text{there is an extension } a'\geq a, \text{ admissible for } i:\phi \text{ such that } \M\models i:\phi[a'].
\end{split}	
\end{equation*}
\end{defn}

DFOL logical consequence bears similarities and differences w.r.t.~logical consequence for first\ftext{-}order logic. Focusing on the similarities\ftext{,} we can observe that if we restrict to a single knowledge base  $i$, and we consider a fixed set of arrow variables, for which we assume the existence of an admissible assignment, then the behaviour of logical consequence in DFOL turns out to be similar to that of first\ftext{-}order logic, as shown by the following proposition:

\begin{prop}[Basic properties of logical consequence]\ 
\label{prop:logical-consequence}
\begin{enumerate}[(i)]
\item\label{th:reflexivity} \textbf{Reflexivity}: $\Gamma,\co{\phi}{i}\brmodels\co{\phi}{i}$;
\item\label{th:weak-monotonicity} \textbf{Weak monotonicity}: if $\Gamma\brmodels\co{\phi}{i}$, and $\Sigma$ is a set of formulas whose arrow variables either occur in
  $\Gamma$ or do not occur in $\co{\phi}{i}$, then $\Gamma,\Sigma\brmodels\co{\phi}{i}$;
\item\label{th:cut} \textbf{Cut}: if $\Gamma\brmodels\co{\phi}{i}$ and $\Gamma,\co{\phi}{i}\brmodels\co{\psi}{j}$, then $\Gamma\brmodels\co{\psi}{j}$; 
\item\label{th:fol-lc-conservative-extension} \textbf{Extension of first\ftext{-}order logical consequence}: 
   Let \IC be an empty set of bridge rules, and $\Gamma$ be a set of $i$-formulas. We have that 
   \begin{equation}
   \label{eq:fol-lc-conservative-extension}
   \Gamma\models\phi \quad \text{if and only if} \quad \co{\bigwedge_{k=1}^{n}\left(\exists
   y_k.y_k=\tovar{x_k}{}\right)}{i},\co{\Gamma}{i}\brmodels\co{\phi}{i}
   \end{equation}
   where \ $\tovar{x_1}{},\dots,\tovar{x_n}{}$ are the arrow variables occurring in $\phi$ but not in $\Gamma$, and $\co{\Gamma}{i}$ is used to denote the set $\{\co{\phi}{i}| \phi \in \Gamma\}$. If there are no arrow variables occurring only in $\phi$ and not in $\Gamma$, then \eqref{eq:fol-lc-conservative-extension} reduces to 
 \begin{center}
	$\Gamma\models\phi$ if and only if $\co{\Gamma}{i}\brmodels\co{\phi}{i}.$
 \end{center}
\end{enumerate}
\end{prop}

\begin{pf} 
	Properties \eqref{th:reflexivity}--\eqref{th:cut} are easy consequence\ftext{s} of Definition \ref{def:logical-consequence}. 
	Concerning item \eqref{th:fol-lc-conservative-extension}, we prove here the simplified version $\Gamma\models\phi$ if and only if $\co{\Gamma}{i}\brmodels\co{\phi}{i}$. The proof of the general case shown in Equation \eqref{eq:fol-lc-conservative-extension} is similar.  
	\begin{itemize}
		\item The fact that $\Gamma\models\phi$ implies $\co{\Gamma}{i}\brmodels\co{\phi}{i}$ is an easy consequence of the fact that each $M_i$ is a set of first\ftext{-}order models $m$ for $L_i$. 
		\item Assume that $\co{\Gamma}{i}\brmodels\co{\phi}{i}$. Since $\IC$ is empty, $\phi$ does not contain new arrow variables, and since $\Gamma$ is a set of $i$-formulas, we can rewrite Definition \ref{def:logical-consequence} as: for all DFOL models $\M$, $M_i \models \co{\Gamma}{i}[a]$ implies $M_i \models \co{\phi}{i}[a]$. Let $m$ be an arbitrary first\ftext{-}order model for $L_i$. Among all the possible DFOL models there is surely one such that $M_i=\{m\}$. Thus $m\models \Gamma$ implies $m\models \phi$ and $\Gamma \models \phi$.
	\end{itemize}	
\end{pf}

The key point in proving that $\co{\Gamma}{i}\brmodels\co{\phi}{i}$ implies  $\Gamma\models\phi$ is the fact that we can consider arbitrary DFOL models, and therefore also models such that $M_i =\{m\}$. This assumption cannot be made when $\IC$ is not empty, as we need to restrict to specific classes of $\IC$-models. In other words, as soon as we consider different local knowledge bases, which interact via bridge rules, the behaviour of logical consequence in DFOL differs from that of logical consequence in first\ftext{-}order logic, even if we restrict to ``safe'' sets of arrow variables or no arrow variables at all. An important difference with first\ftext{-}order logic is given by the fact that the deduction theorem does not hold in the general case:

\begin{prop} 
	\label{prop:no-deduction-theorem}
  Let $\IC$ be an arbitrary set of bridge rules, and $\co{\phi}{i}$ be a formula whose arrow variables occur entirely in $\Gamma$. $\Gamma,\co{\phi}{i}\brmodels\co{\psi}{i}$ does not imply $\Gamma\brmodels\co{\phi\imp\psi}{i}$. 
\end{prop}
\begin{pf}
	Let us assume that $\Gamma,\co{\phi}{i}\brmodels\co{\psi}{i}$ holds and let us pick a $\IC$-model $\M$ such that $\Gamma\brmodels\co{\phi\imp\psi}{i}$ does not hold. 
	In particular let model $\M$ be a \IC-model such that $\M \brmodels \Gamma$ but $\M \not \brmodels \co{\phi}{i}$. Assume in particular that $M_i$ contains two local models $m_1$ and $m_2$ such that $m_1 \models \phi$, $m_2 \models \neg \phi$, and both $m_1$ and $m_2$ satisfy $\neg \psi$. Since $\IC$ is an arbitrary set of bridge rules we are guaranteed that we can perform this construction.   
	Model $\M$ is the counterexample we need to falsify $\Gamma\brmodels\co{\phi\imp\psi}{i}$. In fact, it satisfies $\Gamma$ but falsifies $\phi\imp\psi$ because of $m_1$. 
\end{pf}
Note that, if $\co{\phi}{i}$ is a complete formula, or the class of $\IC$-models are such that all $m \in M_i$ satisfy $\phi \imp \psi$, the counter-example shown in the proof above cannot be built and we can prove that the deduction theorem holds (modulo arrow variables) using property~\eqref{th:fol-lc-conservative-extension} in Proposition \ref{prop:logical-consequence}.
We can therefore conclude that bridge rules, used together with assumptions which consist of partial knowledge, are the reason of the failure of the deduction theorem in DFOL. 

Another important characteristics of logical consequence in DFOL is the fact that it preserves local inconsistency, without making it global. 
\begin{prop}
	\label{prop:inconsistency-does-not-propagate}
	Let $\IC$ be an arbitrary set of bridge rules, $\co{\bot}{i}\not\brmodels\co{\bot}{j}$. 
\end{prop}

Since $\IC$ is an arbitrary set of bridge rules, we can assume that the model used to validate Proposition~\ref{prop:local-inconsistency} is a $\IC$-model. Thus $\co{\bot}{i}\not\brmodels\co{\bot}{j}$. 

Finally, from the definition of admissible assignment, we can see that an arrow variable $\tovar{x}{j}$ which occur in an $i$-formula represents the pre-image (via $\dr^{-1}_{ij}$) of a variable $x$ in $j$, while an arrow variable $\fromvar{y}{i}$ occurring in a formula with index $j$ represents an image of $y$ in
$i$ (again via $\dr_{ij}$).  This means that if $y=\tovar{x}{j}$ holds in $i$ then $\fromvar{y}{i}=x$
holds in $j$. A similar property holds for $\dr_{ji}$.

\begin{prop}
	\label{prop:logical-consequence-arrow-variable}
  $\co{y=\tovar{x}{j}}{i}\brmodels\co{\fromvar{y}{i}=x}{j}$ and $\co{x=\fromvar{y}{i}}{j}\brmodels\co{\tovar{x}{j}=y}{i}$. 
\end{prop}

\onlyintechrep{\begin{pf}
	Let $\M$ be an $\IC$-model. and $a$ be an \ftext{assignment} admissible for $\co{y=\tovar{x}{j}}{i}$ such that $\M \models \co{y=\tovar{x}{j}}{i}[a]$. We need to show that: (i) there exist an assignment $a'$ extension of $a$ admissible for $\co{\tovar{x}{j}=y}{i}$ and (i) $\M \models \co{\fromvar{y}{i}=x}{j}[a']$.
\begin{itemize}
	\item Existence of $a'$. Since $\M \models \co{y=\tovar{x}{j}}{i}[a]$, we have that $a_i(y)=a_i(\tovar{x}{j})$. From the definition of assignment (item (i) in Definition~\ref{def:assignment}) we know that $a_i(\tovar{x}{j}) \in r^{-1}_{ij}(a_j(x))$, that is, $r_{ij}(a_i(\tovar{x}{j}))=a_j(x)$. Let us define $a'$ as the extension of $a$ such that $a'_j(\fromvar{y}{i}) = a_j(x)$. 
	Since $a$ was strictly admissible for $\co{y=\tovar{x}{j}}{i}$, $a'_j(\fromvar{y}{i})$ is the only new value we need to add to $a$ to make it admissible for $\co{\tovar{x}{j}=y}{i}$. We need to show that $a'$ is an assignment, that is, it satisfies condition (ii) in Definition~\ref{def:assignment}. 
	This condition requires that $a'_j(\fromvar{y}{i}) \in r_{ij}(a'_i(y))$. Since we have defined $a'_j(\fromvar{y}{i}) = a_j(x)$ and $a'_i(y) = a_i(y) = a_i(\tovar{x}{j})$, we can rewrite condition (ii) as $a_j(x) \in r_{ij}(a_i(\tovar{x}{j}))$. Since we know (see above) that $r_{ij}(a_i(\tovar{x}{j}))=a_j(x)$, $a'$ satisfies condition (ii) of Definition~\ref{def:assignment}. 
	\item $\M \models \co{\fromvar{y}{i}=x}{j}[a']$. Immediately follows from the definition of $a'$.
\end{itemize}

The proof of statement $\co{x=\fromvar{y}{i}}{j}\brmodels\co{\tovar{x}{j}=y}{i}$ is analogous and is left as an exercise. 
\end{pf}}

Note that the proposition above\onlyinpaper{, whose proof can be found in \cite{DFOL-techrep},} states a logical property of arrow variables which depends upon the semantics of arrow variables, and not upon the form of the domain relation. Additional logical properties involving arrow variables, instead, hold for specific sets of  domain relations. These will be illustrated in the next section. 

Finally, bridge rules enjoy the so\ftext{-}called \emph{directionality property}\ftext{. N}amely they allow to transfer knowledge from the premises to the conclusion with no back-flow of knowledge in the opposite direction. More formally: given a set $\BR$ of bridge rules such that $k$ does not appear in the conclusion of a bridge rule neither as the index of the conclusion no\ftext{r} as an index of an arrow variable, then $\Gamma\models_{\BR} k:\phi$ iff $\Gamma_k\models\phi$.
%
The proof of this statement is given in Section~\ref{sec:calculus} in a proof theoretical manner (see Proposition~\ref{prop:directionality}). 

We conclude the presentation of the semantics of DFOL by showing how we can use it to formalise the Magic box scenario and the Mediator scenario introduced in Section~\ref{sec:examples}.

\begin{exmp}[A formalisation of the magic box]
	\label{ex:formalization-magic-box}
	We start from the languages $L_{1}$ and $L_2$ defined in Example~\ref{ex:languages-example}. We also require that both the observers have complete knowledge on their views and therefore we impose that $L_i^c = L_i$ with $i = 1,2$. Local axioms are used to represent the facts that are true in the views of the observers. Examples of local axioms of
	$\obsuno$ and $\obsdue$ follow, where $\bempty(\positions)$ is a shorthand for $\forall x \neg inbox(x,\positions)$ for a given position ``\positions'', and $l,c,r$ are shorthands for ``$left$'', ``$center$'', and ``$right$'', respectively. 
	{\small
	\begin{align}
	\label{eq:mr1-sees-two-slots}
	 & \co{\forall x \forall y (\inbox(x,y)\imp y=\mathit{l}\vee y=\mathit{r})}{1} \\ 
	\label{eq:mr2-sees-three-slots}
	 & \co{\forall x \forall y (\inbox(x,y)\imp y=\mathit{l}\vee y=\mathit{c}\vee y=\mathit{r})}{2}	\\
		\label{eq:configuration-mr1}
		\begin{split}
			1\!: & \exists x (\inbox(x,r) \con \bempty(l)) \vee  \exists x (\inbox(x,l) \con \bempty(r)) \vee \\
			    & \exists x \exists y (\neg (x = y) \con \inbox(x,l) \con \inbox(y,r))	\vee 
			   (\bempty(l) \con \bempty(r))	
		\end{split}
	\end{align}}%
	Axioms \eqref{eq:mr1-sees-two-slots} and \eqref{eq:mr2-sees-three-slots} describe that $\obsuno$ and $\obsdue$ see two and three slots, respectively. Axiom~\eqref{eq:configuration-mr1} describes all the possible configurations of the slots of the box as seen by $\obsuno$.  
		 
	Bridge rules are used to formalise the relation between $\obsuno$'s
	and $\obsdue$'s knowledge on their respective views. A first group of
	bridge rules formalises that: (i) the rightmost ball seen by $\obsuno$ in the box is seen also by $\obsdue$, and (ii) the leftmost ball seen by $\obsdue$ in the box is seen also by $\obsuno$:  
	\begin{align}
	  \label{eq:mbox-inboxright1-imp-existence2}
	  \co{\inbox(x,r)}{1} &\rightarrow  \co{\exists y \ \inbox(\fromvar{x}{1},y)}{2} \\
	  \label{eq:mbox-inboxleft-and-empty-right1-imp-existence2}
	  \co{\inbox(x,l)\wedge\bempty(r)}{1} & \rightarrow  \co{\exists y \ \inbox(\fromvar{x}{1},y)}{2} \\
	  \label{eq:mbox-inboxleft2-imp-existence1} 
	  \co{\inbox(x,l)}{2} &\rightarrow  \co{\exists y\ \inbox(\fromvar{x}{2},y)}{1} \\
	  \label{eq:mbox-inboxcenter-and-empty-left2-imp-existence1}
	  \co{\bempty(l)\wedge\inbox(x,c)}{2} & \rightarrow  \co{\exists y\ \inbox(\fromvar{x}{2},y)}{1} \\
	  \label{eq:mbox-inboxright-and-empty-left-and-center2-imp-existence1}
	  \begin{split}
		\co{\bempty(l)\wedge\bempty(c) \wedge 
		   \inbox(x,r)}{2} & \rightarrow  \co{\exists y\ \inbox(\fromvar{x}{2},y)}{1}	  	
	  \end{split}  
	\end{align}
	A second group of bridge rules formalise that the two observers agree on the colours of the balls they both see: 
	\begin{align}
	\label{eq:mbox-same-color1}
	i:\textit{black}(\tovar{x}{j}) &\rightarrow j:\textit{black}(x), \quad i\neq j\in\{1,2\} \\
	\label{eq:mbox-same-color2}
	i:\textit{white}(\tovar{x}{j}) &\rightarrow j:\textit{white}(x), \quad i\neq j\in\{1,2\} 
	\end{align}
	The domain relations between $\dom_1$ and $\dom_2$ are used to represent the fact that 
	$\obsuno$ and $\obsdue$ look at the same real world objects. A consequence of this is that the domain relations must be one the inverse of the other. 
	This is formalised by a bridge rule as the one below, whose meaning will be better explained in Section \ref{sec:properties-of-dr}
	 and Figure \ref{fig:dr-bridge-rules} :
	\begin{equation}
	\label{eq:mbox-symmetric-domain-relations}
	i:x=\tovar{y}{j}\rightarrow j:y=\tovar{x}{i}
	\end{equation}
	The DFOL model defined in Example~\ref{ex:dfol-model} satisfies all the bridge rules 
	(\ref{eq:mbox-inboxright1-imp-existence2}\ftext{)}--\ftext{(}\ref{eq:mbox-symmetric-domain-relations}). 
	To show how the satisfiability of bridge rules work\ftext{s,} let us consider bridge rule \eqref{eq:mbox-inboxright1-imp-existence2}. 
	In particular\ftext{,} let us consider an assignment $a$ such as $a_i(x)=c$. In this case, $\M \models \co{\inbox(x,r)}{1}[a]$ since $\npla{c,r} \in inbox^{\I_1}$. We need to show that there is an extension $a'$ of $a$, admissible for $\co{\exists y \ \inbox(\fromvar{x}{1},y)}{2}$\ftext{,} such as $\M$ satisfies it. By observing the domain relation $\dr_{12}$ we can define $a'$ as an extension of $a$ with $a'_j(\fromvar{x}{1})=a$. It is now easy to show our claim. In fact, $\npla{a,l} \in inbox^{\I_2}$. Thus, the formula $\exists y \ \inbox(\fromvar{x}{1},y)$ with $\fromvar{x}{1}$ bound to $a$ is satisfied by $m_2$ and, as a consequence, by $\M$. 
\end{exmp}

\begin{exmp}[A formalisation of the mediator]
\label{ex:formalization-mediator}
Let the languages $L_1, L_2$ and $L_m$ be the ones informally defined in Figure \ref{fig:mediator}. We focus here on the bridge rules able to express the relations between 
the sellers and the mediator, that is, the fact that the latter sells all and only products sold by each of the formers, whose price has been set to a specific value. 

First of all, we need to specify the shape of the domain relation, that is, indicate that fruits are mapped into fruits, numbers into numbers, and so on. Let us focus on fruits which is the peculiarity of this example. The choice made by the mediator is to be able to represent all fruits sold by the two sellers. For the sake of this example, we also have decided that the mediator 
sells apples by their specific variety (similarly to \sellerdue) and that he knows that ``apples'' of \selleruno correspond to both ``Delicious'' and ``GrannySmith'' in his own database. 
We express all these choices by means of the following bridge rules: 

\begin{align}
	\label{eq:Abox1}
	& \co{x=Apple}{1} \rightarrow \co{\fromvar{x}{1}=Delicious \vee \fromvar{x}{1}=\ftext{GrannySmith}}{\mediator}\\
	& \co{x=Lemon}{1} \rightarrow \co{\fromvar{x}{1}=Lemon}{\mediator}\\
	& \co{x=Delicious}{2} \rightarrow \co{\fromvar{x}{2}=Delicious}{\mediator}\\
	& \co{x=\ftext{GrannySmith}}{2}\rightarrow \co{\fromvar{x}{2}=GrannySmith}{\mediator}\\
	\label{eq:Abox4}
	& \co{x=Orange}{2} \rightarrow \co{\fromvar{x}{2}=Orange}{\mediator}
\end{align}

The mediator offers \emph{all the fruits} available in $\selleruno$ (resp. $\sellerdue$) whose price has been set. 
\begin{gather}
  \label{eq:mediator-from-1-to-mediator}
  \co{\available(x,y)\wedge\price(x,z)}{1}\rightarrow
  \co{\offer(\fromvar{x}{1},\fromvar{y}{1},\fromvar{z}{1})}{\mediator}\\
  \label{eq:mediator-from-2-to-mediator}
  \co{\available(x,y,z)}{2}\rightarrow\co{\offer(\fromvar{x}{2},\fromvar{y}{2},k) \con k
  = \fromvar{z}{2}\div \fromvar{y}{2}}{\mediator}
\end{gather}
The mediator sells \emph{only fruits} that are available in $\selleruno$ or in $\sellerdue$;  
\begin{multline}
  \label{eq:mediator-fruits}
  \co{\neg\exists y.\available(\tovar{x}{\mediator},y)}{1}, \co{\neg\exists
  y z.\available(\tovar{x}{\mediator},y,z)}{2} \\
  \rightarrow\co{\neg\exists
  y,z.\offer(x,y,z)}{\mediator} 
\end{multline}

In database terms, the above bridge rules can be read as a query definition for the predicate $\offer$ in the database of \mediator\footnote{\ftext{An investigation on the usage of bridge rules for answering queries in distributed databases can be found in~\cite{Luciano-Serafini:2000uq}}.}. When a user submits the query ${\offer}(x,y,z)$ to \mediator, it rewrites this as two queries. The first one is query ${\available}(x,y)\wedge{\price}(x,z)$, generated by \eqref{eq:mediator-from-1-to-mediator}, and sent to $\selleruno$. 
The second query is $\available(x,y,z)$, generated by \eqref{eq:mediator-from-2-to-mediator} and sent to \sellerdue.
$\selleruno$ and $\sellerdue$ separately evaluate the two queries and send the result back to the mediator using the domain relations shaped by bridge rules \eqref{eq:Abox1}--\eqref{eq:Abox4} to appropriately ``translate'' the result. This reading of bridge rules formalises the GAV (global as view) approach to information integration described in~\cite{ullman1}. Finally, 
bridge rule \eqref{eq:mediator-fruits} formalises a closure condition, that is, the fact that all the data relevant to $\offer$ in \mediator are retrieved from the relations $\available$ (and $\price$) of the sellers' databases. 
Similar combinations of bridge rules to constrain the domain relation and the interpretation of predicates are exploited in \cite{DBLP:conf/semweb/SerafiniT07} to perform instance migration among heterogeneous ontologies by means of 
bridge rules between ontology Aboxes  and ontology Tboxes. 
\end{exmp}

%% file: interpretationConstraints.tex
\section{How to represent distributed knowledge via bridge rules}
\label{sec:interpretation-cosntraints}

In this section we illustrate how to represent important types of relations between local knowledge bases by means of bridge rules. We first investigate how to model specific relations between different domains (Section \ref{sec:properties-of-dr}); we then focus on the usage of bridge rules to represent pairwise semantic mappings (Section~\ref{sec:semantic-mappings})\onlyintechrep{and the join of knowledge from different knowledge sources (Section \ref{sec:join_via_mappings})}; finally\ftext{,} we introduce and investigate the notion of entailment of bridge rules (Section \ref{subsec:derived-bridge-rules}). 

\subsection{Representing specific domain relations}
\label{sec:properties-of-dr}

The definition of domain relation as a generic relation provides DFOL with the capability to represent arbitrary correspondences between systems that have been designed autonomously. Nonetheless, the correlation patterns between domains of different knowledge bases often correspond to well known properties of relations. Examples are isomorphic domains, containment between domains, injective transformations, and so on. As already mentioned in Example~\ref{ex:formalization-mediator}\ftext{,} bridge rules can be used to impose restrictions on the shape of the domain relation in order to capture specific correspondences. 
In this paper we consider the following properties \onlyinpaper{(further examples can be found in~\cite{DFOL-techrep})}:
\begin{itemize}
	\item[$\fun_{ij}$:] $\dr_{ij}$ is a (partial) \emph{function}. In this case, the elements in $\bdom_i$ have at most one corresponding element in $\bdom_j$. This is used, for instance, to express the fact that $\bdom_j$ has a smaller granularity than $\bdom_i$\ftext{. An example of this is} the mediator example, where $\bdom_1$ has a smaller granularity w.r.t. $\bdom_\mediator$ since it describes apples ignoring their different varieties. In this case, we could safely assume that $\dr_{1\mediator}$ satisfies the $\fun_{ij}$ property, while we would not impose it for an hypothetical domain relation $\dr_{\mediator 1}$. 
	\item [$\tot_{ij}$:] $\dr_{ij}$ is \emph{total}. In this case, each element of $\bdom_i$ has a corresponding element in $\bdom_j$, and therefore the entire $\bdom_i$ can be embedded (via $\dr_{ij}$) into $\bdom_j$. 
	\item [$\sur_{ij}$:] $\dr_{ij}$ is \emph{surjective}. In this case, each element of $\bdom_j$ is the corresponding of some object of $\bdom_i$, and the entire $\bdom_j$ can be seen as the transformation of some parts of $\bdom_i$.
	\item [$\inj_{ij}$:] $\dr_{ij}$ is \emph{injective}. In this case, inequality is preserved by $\dr_{ij}$. 
	\onlyintechrep{
	\item [$\congr_{ij}$:] $\dr_{ij}$ is a \emph{congruence}, that is, there is a $K \in \NAT$ and two families $\{\bdom_{i_k}\}_{k\in K}$ and $\{\bdom_{j_k}\}_{k\in K}$ of disjoint subsets of $\bdom_i$ and $\bdom_j$ respectively, such that $\dr_{ij}=\bigcup_{k\in K}(\bdom_{i_k}\times \bdom_{j_k})$. In this case we can partition both $\bdom_i$ and $\bdom_j$ in $K$ subsets such that each one of the $dom_{i_k}$ is completely mapped in the corresponding $dom_{j_k}$. In other worlds, we can find an abstraction of both $\bdom_i$ and $\bdom_j$ composed of $K$ elements such that there is a one to one mapping between the two, or alternatively, we can create a mediator's domain composed exactly of $K$ elements which can be used to relate $\bdom_i$ and $\bdom_j$.
	}
	\item [$\inv_{ij}$:] $\dr_{ij}$ is the \emph{inverse} of $\dr_{ji}$; in this case the transformation from $\bdom_i$ to $\bdom_j$ corresponds to the way in which $\bdom_j$ is transformed into $\bdom_i$.  
	\onlyintechrep{
	\item [$\euc_{ijk}$:] $\dr_{jk}$ is the \emph{Euclidean} composition of $\dr_{ij}$ and $\dr_{ik}$, that is for every $d$ in $\bdom_i$, $d'$ in $\bdom_j$ and $d''$ in $\bdom_k$ if $d$ is related to $d'$ via $\dr_{ij}$ and $d$ is related to $d''$ via $\dr_{ik}$, then $d'$ is related to $d''$ via $\dr_{jk}$. Notationally, we express this as $\dr_{jk} \subseteq \dr_{ij} \Join_i \dr_{ik}$. This property can be useful if we consider $i$ to be the knowledge base of a mediator. In this case the \emph{Euclidean} composition ensures that if $d'$ and $d''$ are mediated into $d$, then there exists also a direct transformation between them.
	}
	\item [$\com_{ijk}$:] $\dr_{ik}$ is the \emph{composition} of  $\dr_{ij}$ and $\dr_{jk}$, that is $\dr_{ik} = \dr_{ij} \circ \dr_{jk}$. This property guarantees that if there is a way of transforming an object $d$ of $\bdom_i$ into an object $d'$ of $\bdom_k$ via $\bdom_j$, then there is also a direct way of transforming $d$ into $d'$ using $\dr_{ik}$ (and vice-versa).
\end{itemize}
As we can see these properties can refer to a single domain relation, as in 
\onlyintechrep{$\fun_{ij}$--$\congr_{ij}$,}
\onlyinpaper{$\fun_{ij}$--$\inj_{ij}$,} to two domain relations, as in the case of $\inv_{ij}$, or to several, as in $\com_{ijk}$\onlyintechrep{ and $\euc_{ijk}$}.

The formalisation of the above properties relies on the usage of arrow variables, together with the equality predicate, to write bridge rules able to constrain the shape of the domain relation. As an example, a model $\M$ satisfies a formula of the form  $\co{x=\tovar{y}{j}}{i}$ (resp. $\co{\fromvar{x}{i}=y}{j}$) exactly when $r_{ij}$ relates the object $a_i(x)$ in $\bdom_{i}$ to the object $a_j(y)$ in $\bdom_j$ as in the graphical representation provided below: 
\begin{center}
	\begin{tikzpicture}[descr/.style={fill=white,inner sep=2pt}] 
		\matrix (m) [matrix of math nodes, column sep=2cm] 
		{a_i(x) & a_j(y) \\ }; 
		\path[->,font=\scriptsize, thick] 
		(m-1-1) edge node[descr] {$ \dr_{ij} $} (m-1-2); 
	\end{tikzpicture} 
\end{center}

A more complex scenario is the one in which $\M$ satisfies the two bridge rules $\co{x=\tovar{y}{j}}{i}[a]$ and
$\co{y=\fromvar{z}{k}}{j}[a]$. This originates the more complex diagram:
\begin{center}
	\begin{tikzpicture}[descr/.style={fill=white,inner sep=2pt}] 
		\matrix (m) [matrix of math nodes, column sep=2cm] 
		{ & a_j(y) & \\
		  a_i(x) & & a_k(z) \\ }; 
		\path[->,font=\scriptsize, thick] 
		(m-2-1) edge node[descr] {$ \dr_{ij} $} (m-1-2) 
		(m-2-3) edge node[descr] {$ \dr_{kj} $} (m-1-2);
	\end{tikzpicture}
\end{center}

Using this graphical notation, we can represent 
\onlyinpaper{e.g., $\inv_{ij}$ and $\com_{ijk}$ as follows,}
\onlyintechrep{$\congr_{ij}$, $\inv_{ij}$, $\com_{ijk}$ and $\euc_{ijk}$ as in Figure~\ref{fig:dr-properties},}
where solid lines imply the existence of the dashed lines.

\onlyinpaper{
\parbox{\textwidth}{
\begin{center}
	\begin{tikzpicture}[descr/.style={fill=white,inner sep=2pt}] 
		\matrix[ampersand replacement=\&] (m) [matrix of math nodes, column sep=.7cm, row sep=.5cm] 
		{ a_i(x) \& \& a_j(y) \\
		\&(\inv_{ij})\& \\ }; 
		\path[->,font=\scriptsize, thick] 
		(m-1-1) edge[bend left=30] node[descr] {$\dr_{ij}$}(m-1-3)
		(m-1-3) edge[dashed, bend left=30] node[descr] {$\dr_{ji}$} (m-1-1);
	\end{tikzpicture} \quad 
	\begin{tikzpicture}[descr/.style={fill=white,inner sep=2pt}] 
		\matrix[ampersand replacement=\&] (m) [matrix of math nodes, column sep=.7cm, row sep=.8cm] 
		{ \& a_j(y) \& \\
		  a_i(x) \& \& a_k(z) \\[-1cm] 
	 	\&(\com_{ijk})\& \\ }; 
		\path[->,font=\scriptsize, thick] 
		(m-2-1) edge node[descr] {$ \dr_{ij} $} (m-1-2)
		(m-1-2) edge node[descr] {$ \dr_{jk} $} (m-2-3)
		(m-2-1) edge[dashed] node[descr] {$ \dr_{ik} $} (m-2-3);
	\end{tikzpicture}
\end{center}}
}

\onlyintechrep{
\begin{figure}[htbp]
	\centering
	\begin{tikzpicture}[descr/.style={fill=white,inner sep=2pt}] 
		\matrix[ampersand replacement=\&] (m) [matrix of math nodes, column sep=.7cm, row sep=.5cm] 
		{ a_i(x) \&\& a_j(v) \\
		  a_i(y) \&\& a_j(w) \\[-.5cm] 
		\&(\congr_{ij})\&\\}; 
		\path[->,font=\scriptsize, thick] 
		(m-1-1) edge (m-1-3)
		(m-1-1) edge (m-2-3)
		(m-2-1) edge (m-1-3);
		\path[->,font=\scriptsize, thick,dashed] 
		(m-2-1) edge node[descr] {$ \dr_{ij} $} (m-2-3);
	\end{tikzpicture}\quad \quad \quad 
	\begin{tikzpicture}[descr/.style={fill=white,inner sep=2pt}] 
		\matrix[ampersand replacement=\&] (m) [matrix of math nodes, column sep=.7cm, row sep=.5cm] 
		{ a_i(x) \&\& a_j(y) \\
		\&(\inv_{ij})\& \\ }; 
		\path[->,font=\scriptsize, thick] 
		(m-1-1) edge[bend left=30] node[descr] {$\dr_{ij}$}(m-1-3)
		(m-1-3) edge[dashed, bend left=30] node[descr] {$\dr_{ji}$} (m-1-1);
	\end{tikzpicture}\\[.5cm]
	\begin{tikzpicture}[descr/.style={fill=white,inner sep=2pt}] 
		\matrix[ampersand replacement=\&] (m) [matrix of math nodes, column sep=.7cm, row sep=1cm] 
		{ \& a_j(y) \& \\
		  a_i(x) \& \& a_k(z) \\[-1cm] 
		 \&(\com_{ijk})\& \\ }; 
		\path[->,font=\scriptsize, thick] 
		(m-2-1) edge node[descr] {$ \dr_{ij} $} (m-1-2)
		(m-1-2) edge node[descr] {$ \dr_{jk} $} (m-2-3)
		(m-2-1) edge[dashed] node[descr] {$ \dr_{ik} $} (m-2-3);
	\end{tikzpicture}\quad \quad \quad 
	\begin{tikzpicture}[descr/.style={fill=white,inner sep=2pt}] 
		\matrix[ampersand replacement=\&] (m) [matrix of math nodes, column sep=.6cm, row sep=.3cm] 
		{ \&\& a_j(y)  \\
		  a_i(x) \\
		  \&\& a_k(z) \\[-.3cm] 
		  \&(\euc_{ijk})\&\\}; 
		\path[->,font=\scriptsize, thick] 
		(m-2-1) edge node[descr] {$ \dr_{ij} $} (m-1-3)
		(m-2-1) edge node[descr] {$ \dr_{ik} $} (m-3-3)
		(m-1-3) edge[dashed,bend left=30] node[descr] {$ \dr_{jk} $} (m-3-3)
		(m-3-3) edge[dashed,bend left=30] node[descr] {$ \dr_{kj} $} (m-1-3);
	\end{tikzpicture}
\caption{\label{fig:dr-properties} Graphical representation of the properties of the domain relation.}

\end{figure}
}

We say that a model $\M$ satisfies 
$\fun_{ij}$--$\com_{ijk}$  
if the domain relations it contains satisfy 
$\fun_{ij}$--$\com_{ijk}$.
\begin{prop}
	\label{prop:bridge-rules-for-domain-relation}
	A model $\M$ satisfies the properties $\fun_{ij}$--$\com_{ijk}$ contained in the left hand side column of Figure \ref{fig:dr-bridge-rules} if and only if it satisfies the corresponding bridge rules on the right hand side column.
\end{prop}

\onlyintechrep{
	\begin{figure}[htbp]
		\newcommand\hgap{\hspace{-\tabcolsep}}
		\newcommand\vgap{\\[-.3cm]}
		\renewcommand{\tabcolsep}{.2cm}
		\renewcommand{\arraystretch}{1.8}
		\centering
		\begin{tabular}{|l|l|}\hline
		\textbf{Property} & 
		\textbf{Bridge Rule} \\ \hline\hline
		$\fun_{ij}: \npla{z,x}, \npla{z,y} \in \dr_{ij}$ implies $x=y$ & $\co{\tovar{x}{j}=\tovar{y}{j}}{i}\rightarrow\co{x=y}{j}$  \\ \hline
		$\tot_{ij}: \forall x\in \bdom_i \exists y \in \bdom_j$ s.t. $\npla{x,y} \in \dr_{ij}$ & $\co{x=x}{i}\rightarrow\co{\exists y \ y=\fromvar{x}{i}}{j}$ \\ \hline
		$\sur_{ij}: \forall y\in \bdom_j \exists x \in \bdom_i$ s.t. $\npla{x,y} \in \dr_{ij}$ & $\co{x=x}{j}\rightarrow\co{\exists y \ y=\tovar{x}{j}}{i}$   \\ \hline
		$\inj_{ij}: x \neq y$ implies $\dr_{ij}(x) \cap \dr_{ij}(y) = \emptyset$& $\co{\tovar{x}{j}\neq \tovar{y}{j}}{i}\rightarrow\co{x\neq y}{j}$  \\ \hline
		$\congr_{ij}:$ \hspace{-.3cm}
		$\left.
		\begin{tabular}{l}
		$\npla{x,v}$\vgap
		$\npla{y,v}$\vgap
		$\npla{x,w}$
		\end{tabular}
		\hspace{-.3cm}
		\right\}
		\in \dr_{ij}$ implies $\npla{y,w} \in \dr_{ij}$ & 
		\hgap
		$\left.
		\begin{tabular}{l}
		$\co{x=\tovar{v}{j}}{i}$\vgap
		$\co{y=\tovar{v}{j}}{i}$\vgap
		$\co{x=\tovar{w}{j}}{i}$
		\end{tabular}
		\hgap \right\}$
		$\rightarrow \co{\fromvar{y}{i}=w}{j}$ \\ \hline
		$\inv_{ij}$: \hspace{-.2cm}
		\begin{tabular}{l}
		$\dr_{ij}\subseteq\dr_{ji}^{-1}$ \vgap
		$\dr_{ji}^{-1}\subseteq\dr_{ij}$
		\end{tabular} 
		& 
		\hgap
		\mbox{\begin{tabular}{c}
		$\co{x=\tovar{y}{j}}{i}\rightarrow\co{y=\tovar{x}{i}}{j}$ \vgap
		$\co{x=\tovar{y}{i}}{j}\rightarrow\co{y=\tovar{x}{j}}{i}$ 
		\end{tabular}}  \\ \hline
		$\euc_{ijk}$:\hspace{-.2cm}
		\begin{tabular}{l}
		$\dr_{jk}\subseteq\dr_{ji}^{-1} \circ \dr_{ik}$\vgap
		$\dr_{ji}^{-1} \circ \dr_{ik}\subseteq\dr_{jk}$ \vgap
		$\dr_{kj}\subseteq\dr_{ki}^{-1}\circ\dr_{ij}$ \vgap
		$\dr_{ki}^{-1}\circ\dr_{ij}\subseteq\dr_{kj}$
		\end{tabular} & 
		\hgap
		\begin{tabular}{l}
		$\co{y=\tovar{z}{k}}{j}\rightarrow\co{\tovar{y}{j}=\tovar{z}{k}}{i}$ \vgap
		$\co{\tovar{y}{j}=\tovar{z}{k}}{i}\rightarrow\co{y=\tovar{z}{k}}{j}$ \vgap
		$\co{y=\fromvar{z}{k}}{j}\rightarrow\co{\tovar{y}{j}=\tovar{z}{k}}{i}$ \vgap
		$\co{\tovar{y}{j}=\tovar{z}{k}}{i}\rightarrow\co{y=\fromvar{z}{k}}{j}$
		\end{tabular} \\ \hline
		$\com_{ijk}$:\hspace{-.2cm}
		\begin{tabular}{l}
		$\dr_{ij}\circ\dr_{jk}\subseteq\dr_{ik}$ \vgap
		$\dr_{ik}\subseteq\dr_{ij}\circ\dr_{jk}$ 
		\end{tabular} & 
		\hgap
		\begin{tabular}{l}
		$\co{\fromvar{x}{i}=\tovar{z}{k}}{j}\rightarrow\co{\fromvar{x}{i}=z}{k}$ \vgap
		$\co{x=\tovar{z}{k}}{i}\rightarrow\co{\fromvar{x}{i}=\tovar{z}{k}}{j}$ 
		\end{tabular} \\ \hline 
		\end{tabular}
		\caption{Bridge rules used to constrain domain relations.}
		\label{fig:dr-bridge-rules}
		\end{figure}
}

\begin{pf}
\onlyintechrep{We first show that if $\M$ satisfies a property among $\fun_{ij}$--$\euc_{ijk}$, then $\M$ satisfies the corresponding bridge rule (\textsc{if} direction); then we show the vice-versa (\textsc{only if} direction).}
\onlyinpaper{We provide the proof for $\fun_{ij}$. All the other cases can be found in \cite{DFOL-techrep}.}

\onlyinpaper{
\textsc{if} Direction. Let us assume that $\dr_{ij}$ is a function and that $\M\models\co{\tovar{x}{j}=\tovar{y}{j}}{i}[a]$; we have to show that  
$\M\models\co{x=y}{j}[a]$. 
From $\M\models\co{\tovar{x}{j}=\tovar{y}{j}}{i}[a]$ we have that $a_i(\tovar{x}{j})=a_i(\tovar{y}{j})$. Since $\dr_{ij}$ is a function then 
   $\dr_{ij}(a_i(\tovar{x}{j}))=\dr_{ij}(a_i(\tovar{y}{j}))$ contains at most one element. This implies that $a_j(x)=a_j(y)$, and therefore that $\M\models\co{x=y}{j}[a]$. 

\textsc{only if} Direction. Suppose that $\M\models\co{\tovar{x}{j}=\tovar{y}{j}}{i}\rightarrow\co{x=y}{j}$ and let us prove that $\dr_{ij}$ 
   is a function. Let $d\in\dom_i$ and suppose by contradiction that $d'\neq d''\in\dr_{ij}(d)$. 
   Consider the assignment $a$ with $a_i(\tovar{x}{j})=a_i(\tovar{y}{j})=d$ and 
   $a_j(x)=d'$ and $a_j(y)=d''$. Obviously, $\M\models\co{\tovar{x}{j}=\tovar{y}{j}}{i}[a]$ 
   but $\M\not\models\co{x=y}{j}[a]$, which contradicts the fact that $\M\models\co{\tovar{x}{j}=\tovar{y}{j}}{i}\rightarrow\co{x=y}{j}$. Thus, $\dr_{ij}$ 
   is a function.
}

\onlyintechrep{		
\begin{itemize}
\item[$\fun_{ij}$] \textsc{if} Direction. Let us assume that $\dr_{ij}$ is a function and that $\M\models\co{\tovar{x}{j}=\tovar{y}{j}}{i}[a]$; we have to show that  
$\M\models\co{x=y}{j}[a]$. 
From $\M\models\co{\tovar{x}{j}=\tovar{y}{j}}{i}[a]$ we have that $a_i(\tovar{x}{j})=a_i(\tovar{y}{j})$. Since $\dr_{ij}$ is a function then 
   $\dr_{ij}(a_i(\tovar{x}{j}))=\dr_{ij}(a_i(\tovar{y}{j}))$ contains at most one element. This implies that $a_j(x)=a_j(y)$, and therefore that $\M\models\co{x=y}{j}[a]$. 

\textsc{only if} Direction. Suppose that $\M\models\co{\tovar{x}{j}=\tovar{y}{j}}{i}\rightarrow\co{x=y}{j}$ and let us prove that $\dr_{ij}$ 
   is a function. Let $d\in\dom_i$ and suppose by contradiction that $d'\neq d''\in\dr_{ij}(d)$. 
   Consider the assignment $a$ with $a_i(\tovar{x}{j})=a_i(\tovar{y}{j})=d$ and 
   $a_j(x)=d'$ and $a_j(y)=d''$. Obviously, $\M\models\co{\tovar{x}{j}=\tovar{y}{j}}{i}[a]$ 
   but $\M\not\models\co{x=y}{j}[a]$, which contradicts the fact that $\M\models\co{\tovar{x}{j}=\tovar{y}{j}}{i}\rightarrow\co{x=y}{j}$. Thus, $\dr_{ij}$ 
   is a function.

\item[$\tot_{ij}$] \textsc{if} Direction. Let us assume that $\dr_{ij}$ is a total relation and that $\M\models\co{x=x}{i}[a]$\footnote{This latter assumption is always true.} with $a$ strictly admissible for $\co{x=x}{i}$. We have to show that there is an extension $a'$ such that $\M\models\co{\exists y.y=\fromvar{x}{i}}{j}[a']$. Since $\dr_{ij}$ is total, $\dr_{ij}(a_i(x))$ is not empty, and in particular it contains an element $d'$ such that we can define an extension $a'$ of $a$ with $a'_j(\fromvar{x}{i})=d'$. Thus, $\M\models\co{\exists y.y=\fromvar{x}{i}}{j}[a']$.

   \textsc{only if} Direction. Suppose that $\M\models\co{x=x}{i}\rightarrow\co{\exists y.y=\fromvar{x}{i}}{j}$ and let us prove that $\dr_{ij}$ is total. Let $d\in\dom_i$, and let $a$ be an assignment that does not assign any arrow variable such that $a_i(x)=d$. Since $\M\models\co{x=x}{i}[a]$ then the bridge rule $\co{x=x}{i}\rightarrow\co{\exists y.y=\fromvar{x}{i}}{j}$ guarantees that $a$ can always be extended to an assignment $a'$ admissible for $\co{\exists y.y=\fromvar{x}{i}}{j}$ such that $ai_j(\fromvar{x}{i}) = d'$ for some $d'\in \bdom_j$. Thus, $d'\in\dr_{ij}(d)$ and $\dr_{ij}$ is total.

\item[$\sur_{ij}$] \textsc{if} Direction. Let us assume that $\dr_{ij}$ is surjective and that $\M\models\co{x=x}{j}[a]$ with $a$ strictly admissible for $\co{x=x}{j}$. The fact that $\dr_{ij}$ is surjective implies that there is a pre-image $d \in \bdom_i$ of $a_j(x)$ such that $\npla{d,a_j(x)}\in\dr_{ij}$. Thus, $a$ can be
  extended to $a'$ with $a'_i(\tovar{x}{j})=d$, which is admissible for $\exists y.y=\tovar{x}{j}$. Thus, $\M\models\co{\exists y.y=\tovar{x}{j}}{i}[a']$. 

  \textsc{only if} Direction. Suppose that $\M\models\co{x=x}{j}\rightarrow\co{\exists y.y=\tovar{x}{j}}{i}$ and let us prove that $\dr_{ij}$ is surjective. Let $d$ be an element of
  $\dom_j$, and $a$ be an assignment with $a_j(x)=d$. Then, $\M\models j:x=x[a]$. From the hypothesis $a$ can be extended to an assignment $a'$ admissible
  for $\exists y.y=\tovar{x}{j}$, that is, an assignment $a'$ such that $a'_i(\tovar{x}{j})=d'$ and $\npla{d',d}\in\dr_{ij}$. Thus $\dr_{ij}$ is surjective.

\item[$\inj_{ij}$] \textsc{if} Direction. Let us assume that $\dr_{ij}$ is injective and that $\M\models\co{\tovar{x}{j}\neq \tovar{y}{j}}{i}[a]$. 
    Since $\dr_{ij}$ is injective and $a_i(\tovar{x}{j})\neq a_i(\tovar{y}{j})$ we have that $\dr_{ij}(a_i(\tovar{x}{j}))\cap 
    \dr_{ij}(a_i(\tovar{y}{j}))=\emptyset$. The facts that $a_j(x)\in\dr_{ij}(a_i(\tovar{x}{j}))$ and 
    $a_j(y)\in\dr_{ij}(a_j(\tovar{y}{j}))$ imply $a_j(x)\neq a_j(y)$, and therefore $\M\models\co{x\neq y}{j}[a]$. 

    \textsc{only if} Direction. Suppose that $\M\models\co{\tovar{x}{j}\neq \tovar{y}{j}}{i}\rightarrow\co{x\neq y}{j}$ and let us prove that $\dr_{ij}$ is surjective. Let $d_1 \neq d_2$ be two distinct elements of $\bdom_i$ and let us assume that $\dr_{ij}$ is not surjective, that is, there is a $d$ in $\dr_{ij}(d_1) \sqcap \dr_{ij}(d_2)$. From this we can define an assignment $a$ with $a_i(\tovar{x}{j})=d_1$, $a_i(\tovar{y}{j})=d_2$, $a_j(x)=a_j(y)=d$ such that $\M\models\co{\tovar{x}{j}\neq \tovar{y}{j}}{i}[a]$. But from the hypothesis we have that $\M \models \co{x\neq y}{j}[a]$, that is $a_j(x) \neq a_j(y)$. This is a contradiction and we can conclude that there is no $d$ in $\dr_{ij}(d_1) \sqcap \dr_{ij}(d_2)$. 

\item[$\congr_{ij}$] \textsc{if} Direction. 
    Let us assume that $\dr_{ij}$ is a congruence and that $\M$, $a$ satisfy $\co{x=\tovar{v}{j}}{i}$, $\co{y=\tovar{v}{j}}{i}$ and $\co{x=\tovar{w}{j}}{i}$. This implies that $a_j(v)\in\dr_{ij}(a_i(x))$, $a_j(v)\in\dr_{ij}(a_i(y))$, and $a_j(w)\in\dr_{ij}(a_i(x))$. This situation corresponds to the solid arrows in Figure \ref{fig:dr-properties}.($\congr_{ij}$).
    From the fact that $\dr_{ij}$ is a congruence we can derive that $a_(w)\in\dr_{ij}(a_i(y))$. This implies that  
    $a$ can be extended to an $a'$ with $a'_j(\fromvar{y}{i})=a_j(w)$. Thus $\M\models\co{\fromvar{y}{i}=w}{j}[a']$.
%
%

    \textsc{only if} Direction. Suppose that $\M\models\co{x=\tovar{v}{j}}{i}, \co{y=\tovar{v}{j}}{i}, \co{x=\tovar{w}{j}}{i} \rightarrow \co{\fromvar{y}{i}=w}{j}$ and let us show that $\dr_{ij}$ is a congruence. 
    For every $d,d'\in\dom_i$ let $d\sim_i d'$ iff $\dr_{ij}(d)=\dr_{ij}(d')$. 
    Similarly for every $d,d'\in\dom_j$ let $d\sim_jd'$ if and only if $\dr_{ij}^{-1}(d)=
    \dr_{ij}^{-1}(d)$. $\sim_i$ ($\sim_j$) is an equivalence relation and $[d]_i$ 
    ($[d]_j$) is the equivalence classes of $d$ w.r.t, $\sim_i$ ($\sim_j$).
    Let $[d]_i$ be an equivalence class such that there is a $d'\in\dr_{ij}(d)$. From the hypothesis we have that $[d]_i\times[d']_j\subseteq\dr_{ij}$. Furthermore, if $[c]_i \neq [d]_i$ 	and $c'\in\dr_{ij}(c), d'\in\dr_{ij}(d)$, then $[c']_j \neq [d']_j$. This implies that $\dr_{ij}$ is a congruence that can be 
    expressed as 
  $$
   \dr_{ij}=\bigcup_{d,d'\in\dr_{ij}} [d]_i\times[d]_j
  $$

\item[$\inv_{ij}$] \textsc{if} Direction. Let us assume that $\dr_{ij} \subseteq \dr^-_{ji}$ and that $\M\models\co{x=\tovar{y}{j}}{i}[a]$. 
	From the definition of assignment we have that $\npla{a_i(\tovar{y}{j}),a_j(y)} \in \dr_{ij}$. From $a_i(x)=a_i(\tovar{y}{j})$ we obtain that $\npla{a_i(x),a_j(y)} \in \dr_{ij}$, and from the fact that $\dr_{ij} \subseteq \dr^-_{ji}$ we have that $\npla{a_j(y),a_i(x)}\in\dr_{ji}$. We can therefore extend $a$ to an assignment $a'$ with $a'_j(\tovar{x}{i})=a_j(y)$, 
  such that $\M\models\co{\tovar{x}{i}=y}{j}[a']$. A similar proof can be shown for the case $\dr^-_{ji} \subseteq \dr_{ij}$ and for the second bridge rule of property $\inv_{ij}$ in Proposition \ref{prop:bridge-rules-for-domain-relation}.  

\textsc{only if} Direction. Suppose that $\M\models\co{x=\tovar{y}{j}}{i}\rightarrow\co{y=\tovar{x}{i}}{j}$ and let us show that $\dr_{ij} \subseteq \dr^-_{ji}$. Let $d,d'$ be two elements such that $d'\in\dr_{ij}(d)$ and such that there is an assignment $a$ with $d=a_i(x)$ and $a_j(y)=d'$. It is easy to see that $\M\models\co{x=\tovar{y}{j}}{i}[a]$ holds. From the hypothesis we know that $a$ can be extended to an assignment $a'$ such that $\M\models\co{\tovar{x}{i}=y}{j}[a']$. This implies that $a'_j(\tovar{x}{i})=d'$. From the definition of extension $a'_i(x)=a_i(x)=d$, and therefore $d \in \dr_{ji}(a'_j(\tovar{x}{i}))$, that is $d\in\dr_{ji}(d')$. A similar proof can be done for the second bride rule of property $\inv_{ij}$.

\item[$\com_{ijk}$] \textsc{if} Direction. Let us assume that $\dr_{ij}\circ\dr_{jk}\subseteq\dr_{ik}$ and that $\M\models\co{\fromvar{x}{i}=\tovar{z}{k}}{j}[a]$. If we assume that $y$ is a new variable such that $\fromvar{x}{i}=\tovar{z}{k}=y$ holds it is easy to see that the domain relations comply with the solid arrows in Figure ~\ref{fig:dr-properties}.($\com_{ijk}$). 
  Since $\dr_{ij}\circ\dr_{jk}\subseteq\dr_{ik}$, then $\npla{a_i(x), a_k(z)}\in\dr_{ik}$ as indicated by the dashed arrow in Figure~\ref{fig:dr-properties}.($\com_{ijk}$).  
  This means that $a$ can be extended to an assignment $a'$ with $a'_k(\fromvar{x}{i})=a_k(z)$. 
  This implies that $\M\models\co{\fromvar{x}{i}=z}{i}[a']$.
  The proof for the case $\dr_{ik}\subseteq \dr_{ij}\circ\dr_{jk}$ is analogous.	
   
	\textsc{only if} Direction. Suppose that $\M\models\co{\fromvar{x}{i}=\tovar{z}{k}}{j}\rightarrow\co{\fromvar{x}{i}=z}{k}$ and let us show
   that $\dr_{ij}\circ\dr_{jk}\subseteq\dr_{ik}$, that is given an element $d'$ in $\dr_{jk}(\dr_{ij}(d))$ we have that $d'$ belongs to $\dr_{ik}(d)$. 
	By definition, $d'\in\dr_{jk}(\dr_{ij}(d))$ iff
   there is a $d''\in\dom_j$ such that $d'\in\dr_{jk}(d'')$ and
   $d''\in\dr_{ij}(d)$. Let $a$ be an assignment with $a_i(x)=a_i(\tovar{y}{j})=d$, 
   $a_j(y)=a_j(\tovar{z}{k})=d''$ and $a_k(z)=d'$. This assignment is such that $\M\models\co{\fromvar{x}{i}=\tovar{z}{k}}{i}[a]$. From the hypothesis, $a$
   can be extended to an assignment $a'$ such that $\M\models\co{\fromvar{x}{i}=z}{j}[a']$. This means that
   $d'=a_k(z)=a'_k(z)=a'_k(\fromvar{x}{i})\in\dr_{ik}(d)$ and this ends the proof. The proof for the case $\dr_{ik}\subseteq \dr_{ij}\circ\dr_{jk}$ is analogous.	
   %

\item[$\euc_{ijk}$] The proof is similar to the one for $\com_{ijk}$.
\end{itemize}
}
\end{pf}

From now on we use a label, say $\fun_{ij}$ to refer to both the property of the domain relation and the corresponding bridge rule(s). The context will always make clear what we mean. 

\subsection{Representing semantic mappings}
\label{sec:semantic-mappings}
Bridge rules can be used to formalise the important notion of \emph{semantic mapping} between knowledge bases. Semantic mappings typically involves two knowledge bases only. In this Section we therefore restrict to \emph{pairwise bridge rules}.  

\begin{defn}[Pairwise bridge rule]
\index{Interpretation constraint}
\label{def:pairwise-interpretation-constraint}
A \emph{pairwise bridge rule from $i$ to $j$}, or simply a 
bridge rule from $i$ to $j$, is a bridge rule of the form:
\begin{equation}
\label{eq:pairwise-interpretation-constraint}
\co{\phi(x_1, \ldots x_n, \tovar{y_1}{j},\ldots,\tovar{y_m}{j})}{i}\rightarrow\co{\psi(\fromvar{x_1}{i}, \ldots \fromvar{x_n}{i}, y_1, \ldots, y_m)}{j}
\end{equation}
\end{defn}

Pairwise bridge rules can be used to model different forms of mappings between knowledge sources. A proof of that is the fact that almost all the encodings of different formalisms into DFOL shown in Section \ref{sec:rel-work} make use of pairwise bridge rules. A typical example of pairwise bridge rules are ontology mappings. 
Ontology mapping languages such as Distributed Description Logics (DDL)~\cite{SerafiniBorgidaTamilin-IJCAI05}, $\epsilon$-connections~\cite{e-connections,cuenca1}, and Package-based Description Logics (P-DL)~\cite{Bao:2009:PDL:1560559.1560578} enable the representation of mappings between pairs of ontologies which can be encoded in DFOL as shown in  section~\ref{sec:rel-work} using and extending the work in~\cite{SerafiniStuckenschmidtWache-IJCAI05}\footnote{For a survey on the usage of semantic mappings as a way of matching heterogeneous ontologies see \cite{ontology-matching-EuzenatShvaiko07}.}. 
To briefly illustrate how pairwise bridge rules capture ontology mappings let us consider DDL into and onto mappings: 
\begin{center}
	$\co{C}{i} \into \co{D}{j} \qquad \qquad \co{C}{i}  \onto \co{D}{j}$
\end{center}
used to express that concept $C$ in ontology $O_i$ is mapped into (onto) concept $D$ in ontology $O_j$. As shown in ~\cite{SerafiniStuckenschmidtWache-IJCAI05}, these expressions can be represented by means of pairwise mappings of the form 
\begin{center}
	$\co{C(\tovar{x}{j})}{i} \rightarrow \co{D(x)}{j} \qquad \qquad 
	\co{D(x)}{j}  \rightarrow \co{C(\fromvar{x}{j})}{i}$.
\end{center}

Another typical example of pairwise mappings are mappings occurring in database integration. Here, the work in \cite{calvanese-pods-2004,franconi1} introduces \emph{peer-to-peer} mappings as expressions of the form $cq_1\leadsto
cq_2$ where $cq_1$ and $cq_2$ are conjunctive queries in two distinct knowledge bases. The intuitive meaning of
$cq_1\leadsto cq_2$, is that the answer of the query $cq_1$ to the
knowledge base $KB_1$ must be contained in the answer of $cq_2$
submitted to $KB_2$. We can easily observe that this is similar to the intuitive reading of bridge rules \eqref{eq:simple-br1}--\eqref{eq:simple-br4} in terms of query containment provided at page \pageref{eq:simple-br1}. Other examples of pairwise expressions used to semantically map two databases can be found in
\cite{catarci1,ceri2,levy1-information-manifold,ullman1,gupta1,grefen2}.
Finally, the concept of infomorphism defined by Barwise and Seligman in \cite{barwise-seligman-1997-if} can be formalised via a set of
pairwise bridge rules and one domain relation. Again an encoding of some of these approaches in DFOL is contained in Section~\ref{sec:rel-work}. 

A final instance of DFOL pairwise bridge rule is  $\co{\bot}{j}\rightarrow\co{\bot}{i}$. This rule, called \emph{inconsistency propagation} rule and denoted with $\incp_{ji}$, forces inconsistency to propagate from a source knowledge base $j$ to a target knowledge base $i$. This rule can be  used to enforce the propagation of local inconsistency when needed, since in DFOL $\co{\bot}{j}$ does not necessarily propagate inconsistency to other knowledge bases (see Proposition~\ref{prop:inconsistency-does-not-propagate}).

\subsection{Joining knowledge through mappings} 
\label{sec:join_via_mappings}
While pairwise bridge rules focus on ``point-to-point'' mappings between two knowledge sources, DFOL bridge rules enable to encode also more complex relations involving an arbitrary number of knowledge bases.

\begin{figure}[h]
  \centering
    \includegraphics[width=.8\textwidth]{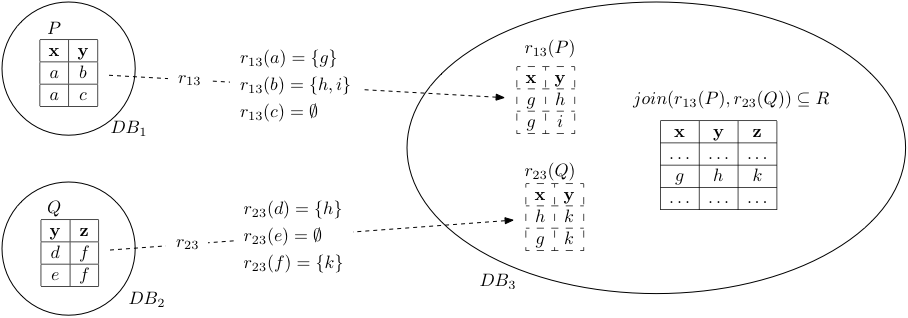}
  \caption{Joining distributed tables.}
  \label{fig:join}
\end{figure}

Bridge rules can be used to express the fact that a certain combination of knowledge coming from $i_1$,\ldots, $i_n$ source knowledge bases entails some other knowledge in a target knowledge base $i$. As an example, bridge rule
\begin{equation}
\label{eq:complex-ic}
\co{P(\tovar{x}{3},\tovar{y}{3})}{1},\co{Q(\tovar{y}{3},\tovar{z}{3})}{2}
\rightarrow\co{R(x,y,z)}{3}
\end{equation} whose graphical representation is provided in Figure~\ref{fig:join},
can be read as a mapping from the join between relation $P(x,y)$ in 1 and $Q(y,z)$ in 2, into $R(x,y,z)$ in $3$.
Indeed bridge rule \eqref{eq:complex-ic} is satisfied if
$
\dr_{13}(\ext{P}{1})\bowtie\dr_{23}(\ext{Q}{2})\subseteq\ext{R}{3}
$.

\subsection{Entailing bridge rules}
\label{subsec:derived-bridge-rules}

A logic based formalisation of the notion of mapping provides the basis to introduce the notion of entailment (logical consequence) between mappings. Entailment between mappings is important as it enables to prove that a mapping is redundant (as it can be derived from others), or that a set of mappings is inconsistent. Thus, it enables to compute sets of minimal mappings between e.g., ontologies and it can provide the basis for mapping debugging / repair, as shown for instance, in the work of Meilicke et al.~\cite{ChristianMeilicke10012009} and the one of Wang and Xu~\cite{Wang:2008jk}.

DFOL provides a precise characterisation of when bridge rules are entailed by others. For instance, to say that the bridge rule $1:A \rightarrow 3:C$ is a logical consequence of $1:A \rightarrow 2:B$ and $2:B\rightarrow 3:C$.  In this section we provide a precise definition of entailment between bridge rules and we study the general properties of such an entailment.

\begin{defn}[Entailment of bridge rules] 
	\label{def:entailment-bridge-rules}
$\co{\phi_1}{i_1},\ldots,\co{\phi_n}{i_n}\rightarrow\co{\phi}{i}$ is entailed by
a set of bridge rules $\BR$, in symbols $\BR\models
\co{\phi_1}{i_1},\ldots,\co{\phi_n}{i_n}\rightarrow\co{\phi}{i}$, if
$\co{\phi_1}{i_1}\ldots\co{\phi_n}{i_n}\models_{\BR}\co{\phi}{i}$. 
\end{defn}

The following proposition illustrates the effects on bridge rule entailment of the main operations we can perform on mappings, that is: conjunction, disjunction, existential / universal restriction, composition, instantiation and inversion of mappings. 

\begin{prop}
	\label{prp:entailment-of-br} The following entailments of bridge rules hold: 

\begin{description}
\item \textbf{Conjunction}
    \begin{enumerate}
    \item $\co{\phi}{i}\rightarrow\co{\psi}{j},\ 
           \co{\phi'}{i}\rightarrow\co{\psi'}{j}\models
           \co{\phi\wedge\phi'}{i}\rightarrow\co{\psi\wedge\psi'}{j}$ 
           if $\phi$ and $\phi'$ do not have arrow variables in common. 
    \item If $\fun_{ij}$ holds, then $\co{\phi}{i}\rightarrow\co{\psi}{j},\
           \co{\phi'}{i}\rightarrow\co{\psi'}{j}\models
           \co{\phi\wedge\phi'}{i}\rightarrow\co{\psi\wedge\psi'}{j}$ 
    \end{enumerate}
\item \textbf{Disjunction}
     \begin{enumerate}
     \item $\co{\phi_1(x)}{i}\rightarrow\co{\psi_1(\fromvar{x}{i})}{j},\
            \co{\phi_2(x)}{i}\rightarrow\co{\psi_2(\fromvar{x}{i})}{j}
            \models 
            \co{\phi_1(x)\vee\phi_2(x)}{i}
            \rightarrow
            \co{\psi_1(\fromvar{x}{i})\vee\psi_2(\fromvar{x}{i})}{j}$, if 
            at least one among $\phi_1(x)$ or $\phi_2(x)$ is a complete formula. 
     \end{enumerate}
\item \textbf{Existential and universal quantification}
     \begin{enumerate}
     \item  $\co{\phi(x)}{i}\rightarrow\co{\psi(\fromvar{x}{i})}{j}\models
             \co{\exists x\phi(x)}{i}\rightarrow\co{\exists x \psi(x)}{j}$
            if $\phi$ is a complete formula.
     \item  If $\sur_{ij}$ holds, then $\co{\phi(\tovar{x}{j})}{i}\rightarrow\co{\psi(x)}{j}\models
             \co{\forall x\phi(x)}{i}\rightarrow\co{\forall x \psi(x)}{j}$
     \end{enumerate}
\item \textbf{Composition} If $\com_{ijk}$ holds, then:
     \begin{enumerate}
     \item $\co{\phi(x)}{i}\rightarrow\co{\psi(\fromvar{x}{i})}{j}, 
            \co{\psi(x)}{j}\rightarrow\co{\theta(\fromvar{x}{j})}{k}\models
            \co{\phi(x)}{i}\rightarrow\co{\theta(\fromvar{x}{i})}{k}$
     \item $\co{\phi(\tovar{x}{j})}{i}\rightarrow\co{\psi(x)}{j}, 
            \co{\psi(\tovar{x}{k})}{j}\rightarrow\co{\theta(x)}{k}\models
            \co{\phi(\tovar{x}{k})}{i}\rightarrow\co{\theta(x)}{k}$
     \end{enumerate}
\item \textbf{Instantiation}
     \begin{enumerate}
     \item $\co{x=t}{i}\rightarrow\co{\fromvar{x}{i}=s}{j},\ 
            \co{\phi(\tovar{x}{j})}{i}\rightarrow\co{\psi(x)}{j} \models
            \co{\phi(t)}{i}\rightarrow\co{\psi(s)}{j}$, with $t$ 
            complete ground term of $L_i$. 
     \end{enumerate}
\item \textbf{Inversion} If $\fun_{ij}$ and $\inv_{ji}$ hold: 
     \begin{enumerate}
     \item $\co{\phi(x)}{i}\rightarrow\co{\psi(\fromvar{x}{i})}{j}\models
           \co{\neg\psi(\fromvar{x}{i})}{j}\rightarrow\co{\neg\phi(x)}{i}$, if
           $\phi(x)$ is a complete formula. 
     \end{enumerate}
\end{description}
\end{prop}

\begin{pf}
	\onlyinpaper{We show here only the case of Conjunction. The remaining cases are shown in \cite{DFOL-techrep}.
	
	Suppose that $\M\models i:\phi\rightarrow j:\psi$, $\M\models i:\phi'\rightarrow j:\psi'$ and $\M\models i:\phi\wedge\phi'[a]$. 
		  Then $a$
		  can be extended to $a'$ and $a''$ admissible for
		  $j:\psi$ and $j:\psi'$ respectively, and such that $m\models\psi[a']$ and
		  $m\models\psi'[a'']$ for all $m\in M_j$.  If either (case 1) the
		  arrow variables of $\psi$ and $\psi'$ are disjoint, or (case 2)
		  $\dr_{ij}$ is functional, then $a'\cup a''$ is an extension of
		  $a$, admissible for $\psi\con\psi'$ and such that
		  $m\models\psi\con\psi'[a'\cup a'']$.
	}
	
	\onlyintechrep{
	\ \\ 
	\begin{itemize}
		\item \textbf{Conjunction}. Suppose that $\M\models i:\phi\wedge\phi'[a]$, Since $\M\models
			  i:\phi\rightarrow j:\psi$ and $i:\phi'\rightarrow j:\psi'$, then $a$
			  can be extended to $a'$ and $a''$ admissible for
			  $j:\psi$ and $j:\psi'$ respectively, and such that $m\models\psi[a']$ and
			  $m\models\psi'[a'']$ for all $m\in M_j$.  If either (case 1) the
			  arrow variables of $\psi$ and $\psi'$ are disjoint, or (case 2)
			  $\dr_{ij}$ is functional, then $a'\cup a''$ is an extension of
			  $a$, admissible for $\psi\con\psi'$ and such that
			  $m\models\psi\con\psi'[a'\cup a'']$.
  		\item \textbf{Disjunction}.
			We prove the case of $\phi_1(x)$ complete formula, the other
			    case is specular. Suppose that
			    $\M\models\co{\phi_1(x)\vee\phi_2(x)}{i}[a]$. Since $\phi_1(x)$ is 
			    complete then either $\M\models\co{\phi_1(x)}{i}[a]$ or
			    $\M\models\co{\neg\phi(x)}{i}[a]$. In the first case since
			    $\M\models\co{\phi_1(x)}{i}\rightarrow\co{\psi_1(\fromvar{x}{i})}{j}$,
			     $a$ can be extended to $a'$ such that
			    $\M\models\co{\psi_1(\fromvar{x}{i})}{j}[a']$ and therefore 
			     $\M\models\co{\psi_1(\fromvar{x}{i})\vee\psi_2(\fromvar{x}{i})}{j}[a']$. 
			    In the second case, 
			    $\M\models\co{\phi_2(x)}{i}[a]$, and since, 
			    $\M\models\co{\phi_2(x)}{i}\rightarrow\co{\psi_2(\fromvar{x}{i})}{j}$,
			    $a$ can be extended to $a'$ such that 
			     $\M\models\co{\psi_1(\fromvar{x}{i})\vee\psi_2(\fromvar{x}{i})}{j}[a']$.

		\item \textbf{Composition}.
		\begin{enumerate}
		\item $\M\models\co{\phi(x)}{i}[a]$ implies that 
		$a$ can be extended to $a'$ such that $\M\models\co{\psi(\fromvar{x}{i})}{j}[a']$. 
		Since $\fromvar{x}{i}$ is the only free variable of $\co{\psi(\fromvar{x}{i})}{j}$, then 
		$a'$ is also strictly admissible. Let $a''$ be obtained from $a'$ 
		by setting $a''_j(x)=a'_j(\fromvar{x}{i})$ and $a''_i(x)$ as undefined. 
		$a''$ is strictly admissible for $\co{\psi(x)}{j}$ and therefore it can be extended 
		to $a'''$, such that $\M\models\co{\theta(\fromvar{x}{j})}{k}[a''']$. Let $a^*$ be the assignment 
		obtained by extending $a$ with $a^*_k(\fromvar{x}{i})=a'''_k(\fromvar{x}{j})$. 
		The fact that $\dr_{ik}=\dr_{ij}\circ\dr_{jk}$ implies that $a_i(x)\in\dr_{ik}(a^*(\fromvar{x}{i}))$. 
		Furthermore, $\M\models\co{\theta(\fromvar{x}{j})}{k}[a']$ implies that 
		$\M\models\co{\theta(\fromvar{x}{j})}{k}[a^*]$.
		\item $\M\models\phi(\tovar{x}{k})[a]$, implies that $a_i$ is defined 
		  on $\tovar{x}{k}$ and that $(a_i(\tovar{x}{k}),a_k(x))\in\dr_{ik}$. By condition 
		  $\com_{ijk}$ there is a $d\in\dom_j$ such that $(a(\tovar{x}{k}),d)\in\dr_{ij}$ 
		  and $(d,a_k(x))\in\dr_{jk}$. Let us assume, without loss of generality that $a_j(x)=d$. 
		  Let $a'$ be an extension of $a$ with $a'_i(\tovar{x}{j})=a_i(\tovar{x}{k})$ and $a_j(x)=d=a'_j(\tovar{x}{k})$. 
		  The fact that $M_i\models\phi(\tovar{x}{k})[a]$ implies that 
		  $M_i\models\phi(\tovar{x}{j})[a']$. The fact that 
		  $\M\models i:\phi(\tovar{x}{j})\rightarrow j:\psi(x)$ implies that 
		  $M_j\models\psi(x)[a']$, and since $a'_j(x)=a'_j(\tovar{x}{k})$ we have that 
		  $M_j\models\psi(\tovar{x}{k})[a']$. The fact that 
		  $\M\models j:\psi(\tovar{x}{k})\rightarrow k:\theta(x)$ implies that 
		  $M_k\models\theta(x)[a']$, which, in turn means that $M_k\models\theta(x)[a]$. 
		\end{enumerate}

		\item \textbf{Existential and universal quantification}.
		\begin{enumerate}
		\item Suppose that $\M\models\co{\exists x\phi(x)}{i}$, then since
		  $\phi$ is complete, there is an assignment $a$, defined only on
		  $a_i(x)$ such that $\M\models\co{\phi(x)}{i}[a]$. This implies that
		  $a$ can be extended to $a'$, such that $\M\models\co{\psi(\fromvar{x}{i})}{j}[a']$. 
		  This trivially implies that $\M\models\co{\exists x\psi(x)}{j}$.
		\item Let $d\in\dom_j$. The fact that $\dr_{ij}$ is surjective implies 
		  that there is a $d'\in\dom_i$ with $\npla{d,d'}\in\dr_{ij}$. 
		  Let $a$ be an assignment with $a_i(\tovar{x}{j})=d'$ and $a_j(x)=d$. 
		  This assignment is admissible for $i:\phi(\tovar{x}{j})$. The fact 
		  that $\M\models\co{\forall x\phi(x)}{i}$, implies that 
		  $\M\models\co{\phi(\tovar{x}{j})}{i}[a]$. The fact that 
		  $\M$ satisfies that bridge rule 
		  $\co{\phi(\tovar{x}{j})}{i}\rightarrow\co{\psi(x)}{j}$ implies that 
		  for all $m\in M_j$, $m\models\psi(x)[a]$. We can therefore conclude that 
		  each $m\models\forall x\psi(x)$. 
		\end{enumerate}
		\item \textbf{Instantiation}.
		If $\M\models\co{\phi(t)}{i}$ (no assignment is necessary as $\phi(t)$
		does not contain any free variable) if $\M\models\co{\phi(x)}{i}[a]$
		where $a_i(x)$ is equal to the interpretation of $t$ in all the models
		of $S_i$. Such a unique value exists since $t$ is a complete
		term. Furthermore $\M\models\co{x=t}{i}[a]$. From the fact that
		$\M\models\co{x=t}{i}\rightarrow\co{\fromvar{x}{i}=s}{j}$, $a$ can be
		extended to $a'$, where $a'_j(x)$ is equal to the interpretation of
		$s$ in all the local models of $S_i$. 
		Let $a''$ be the assignment that assigns $a_i''(\tovar{x}{j})=a_i(x)$ and 
		$a''_j(x)=a'_j(\fromvar{x}{i})$. $a''$ is strictly admissible for
		$\co{\phi(\tovar{x}{j})}{i}$, and $\M\models\co{\phi(\tovar{x}{j})}{i}[a'']$. 
		The fact that
		$\M\models\co{\phi(\tovar{x}{j})}{i}\rightarrow\co{\psi(x)}{j}$, implies
		that $\M\models\co{\psi(x)}{j}[a'']$ and since $a''_j(x)$ is equal to
		the interpretation of $s$ in all the local models of $S_i$,
		$\M\models\co{\psi(s)}{j}$.

		\item \textbf{Inversion}.
		If $\M\models\co{\neg\psi(\fromvar{x}{i})}{j}[a]$, then either
		$S_i=\emptyset$ and $\M\models\co{\bot}{j}$, or
		$\M\not\models\co{\psi(\fromvar{x}{i})}{j}[a]$. In the first case, 
		since $\M\models\co{\bot}{j}\rightarrow\co{\bot}{i}$, we have that
		$\M\models\co{\bot}{i}$ which implies that
		$\M\models\co{\phi(x)}{i}[a]$. In the second case, let us suppose 
		by contradiction that $\M\not\models\co{\neg\phi(x)}{i}[a]$. 
		Since $\phi(x)$ is a complete formula 
		$\M\models\co{\phi(x)}{i}[a]$. This means that 
		$\M\models\co{\phi(x)}{i}[a']$. where $a'$ is the restriction 
		of $a$ to the value of $a_i(x)$. 
		Since $\M\models\co{\phi(x)}{i}\rightarrow\co{\psi(\fromvar{x}{i})}{i}$, 
		there is an extension $a''$ to $a'$, such that
		$\M\models\co{\psi(\fromvar{x}{i})}{i}[a'']$, 
		The fact that $\dr_{ij}$ is a function implies that
		$a_j(\fromvar{x}{i})=a''_j(\fromvar{x}{i})$. This implies that
		$\M\models\co{\psi(\fromvar{x}{i})}{j}[a]$, which contradict the initial
		hypothesis.
	\end{itemize}
	}
\end{pf}

To show the usefulness of bridge rules entailment consider a simple scenario composed of three ontologies $O_1, O_2$, and $O_3$, pairwise connected by means of the following DDL mappings: 
\begin{align}
	\co{AcademicPaper}{1} &\into \co{AcademicPaper}{2}\\
	\co{Document}{2} &\into \co{Document}{3}
\end{align}
and where $O_2$ contains the following terminological axiom $\co{AcademicPaper}{2} \isa \co{Document}{2}$. If we translate the DDL formulas into corresponding DFOL statements as follows:  
\begin{align}
	& \co{AcademicPaper(\tovar{x}{2})}{1} \rightarrow \co{AcademicPaper(x)}{2}\\
	& \co{Document(\tovar{x}{3})}{2} \into \co{Document(x)}{3}\\
	& \co{\forall x.AcademicPaper(x) \imp Document(x)}{2}
\end{align}
and we impose $\com_{123}$ between the three ontologies we can use a slight modification of the proof of \textbf{Composition} above (item 2) to show that $\co{AcademicPaper(\tovar{x}{3})}{1} \rightarrow \co{Document(x)}{3}$ \ftext{holds. This}, in turn, can be translated into the DDL mapping $\co{AcademicPaper{}}{1} \into \co{Document}{3}$. We have intentionally chosen a simple scenario. Nonetheless, being able to compute this inferred mapping may be crucial in the presence of a rich network of mappings containing also assertions $\co{AcademicPaper}{1} \into \co{RethoricalWriting}{3}$ and $\co{Document \isa \neg RethoricalWriting}{3}$. In that case mapping entailment would enable us to spot an inconsistent set of mappings, paving the way to techniques of mapping debugging / repair~\cite{ChristianMeilicke10012009, Wang:2008jk}.

%% file: calculus.tex
\section{Logical reasoning for the bridge rules}
\label{sec:calculus}

In this section we define a Natural Deduction (ND) Calculus for DFOL: given a set of bridge rules $\IC$ we define a calculus $\MC{\IC}$ which is 
strongly sound and complete with respect to the notion $\models_{\BR}$ of logical consequence w.r.t. $\BR$. 
The calculus provides a proof-theoretic counterpart of the notion of entailment between bridge rules introduced in Section \ref{subsec:derived-bridge-rules}, and can be therefore used to support formal reasoning in DFOL. By applying a finite set of inference rules, one can prove, for instance, 
that a set of bridge rules is consistent, or that a bridge rule is redundant being derivable from others, or that two set\ftext{s} of bridge rules are equivalent, and so on.

%

We follow the approach of \emph{Multi Language Systems} (ML systems) \cite{giunchiglia38,serafini8} and see a deduction in DFOL as composed of a set of \emph{local deductions}, which represent reasoning in a single theory, glued together by the applications of bridge rules, which enable the transfer of truth from a local knowledge base to another. For instance, the bridge rule $\co{\phi(x)}{i}\rightarrow\co{\psi(\fromvar{x}{i})}{i}$ can be read as 
\begin{quote}
	``if a certain object $x$ has the property $\phi$ in $i$, then, it has a translation
	$\fromvar{x}{i}$ in $j$ which has the property $\psi$''.
\end{quote}

\subsection{A Multi Language System for DFOL}

A \emph{ML system} is a triple $\npla{\{L_i\},\{\Omega_i\},\Delta}$  
where $\{L_i\}$ is a family of languages, $\{\Omega_i\}$ is a family
of sets of axioms, and $\Delta$ is a set of inference rules. $\Delta$
contains two kinds of inference rules: rules with premises and conclusions in the
same language, and rules with  premises  and  conclusions belonging  to  different languages. 

Derivability  in   a  ML  system  is  a   generalisation  of  derivability  in  a \ftext{N}atural \ftext{D}eduction  system.

In adapting the original definition of ML system given in \cite{giunchiglia38,serafini8} to the case of DFOL we require each $L_i$ to be a first\ftext{-}order language with equality. This can be axiomatised by setting $\Omega_i$ as the set of classical Natural Deduction axioms given in~\cite{prawitz1}, and the rules in $\Delta$ that take care of connectives, quantifiers, and equality to mimic the inference rules given in~\cite{prawitz1}. As we will see, we have to slightly modify the applicability conditions of these rules in order to deal with arrow variables in a proper manner.
Moreover, $\Delta$ has to contain the Natural Deduction version of the DFOL bridge rules introduced in Definition \ref{def:bridge-rule}, and of the logical properties of arrow variables stated in Proposition \ref{prop:logical-consequence-arrow-variable}.

Notationally, we use $\phi^t_x$ to indicate the result of replacing $t$ for all the free occurrences of $x$ in $\phi$, provided that $x$ does not occur free in
the scope of a quantifier of some variable of $t$.

\newcommand{\vgap}{.5cm}

\begin{defn}
  \label{def:mlsystem}
  The ML system $\MC{\IC}$ for a DFOL with languages $\{L_i\}$ and bridge rules $\IC$ is the triple 
  $\npla{\{L_i\},\{\Omega_i\},\Delta}$, where $\Omega_i$ is empty and $\Delta$ contains the following inference rules:
  
  \begin{mdframed}
  \small
  $$  
  \begin{array}{c@{\hspace{.5cm}}c@{\hspace{.7cm}}c@{\hspace{.5cm}}c}
  \infer[\impi{i}]
    {\co{\phi\imp \psi}{i}}
    {\infer*
      {\co{\psi}{i}}
      {[\co{\phi}{i}]}}  
  &  
  \infer[\impe{i}]
   {\co{\psi}{i}}
   {\co{\phi}{i} & \co{\phi\imp \psi}{i}} 
  &
   \infer[\andi{i}]
     {\co{\phi\con \psi}{i}}
     {\co{\phi}{i} &\co{\psi}{i}}  
  &  
   \infer
     {\co{\phi}{i}}
     {\co{\phi\con \psi}{i}} \hspace{.4cm}
   \infer[\ande{i}]
     {\co{\psi}{i}}
     {\co{\phi\con \psi}{i}} 
  \end{array}
  $$
  $$
  \begin{array}{c@{\hspace{1cm}}c@{\hspace{1cm}}c@{\hspace{1cm}}c}
   \infer
     {\co{\phi\dis \psi}{i}}
     {\co{\phi}{i}} \hspace{.5cm}
   \infer[\ori{i}]
     {\co{\psi\dis \phi}{i}}
     {\co{\phi}{i}} 
   &
    \infer[\ore{ji}]
  {\co{\theta}{i}}
  {\co{\phi\dis\psi}{j} &
    \infer*
    {\co{\theta}{i}}
    {[\co{\phi}{j}]} &
    \infer*
    {\co{\theta}{i}} 
    {[\co{\psi}{j}]}}
	&
    \infer[\bot_i]
    {\co{\neg \phi}{i}}
    {\infer*
      {\co{\bot}{i}}
      {[\co{\phi}{i}]}}
 \end{array}
 $$
%
%
$$
 \begin{array}{c@{\hspace{1cm}}c@{\hspace{1cm}}c@{\hspace{1cm}}c}
    \infer[\alli{i}]
       {\co{\forall x \phi}{i}}
       {\co{\phi}{i}}
   &
   \infer[\alle{i}]
       {\co{\phi^t_x}{i}}
       {\co{\forall x \phi}{i}} 
  &
   \infer[\exi{i}]
       {\co{\exists x \phi}{i}}  
       {\co{\phi^t_x}{i}}
   &
     \infer[\exe{ji}]
      {\co{\psi}{i}}
      {\co{\exists x \phi}{j} & 
       \infer*
        {\co{\psi}{i}}
        {[{\co{\phi}{j}}]}} \\[\vgap]
\end{array}
$$
$$
  \begin{array}[t]{c@{\hspace{1cm}}c@{\hspace{1cm}}c}
  \infer[\eqi{i}]
     {\co{t=t}{i}}
     {\co{\phi_1}{i}, \ldots \co{\phi_n}{i}}
  & 
\infer[\eqe{i}]
     {\co{\phi^u_x}{i}}
     {\co{\phi^t_x}{i} & \co{t=u}{i}}
  &
     \infer[\cut{ji}]
      {\co{\psi}{i}}
      {\co{\phi}{j} & 
       \infer*
        {\co{\psi}{i}}
        {[\co{\phi}{j}]}}
	\end{array}
$$ 
\centerline{\text{$i$-rules: rules for connectives, quantifiers, equality, and Cut.}}
\end{mdframed}

\begin{mdframed}
\small
$$
  \begin{array}{c@{\hspace{1cm}}c@{\hspace{1cm}}c@{\hspace{1cm}}c}
  \infer[\fromI{i}I_{ij}]
     {j:\fromvar{x}{i}=y} 
     {i:x=\tovar{y}{j}} 
  & 
  \infer[\toI{i}I_{ij}]
     {j:\tovar{x}{i}=y}
     {i:x=\fromvar{y}{j}}
  \end{array}
  $$
  \begin{center}
	  $
	  \vcenter{\infer[\IC]
   		{i:\phi}
   		{\co{\phi_1}{i_1} & \ldots & \co{\phi_n}{i_n}}
		}
	$ \quad 
	for each $\co{\phi_1}{i_1},\ldots,\co{\phi_n}{i_n} \rightarrow\co{\phi}{i}$ in  $\IC$
	\end{center}
	
\centerline{\text{$b$-rules: rules for arrow variables and bridge rules.}}
\end{mdframed}
\end{defn}

A \emph{formula tree} in $\MC{\BR}$ is a tree $\Pi$ which is constructed starting from a set of assumptions and axioms by applying the $i$-rules and $b$-rules given above. The occurrence of an arrow variable in a node $\co{\phi}{i}$ of a formula tree $\Pi$ is called \emph{existential} if this arrow variable does not occur in the assumptions from which $\co{\phi}{i}$ depends on. Given a formula tree $\Pi$ with root $i:\phi$, an assumption $j:\psi$ is called \emph{local assumption} if 
$i=j$ and the branch from $i:\phi$ to $j:\psi$ contains only applications of $i$-rules. An assumption is \emph{global} if it is not local. A set of assumptions is local iff all the assumptions it contains are local. It is global otherwise. The distinction between local and global assumptions is necessary to correctly characterise the notion $\models_{\BR}$ of DFOL logical consequence where, as we have seen in Proposition~\ref{prop:no-deduction-theorem}, the deduction theorem only holds with complete formulas or local assumptions. This distinction will become clearer in discussing restriction R\ref{item:restr-disch-complete-global-assumptions} introduced in the next definition. We only remark here that an application of a $b$-rule makes all the assumptions become global, and this reflects the fact that the satisfiability of bridge rules is defined over sets of local models, instead of a single model.

\begin{defn}[{Derivability}]
$i:\phi$ is \emph{derivable} in $\MC{\BR}$ from a set of global assumptions $\Gamma$ and a set of local assumptions $\Sigma$, in symbols $(\Gamma,\Sigma)\vdash_{\BR} i:\phi$, 
if there is a formula tree $\Pi$ with root $i:\phi$, global assumptions $\Gamma$ and local assumptions $\Sigma$ such that
the following restrictions on the application of the rules in $\Delta$ are satisfied: 

\begin{enumerate}[{R}1.]
\item \label{item:no-existential-arrow-variables} The only rules whose premises can contain existential variables are $\cut{ji}$, $\ore{ji}$, and $\exe{ji}$.
\item \label{item:restr-introduce-arrow-variable} The only rules that can introduce new existential variables are $\toI{i}I_{ij}$, $\fromI{i}I_{ij}$ and $\BR$. In addition, the arrow variables contained in the conclusions of $\toI{i}I_{ij}$ and $\fromI{i}I_{ij}$ must be existential. 
\item \label{item:restr-disch-complete-global-assumptions} The application of $\impi{i}, \bot_{i}, \exe{ji}, \cut{ji}$ can discharge only assumptions that are either local or complete formulas. The application of $\ore{ji}$  can discharge only assumptions that are either local or such that at least one is a complete formula.
\item \label{item:restriction-cut} $\cut{ji}$ and $\exe{ji}$ can be applied only if the existential variables in $\co{\phi}{j}$ do not occur in any other assumption employed in the derivation of $\co{\psi}{i}$. $\ore{ji}$ can be applied only if the existential variables in $\co{\phi\dis\psi}{j}$ do not occur in any other assumption employed in the derivation of $\co{\theta}{i}$. 
\item \label{item:restriction-alli} $\alli{i}$ can be applied only if $x$ does not occur free in any assumption with index $i$, and $\tovar{x}{i}$ and $\fromvar{x}{i}$ do not appear in any assumption with index $j\neq i$.
\item \label{item:restriction-exe} $\exe{ji}$ can be applied only if $x$ does not occur free in any assumption with index $j$ different from $\co{\phi}{j}$. Moreover, if  $j=i$ then $x$ cannot occur free in $\co{\psi}{i}$, otherwise if $j \neq i$, then $\fromvar{x}{j}$ and $\tovar{x}{j}$ cannot occur in $\co{\psi}{i}$ or in any assumption employed to derive it. 
  \end{enumerate}
\end{defn}

$i$-rules $\impi{i}$--$\exe{ji}$ provide the DFOL version of Natural Deduction rules for logical connectives and quantifiers, respectively, while $i$-rules $\eqi{i}$ and $\eqe{i}$ are the DFOL version of Natural Deduction rules for the equality predicate. 
If we ignore the label of the formulae and restrictions R\ref{item:no-existential-arrow-variables}--R\ref{item:restriction-exe} (which will be illustrated in detail later), the shape of the inference rules for connectives, quantifie\ftext{r}s, and equality is the same as the ones of first\ftext{-}order logic with equality. 
Rules $\fromI{i}I_{ij}$ and $\toI{i}I_{ij}$ are the proof theoretical counterpart of Property~\ref{prop:logical-consequence-arrow-variable}. In particular, $\fromI{i}I_{ij}$ states that $x$ and $y$ belong to the domain relation $\dr_{ij}$, while $\toI{i}I_{ij}$ states that $x$ and $y$ belong to the domain relation $\dr_{ji}$.  
Rule \IC provides an axiomatisation of the propagation of knowledge enforced by bridge rule  $\co{\phi_1}{i_1},\ldots,\co{\phi_n}{i_n} \rightarrow\co{\phi}{i}$. 
Finally, $\cut{ji},\ore{ji}$, and $\exe{ji}$ together with restrictions R\ref{item:no-existential-arrow-variables} and R\ref{item:restriction-cut} regulate the usage of arrow variables within deduction trees and will be illustrated further in the remaining of the section. 

Restrictions R\ref{item:no-existential-arrow-variables}--R\ref{item:restriction-exe} are used to model the behaviour of local assumptions, global assumptions, and arrow variables. While restrictions R\ref{item:restriction-alli} and R\ref{item:restriction-exe} extend the restrictions of the FOL Natural Deduction rules $\alli{i}$ and $\exe{ji}$ to take into account the occurrence of arrow variables, restrictions R\ref{item:no-existential-arrow-variables}--R\ref{item:restriction-cut} are proper to  DFOL and deserve some explanation. 
Restriction R\ref{item:no-existential-arrow-variables} states that we cannot freely make inferences from inferred facts that contain existential arrow variables. 
In fact, existential arrow variables have, as their name suggest, an existential meaning. As a consequence\ftext{,} the same existential arrow variable occurring in, say, two different inferred formulae is not guaranteed to denote the same element of the domain in the proof. Therefore a way to control their usage in the proof tree is needed. To further clarify this point consider the following proof:
\begin{equation}
	\label{proof:andi}
	\infer[\andi{i}]
	  {\co{\psi_1 \con \psi_2 (\fromvar{x}{j})}{i}}
	  {\infer*
	    {\co{\psi_1(\fromvar{x}{j})}{i}}
	    {\Gamma_1}
	   \quad
	   \infer*
	    {\co{\psi_2(\fromvar{x}{j})}{i}}
	    {\Gamma_2}
	}
\end{equation}
where the application of $\andi{i}$ violates R\ref{item:no-existential-arrow-variables}. 
In this case the application of $\andi{i}$ allows to infer $\Gamma_1,\Gamma_2 \der{\IC} \co{\psi_1 \con \psi_2 (\fromvar{x}{j})}{i}$ from  $\Gamma_1\der{\IC}\co{\psi_1(\fromvar{x}{j})}{i}$ and $\Gamma_2 \der{\IC}\co{\psi_2 (\fromvar{x}{j})}{i}$. This inference is unsound. 
In fact, $\Gamma_1\models_{\IC}\co{\psi_1(\fromvar{x}{j})}{i}$ and $\Gamma_2 \models_{\IC}\co{\psi_2 (\fromvar{x}{j})}{i}$ guarantee that if $\M$ satisfies both $\Gamma_1[a]$ and $\Gamma_2[a]$, then there are two extensions $a'$ and $a''$ of $a$, admissible for $\co{\psi_1(\fromvar{x}{j})}{i}$ and $\co{\psi_2(\fromvar{x}{j})}{i}$ respectively, such that $\M$ satisfies both $\co{\psi_1(\fromvar{x}{j})}{i}[a']$ and $\co{\psi_2(\fromvar{x}{j})}{i}[a'']$. This unfortunately does not guarantee the existence of an extension $\overline{a}$ of $a$ admissible for $\co{\psi_1 \con \psi_2 (\fromvar{x}{j})}{i}$ such that $\M \models \co{\psi_1 \con \psi_2 (\fromvar{x}{j})}{i}[\overline{a}]$. In fact, assume that $a'(\fromvar{x}{j}) = d'$, $a'(\fromvar{x}{j}) = d''$, with $d'\neq d''$ where $d'$ is the only element of $\bdom_i$ in the interpretation of $\phi_1$ and $d''$ is the only element of $\bdom_i$ in the interpretation of $\phi_2$. It is easy to see that for such a model  $\Gamma_1\models_{\IC}\co{\psi_1(\fromvar{x}{j})}{i}$ and $\Gamma_2 \models_{\IC}\co{\psi_2 (\fromvar{x}{j})}{i}$, but $\Gamma_1,\Gamma_2 \not\models_{\IC}\co{\psi_1 \con \psi_2 (\fromvar{x}{j})}{i}$. 
To avoid unsound inferences of this kind we provide the ability to infer from formulas containing arrow existential variables using only rules which: (i) combine different proof trees, and (ii) infer one of the premises of the rule, possibly discharging assumptions, as in the case of $\ore{ji}, \exe{ji}$ and $\cut{ji}$. 

$\cut{ji}$ is the rule that takes mostly care of existential arrow variables in proofs. The idea here is that if we have an inference $\Pi$ of $\co{\alpha}{k}$ from $\Gamma$ which makes use of an inference rule whose premises contain $\co{\phi(\bothvar{x})}{i}$, with $\bothvar{x}$ existential arrow variable, then we can split this inference in two parts $\Pi_1, \Pi_2$ and then ``glue'' them with an application of $\cut{}$ as depicted below: 
\begin{center}
\includegraphics[width=.6\textwidth]{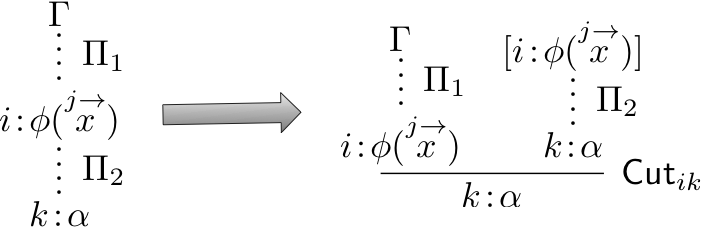}	
\end{center}
Restriction R\ref{item:restriction-cut} ensures that we can perform this ``gluing'' only for sound deductions. For instance, we can use the $\cut{}$ rule to enable a sound application of $\andi{i}$ as in the following proof tree  
\begin{equation}
	\label{eq:cut}
\infer[\cut{ii}]
  {\co{\psi_1(\fromvar{x_1}{j}) \con \psi_2 (\fromvar{x_2}{j})}{i}}
  {\infer*
	{\co{\psi_1(\fromvar{x_1}{j})}{i}}
	{\Gamma_1}
   	\quad
   	\infer[\cut{ii}]
	{\co{\psi_1(\fromvar{x_1}{j}) \con \psi_2 (\fromvar{x_2}{j})}{i}}
	{\infer*
		{\co{\psi_2(\fromvar{x_2}{j})}{i}}
		{\Gamma_2} 
	 	\quad
	  	\infer[\andi{i}]
			{\co{\psi_1(\fromvar{x_1}{j}) \con \psi_2 (\fromvar{x_2}{j})}{i}}
			{[\co{\psi_1(\fromvar{x_1}{j})}{i}] \quad [\co{\psi_2(\fromvar{x_2}{j})}{i}]}
	}}
\end{equation}
while we cannot use the $\cut{}$ rule to enable an unsound application of $\andi{i}$ to obtain $\co{\psi_1 \con \psi_2 (\fromvar{x}{j})}{i}$ from $\co{\psi_1(\fromvar{x}{j})}{i}$ and $\co{\psi_2(\fromvar{x}{j})}{i}$ as in proof \eqref{proof:andi}. The key point in proof \eqref{eq:cut} is obviously the occurrence of two distinct arrow variables $\fromvar{x_1}{j}, \fromvar{x_2}{j}$, which rules out the scenario described in explaining proof \eqref{proof:andi}.

Restriction R\ref{item:restr-introduce-arrow-variable} regulates (prevents) the introduction of new existential variables in the proof. In fact, we must avoid the introduction of terms (existential variables, in this case) which may not denote any element. 
Consider, for instance, the following unrestricted application of $\ori{i}$ 
$$\infer[\ori{i}]
  {\co{\phi\vee \psi(\tovar{x}{j})}{i}}
  {\co{\phi}{i}}
$$
with $\tovar{x}{j}$ new existential variable. This inference is unsound. In fact, given an assignment $a$ for $\co{\phi}{i}$, we cannot guarantee the existence of an extension $a'$ admissible for $\tovar{x}{j}$ in $i$ (a trivial counter-model is the one with $\dr_{ij}=\emptyset$). However, if $\co{\phi}{i}$ depends upon an assumption $\co{\gamma(\tovar{x}{j})}{i}$, then the application of $\ori{i}$ satisfies restriction R\ref{item:restr-introduce-arrow-variable} (as $\tovar{x}{j}$ is not existential anymore) and the inference of $\co{\phi\vee \psi(\tovar{x}{j})}{i}$ from $\co{\phi}{i}$ is sound. In this case\ftext{,} assumption $\co{\gamma(\tovar{x}{j})}{i}$ forces $a$ to be already admissible for $\tovar{x}{j}$ removing the obstacle shown above. 
Definition \ref{def:satisfiability-bridge-rules} and Proposition \ref{prop:logical-consequence-arrow-variable} instead ensure that $\toI{i}I_{ij}$, $\fromI{i}I_{ij}$ and $\BR$ can safely introduce new existential arrow variables. 

Restriction R\ref{item:restr-disch-complete-global-assumptions} reflects the fact that $\models_{\BR}$ is defined over sets of local models, rather than a single model, and that this can cause the failure of the deduction theorem, as seen in Proposition~\ref{prop:no-deduction-theorem}. 
Thus, to ensure soundness of the inference rules we have to force global assumptions to be complete (in $\ore{ji}$ at least one of the disjuncts to be complete). If proofs consist only of local assumptions, then the requirement of being a complete formula can be dropped. In this case, in fact, $\models_{\BR}$ reduces to first\ftext{-}order logical consequence (modulo arrow variables) as illustrated by property \eqref{th:fol-lc-conservative-extension} in Proposition \ref{prop:logical-consequence}. 

We conclude the formal presentation of the DFOL calculus by proving that bridge rules are directional: 
\begin{prop}[Directionality]
	\label{prop:directionality}
	Given a set $\BR$ of bridge rules such that $k$ does not appear in the conclusion of a rule neither as the index of the conclusion nor as an index of an arrow variable, then $\Gamma\vdash_{\BR} k:\phi$ iff $\Gamma_k\vdash\phi$
	\end{prop}
The proof easily follows from the observation that the ML system $\MC{\IC}$ does not contain any deduction rule which enables to infer a formula in $k$ (apart from local inference rules) unless $k$ appears in the conclusion of a bridge rule as the index or as an index of an arrow variable. By showing that the DFOL calculus is a sound and complete axiomatisation of the notion of logical consequence of DFOL (Section~\ref{sec:soundnesscompleteness}) we can transfer the result of Proposition~\ref{prop:directionality} to easily show that, for the specific set $\BR$ of Proposition~\ref{prop:directionality}, $\Gamma\models_{\BR} k:\phi$ iff $\Gamma_k\models\phi$.

We illustrate now the usage of the calculus by applying it to the Magic box scenario. 
For the sake of presentation we present the proof using a linear notation (similar to the Lemmon-style for ND~\cite{LemmonLogic}) rather than a tree-based one. In this notation, each line of the deduction (the deduction step) contains a label, the inferred formula, the set of assumptions from which the inferred formula depends upon, and the inference rule used in the deduction step.
\onlyinpaper{Additional examples of proofs, which show how the calculus can be used to infer the statements corresponding to the entailed bridge rules of Proposition~\ref{prp:entailment-of-br} can be found in~\cite{DFOL-techrep}.}
\onlyintechrep{Additional examples of proofs, which show how the calculus can be used to infer the statements corresponding to the entailed bridge rules of Proposition~\ref{prp:entailment-of-br} can be found in Appendix~\ref{sec:appendix-deductions}.}

\begin{exmp}
	\label{ex:calculus-magic-box}
Let us consider the formalisation of the magic box presented in Example \ref{ex:formalization-magic-box}. Figure~\ref{fig:mbox-deduction} shows a proof where we use the $b$-rules derived from bridge rules~\eqref{eq:mbox-inboxright1-imp-existence2} and \eqref{eq:mbox-inboxleft-and-empty-right1-imp-existence2} to prove that if $\obsuno$ sees a ball in the box, then $\obsdue$ sees a ball in the box too, that is,: 
$$
\co{\exists x \exists y \ \inbox(x,y)}{1}\vdash\co{\exists x\exists y \ \inbox(x,y)}{2}
$$
Notationally, we use $\IC_{(n)}$ to denote the $b$-rule corresponding to the bridge rule in Equation $(n)$. We also abbreviate ``\textit{left}'' to ``\textit{l}'' and ``\textit{right}'' to ``\textit{r}''.
\newcommand{\sinistra}{\textit{l}}
\newcommand{\destra}{\textit{r}}

\begin{figure}[ht]
   $$
   \begin{ndproof}
     (1) & \co{\exists x\exists y \ \inbox(x,y)}{1} 
         & (1) & & \shortreason{Assumption} \\
     (2) & \longformula{$1:\exists x (\inbox(x,\destra)  \vee$ \\ 
           \hspace*{3pt}  $(\inbox(x,\sinistra) \wedge \bempty(\destra)))$}
         & (1) & & \shortreason{From (1) and local axioms using local rules in $1$} \\ 
     (3) & \longformula{$1:\inbox(x,\destra)  \vee$ \\
           \hspace*{3pt}  $(\inbox(x,\sinistra)\wedge\bempty(\destra))$}
         & (3) & & \shortreason{Assumption} \\
     (4) & 1:\inbox(x,\destra)
         & (4) & & \shortreason{Assumption} \\
     (5) & \co{\exists y \ \inbox(\fromvar{x}{1},y)}{2} 
         & & (4) & \shortreason{From (3) by $\IC_{(\ref{eq:mbox-inboxright1-imp-existence2})}$} \\
     (6) & \co{\exists y \ \inbox(\fromvar{x}{1},y)}{2} 
         & (6) & & \shortreason{Assumption} \\
	 (7) & \co{\exists x \exists y \ \inbox(x,y)}{2} 
	     & (6) & & \shortreason{From (6) by $\exi{2}$} \\
	 (8) & \co{\exists x \exists y \ \inbox(x,y)}{2} 
	     & & (4) & \shortreason{From (5) and (7) by $\cut{12}$} \\
     (9) & 1:\inbox(x,\sinistra)\wedge\bempty(\destra)
         & (9) & & \shortreason{Assumption} \\
    (10) & \co{\exists y.\inbox(\fromvar{x}{1},y)}{2} 
         & & (9) & \shortreason{From (9) by $\IC_{(\ref{eq:mbox-inboxleft-and-empty-right1-imp-existence2})}$} \\
    (11) & \co{\exists y \ \inbox(\fromvar{x}{1},y)}{2} 
         & (11) & & \shortreason{Assumption} \\
    (12) & \co{\exists x \exists y \ \inbox(x,y)}{2} 
	     & (11) & & \shortreason{From (11) by $\exi{2}$} \\
	(13) & \co{\exists x \exists y \ \inbox(x,y)}{2} 
		 & & (9) & \shortreason{From (9) and (11) by $\cut{12}$} \\
    (14) & \co{\exists x \exists y \ \inbox(x,y)}{2} 
         & & (3) & \shortreason{From (3), (8), and (13) by $\ore{12}$ discharging (4) and (9)} \\
    (15) & \co{\exists x \exists y \ \inbox(x,y)}{2} 
         & & (1) & \shortreason{From (2) and (14) by $\exe{12}$ discharging (3)} \\ \hline
   \end{ndproof}
   $$
   \caption{A derivation of $\co{\exists x \exists y \ \inbox(x,y)}{2}$ from $\co{\exists x \exists y \ \inbox(x,y)}{1}$.}
\label{fig:mbox-deduction}
\end{figure}

The deduction starts from the assumption that, according to $\obsuno$, there is a ball in the box (assumption (1)). From this fact, we can use local axioms and inference rules in the knowledge base of $\obsuno$ to prove that the ball is in the right hand side slot or it is in the left hand side slot and the right hand side one is empty (these steps are omitted for the sake of presentation as our focus is on the usage of $b$-rules). We now reason by cases, considering first the case in which the ball is on the right hand side (step 4) and then the case in which the ball \ftext{is} in the left hand side slot and the right hand side is empty (step 9). In both cases we can infer that $\obsdue$ also sees a ball (steps (8) and (12), respectively). This is done by using the $b$-rules $\IC_{(\ref{eq:mbox-inboxright1-imp-existence2})}$ and $\IC_{(\ref{eq:mbox-inboxleft-and-empty-right1-imp-existence2})}$, and by using the $\cut{}$ rule to handle the existential arrow variables introduced by $\IC_{(\ref{eq:mbox-inboxright1-imp-existence2})}$ and $\IC_{(\ref{eq:mbox-inboxleft-and-empty-right1-imp-existence2})}$. We can therefore use the ``or elimination'' rule to infer that $\obsdue$ also sees a ball directly from the original ``or'' formula assumed in step (3), and then from assumption (1) by an application of an ``exist elimination'' rule. 
\end{exmp}

\onlyintechrep{
\begin{exmp}
In Figure \ref{fig:conjunction-deduction} we show a proof of $j:\psi\wedge\psi'(\fromvar{x}{i})$ from $i:\phi\wedge\phi'(x)$, using the bridge rules 
\begin{align}
	& \tag{$\fun_{ij}$} \co{\tovar{x}{j}=\tovar{y}{j}}{i}\rightarrow\co{x=y}{j} \\
	\label{eq:br1}& i:\phi(x)\rightarrow j:\psi(\fromvar{x}{i})\\
	\label{eq:br2}& i:\phi'(x)\rightarrow j:\psi'(\fromvar{x}{i})
\end{align}
 Note that this proof constitutes an example of how the calculus can be used to infer statements corresponding to the entailed bridge rules of Proposition~\ref{prp:entailment-of-br}, where the the current example corresponds to the case of Conjunction. All the remaining cases of Proposition~\ref{prp:entailment-of-br} are shown in~\ref{sec:appendix-deductions}.

\begin{figure}[htbp]
	\centering
   \scalebox{0.82}{
	$
	\begin{ndproof}
	    (1) & i:\phi(x)\wedge\phi'(x) & (1) & & \reason{Assumption} \\ 
		(2) & i:\phi(x) & (1) & & \reason{From (1) by $\ande{i}$} \\ 
		(3) & j:\psi(\fromvar{x}{i}) & & (1) & \reason{From (2) by $\BR_{\eqref{eq:br1}}$} \\ 
		(4) & i:\phi'(x) & (1) & & \reason{From (1) by $\ande{i}$} \\
		(5) & i:x=z & (5) & & \reason{Assumption} \\
		(6) & i:\phi'(z) & (1)(5)& & \reason{From (4), (5) by $\eqe{i}$} \\		
		(7) & j:\psi'(\fromvar{z}{i}) & & (1)(5) & \reason{From (6) by $\BR_{\eqref{eq:br2}}$} \\		
		(8) & j:\psi(\fromvar{x}{i}) & (8) & & \reason{Assumption} \\
		(9) & j:\psi'(\fromvar{z}{i}) & (9) & & \reason{Assumption} \\
		(10) & 	j:\psi(\fromvar{x}{i}) \con \psi'(\fromvar{z}{i}) & (8)(9) & & \reason{From (8) and (9) by $\andi{j}$} \\	
		(11) & 	j:\psi(\fromvar{x}{i}) \con \psi'(\fromvar{z}{i}) & (9) & (1) & \reason{From (3) and (10) by $\cut{ji}$ discharging (8)} \\	
		(12) & 	j:\psi(\fromvar{x}{i}) \con \psi'(\fromvar{z}{i}) & & (1) & \reason{From (7) and (11) by $\cut{ji}$ discharging (9)} \\	
		(13) & j:\fromvar{x}{i}=y & (13) & & \reason{Assumption} \\ 
		(14) & i:x=\tovar{y}{j} & & (13) & \reason{From (13) by $\toI{j}I_{ji}$} \\
		(15) & j:\fromvar{z}{i}=w & (15) & & \reason{Assumption} \\ 
		(16) & i:z=\tovar{w}{j} & & (15) & \reason{From (15) by $\toI{j}I_{ji}$} \\ 
	    (17) & i:\tovar{y}{i}=\tovar{w}{i} & (5) & (13)(15) & \reason{from (5), (14) and (16) by $\eqe{i}$ and applications of $\cut{}$ to handle existential arrow variables} \\ 
		(18) & j:y=w & & (5)(13)(15) & \reason{From (17) by $\IC_{\fun_{ij}}$}  \\ 
		(19) & j:\fromvar{x}{i}=\fromvar{z}{i} & &(5)(13)(15) &  \reason{From (13), (15) and (19) by $\eqe{j}$}\\
		(20) & j:\psi\con \psi'(\fromvar{x}{i}) & & (1)(5)(13)(15) & \reason{From (12) and (19) by $\eqe{j}$} \\ 	
		(21) & j:\fromvar{x}{i}=\fromvar{x}{i}& & (1) & \reason{From (3) by $\eqi{j}$} \\ 
		(22) & j:\exists y \ \fromvar{x}{i}=y & & (1) & \reason{From (21) by $\exi{j}$} \\ 
		(23) & j:\fromvar{z}{i}=\fromvar{z}{i} & & (1)(5) & \reason{From (7) by $\eqi{j}$} \\ 
		(24) & j:\exists w \ \fromvar{z}{i}=w & & (1) & \reason{From (23) by $\exi{j}$} \\ 
		(25) & j:\psi\con \psi'(\fromvar{x}{i}) & &(1)(5)(15) &  \reason{From (20) and (21) by $\exe{j}$ discharging (13)}\\		
		(26) & j:\psi\con \psi'(\fromvar{x}{i}) & &(1)(5) &  \reason{From (24) and (25) by $\exe{j}$ discharging (15)}\\ 
		(27) & i:z=z & & & \reason{By $\eqi{i}$}\\
		(28) & i:\exists x \ x=z & & & \reason{From (27) by $\exi{i}$}\\
		(29) & j:\psi\con \psi'(\fromvar{x}{i}) & & (1) &  \reason{From (26) and (27) by $\exe{ji}$ discharging (5)}\\ \hline
	\end{ndproof}
	$
	}
	\caption{An articulated DFOL derivation.}
\label{fig:conjunction-deduction}
\end{figure}

   Notice that, in the deduction shown in Figure~\ref{fig:conjunction-deduction} we need to rename the variable $x$ with a fresh variable $z$ in step (5) in order to be able to correctly apply the cut rule (see Restriction~\ref{item:restriction-cut}) to infer $j:\psi(\fromvar{x}{i}) \con \psi'(\fromvar{z}{i})$ and to avoid problems as the ones illustrated with the unsound application of the $\andi{i}$ rule in Equation \eqref{proof:andi}. The $b$-rule $\IC_{\fun_{ij}}$, which correspond to functional domain relations, is then used to infer $j:\fromvar{x}{j} = \fromvar{z}{j}$ (step (18)) and this enables the replacement of $\fromvar{z}{i}$ with $\fromvar{x}{i}$ in $j:\psi'$ in order to obtain the desired formula. 
\end{exmp}
}


%% file: soundAndcomplete.tex
\subsection{Soundness and Completeness}
\label{sec:soundnesscompleteness}

The goal of this section is to show that the calculus defined in Section~\ref{sec:calculus} for a given set $\IC$ of bridge rules 
is sound and complete with respect to the class of $\IC$-models defined in Section~\ref{sec:semantics}.
In  \ref{sec:soundness} we prove the Soundness Theorem and in  \ref{sec:completeness} the Completeness Theorem. The main body of \ref{sec:completeness} concentrates on a method for constructing $\IC$-canonical models.

Before stating the correspondence between $\vdash_{\IC}$ and $\brmodels$ which we are going to prove in this section we need to introduce some notation. 
Given a set of formulas $\Sigma$, we use $e(\Sigma)$ to denote the set of formulas $\co{\exists y \ y=\bothvar{x}}{i}$ such that $\bothvar{x}$ is an arrow variable that occurs in a formula in $\Sigma$. $e(\Sigma)$ intuitively contains the statements of existence for all the arrow variables in $\Sigma$. 

\begin{thm}[Soundness and Completeness theorem]
\label{theo:soundness-and-completeness}
$$
(\Gamma,\Sigma)\vdash_{\IC}\co{\phi}{i} \Longleftrightarrow
\Gamma,e(\Sigma)\models_{\IC} \co{\bigwedge_{i:\sigma \in \Sigma}\sigma\imp\phi}{i}
$$
where $\Gamma$ is the set of global assumptions and $\Sigma$ is the set of local assumptions of the formula tree $\Pi$ with root $\co{\phi}{i}$.
\end{thm}

This theorem states that the calculus defined in the previous section computes a derivability relation $\vdash_{\IC}$ which corresponds to the consequence relation $\brmodels$ between 
the global assumptions $\Gamma$ of $\vdash_{\IC}$ and the logical implication of $\co{\phi}{i}$ from the conjunction of all the local assumptions in $\Sigma$ (that is, $\co{\bigwedge_{i:\sigma \in \Sigma}\sigma\imp\phi}{i}$), modulo the existence of all the arrow variables in $\Sigma$ (that is, $e(\Sigma)$). Its proof is a direct consequence of the proofs of Soundness and Completeness \onlyinpaper{summarised}\onlyintechrep{that can be found} in \ref{sec:soundness} and \ref{sec:completeness}, respectively\onlyinpaper{, and provided in full in \cite{DFOL-techrep}}.


%% file: relatedwork.tex
\newcommand\name[1]{\ulcorner#1\urcorner}
\def\D{\mathcal{D}}
\def\K{\mathcal{K}}
\def\L{\mathcal{L}}
\def\QLCM{\mathcal{M_{\text{QLC}}}}
\def\QMLM{\mathcal{M_{\text{QML}}}}
\def\CPM{\mathcal{M_{\text{CP}}}}

\def\BRQLC{\BR_{\text{QLC}}}
\def\BRQML{\BR_{\text{QML}}}
\def\BRCP{\BR_{\text{CP}}}
\def\R{\mathcal{R}}
\def\W{\mathcal{W}}
\def\V{\mathcal{V}}
\def\ist{\textit{ist}}
\def\name#1{\text{`}#1\text{'}}
\def\tauQML{\tau_{\text{QML}}}
\def\tauQLC{\tau_{\text{QLC}}}
\def\tauCP{\tau_{\text{CP}}}
\def\C{\mathcal{C}}
\def\T{\mathcal{T}}

\section{\ftext{Analysing} formalisms for distributed knowledge through DFOL} 
\label{sec:rel-work} 

The need to represent and reason about distributed and context\rtext{-}dependent knowledge able to deal with semantic heterogeneity
has fostered the development of various logical formalisms. Areas such as the Semantic Web, Databases, Linked Data, and Peer-to-Peer systems have seen a quest for logics able to represent and reason about knowledge contained in sets of different knowledge bases \ftext{that describe} overlapping knowledge by means of heterogeneous schemata. Examples are: mappings between overlapping ontologies or DB schemas, or relations between different contexts.  All these formalisms make (implicit or explicit) assumptions about the following questions: (i) what is the structure (hierarchical, peer-to-peer, mediator based) in which the different local knowledge bases are embedded?; (ii) which is the type of knowledge that can be represented in each KB (e.g., only local knowledge, views on knowledge of other KB\ftext{s}, \dots)?; (iii) what type of domain is used to interpret the local knowledge (i.e., local domain or global domain)?; (iv) are there \ftext{any} relations between local domains and which ones (e.g., intersection, mapping, subset, identity, \ldots)?; (v) what are the relations between local truth \ftext{in different KBs}? In this section, we consider a significant number of the most relevant first\ftext{-}order logic based frameworks for the representation of static \ftext{and} semantically heterogeneous distributed knowledge-base systems and show how their encoding in DFOL allows us to make \ftext{these assumptions explicit}\footnote{For space reason, the related formalisms will be described informally. Sometimes we have simplified
    \ftext{them}. Nevertheless\ftext{,} \ftext{our} description\ftext{s are} consistent with the original \ftext{formulations}.}. 
Finally, we briefly discuss the relationship between DFOL and non-monotonic extensions of the original multi-context systems (MCS) introduced in~\cite{giunchiglia38,ghidini10} focusing especially on the equilibria\ftext{-}based MCS introduced in~\cite{brewka-and-eiter-2007}.

\subsection{Quantified Modal Logics} 
\label{sec:QML}
Quantified modal logic \ftext{(}QML\ftext{)}~\cite{qml-Garson1984} extends \ftext{a} \ftext{first-order language with modal operators. The semantics of QML is based on possible worlds. In its general form, the semantics of non logical symbols depends upon the possible worlds. Several important issues in QML arise from the combined semantics of quantifiers and modal operators. These issues have originated the development of different semantics for QML~\cite{qml-Garson1984}, which can be represented in DFOL using different bridge rules. In the remaining of this section, we provide an example of how to represent different QML semantics via DFOL bridge rules by focusing on its original Kripke semantics (Section~\ref{sec:QML-kripke}) and on the more recent counterpart semantics (Section~\ref{sec:QML-counterpart}).} \rtext{For the sake of simplicity we restrict our comparison to QML without equality}.

\subsubsection{Kripke semantics for QML}
\label{sec:QML-kripke}
\ftext{A QML language is obtained by extending a first-order language $\L$} with the modal operator $\Box$. The \ftext{simplest} semantics for QLM \ftext{is based on possible worlds}. A QML model $\QMLM$ is a 4-tuple $\langle \W,\R,\D,\I \rangle$, where $\W$ is a non empty set of worlds, $\R$ a binary relation on $\W$, $\D$ is a function that associates to each $w$ a non empty set $\D(w)$, \ftext{satisfying} $w\R v\Rightarrow \D(w)\subseteq\D(v)$, and $\I$ is a function that associates to each $w$ an interpretation $\I(w)$ of a first\ftext{-}order language $\L$ on the domain $\D(w)$.  
Satisfiability is defined as usual on atomic formulas and propositional connectives. Universal quantification is interpreted w.r.t.~the domain of the current world:
$$
\M,w\models\forall x.\phi[a] \mbox{ iff } \M,w\models\phi[a(x:=d)] \mbox{ for all $d\in\D(w)$}
$$
where $a[x:=d]$ denotes the assignment obtained by setting $a(x)=d$ in $a$. Modal formulas are interpreted as follows\ftext{:}
\begin{equation}
	\label{eq:box}
\M,w\models\Box\phi \mbox{ iff } \M,v\models\phi \mbox{ for all $v$ with $w\R v$}
\end{equation}
QML can be translated in a DFOL on a countable set $I=\{0,1,2,\dots\}$ of indices. Each $L_i$ (the language associated to the index $i\in I$) is obtained by \ftext{extending $\L$} with an $n$-ary predicate $\Box\name{\phi}(\bx)$ for every formula $\phi(\bx) \in L_{i-1}$ that contains $n$ distinct free variables $\bx=\left<x_1,\dots,x_n\right>$. 
Intuitively $\Box\name\phi(t_1,\dots,t_n)$ denotes the proposition stating that the tuple of objects denoted by $\langle t_1,\dots,t_n\rangle$ has necessarily the property denoted by the $L_{i-1}$-formula $\phi(x_1,\dots,x_n)$. Notice that the formulas $\Box\name\phi(t_1,\dots,t_n)$ and $\Box\name{\phi(t_1,\dots,t_n)}$ are syntactically and semantically different.  The first is the atomic formula obtained by applying the $n$-ary predicate $\Box\name\phi$ to $\left<t_1,\dots,t_n\right>$, while the second is a $0$-ary predicate, i.e., an atomic proposition.  This difference corresponds to the two readings of the modal formula $\Box\phi(t_1,\dots,t_n)$ called ``de re'' (the former) and ``de dicto'' (the latter).

The translation $\tauQML$ from QML formulas into $L_i$ formulas is defined as follows: $\tauQML$ is the identity transformation on formulas with no modal operators, and it distributes over connectives and quantifiers.
If $\phi$ is a formula with $n$ distinct free variables that contains at most $i-1$ nested modal operators, then 
\ftext{$\tauQML(\Box\phi)=\Box\name{\tauQML(\phi)}(x_1,\dots,x_n)$}. 
\ftext{To provide the DFOL version of the semantics of $\Box$ defined in~\eqref{eq:box}, we use the following bridge rules, where $\bx=\left<x_1,\dots,x_n\right>$, $\tovar{\bx}{i}=\langle\tovar{x_1}{i},\dots,\tovar{x_n}{i}\rangle$ and $\fromvar{\bx}{i+1}=\langle\fromvar{x_1}{i+1},\dots,\fromvar{x_n}{i+1}\rangle$: 
\begin{align}
\label{eq:box-in}
i+1:\Box\name\phi(\tovar{\bx}{i}) & 
\rightarrow i:\phi(\bx) \\
\label{eq:box-out}
i:(\bigwedge_{i=1}^k\phi_i\imp\psi)(\fromvar{\bx}{i+1})
                             & \rightarrow
                               i+1:(\bigwedge_{i=1}^k\Box\name{\phi_i}\imp\Box\name\psi)(\bx)
 \\
\label{eq:increasing-domain}
& \rtext{\rightarrow i:\fromvar{x}{i+k}=\fromvar{x}{i+k}, \quad k>0}
\end{align}
}
\ftext{The set of bridge rules \eqref{eq:box-in}--\eqref{eq:increasing-domain}, called $\BRQML$, allows one to prove the DFOL translation of the (K) axiom for QML (i.e., $\forall x(\Box(\phi\imp\psi)\imp\Box\phi\imp\Box\psi)$)
and of the Barcan formula (i.e.,  $\forall x\Box\phi(x)\imp\Box\forall
x\phi(x)$). This implies that DFOL is stronger than QML. The opposite relation from QML to DFOL is stated in the following theorem.}

\begin{thm}\label{th:modal1}
\rtext{Let $\QMLM=\left<\W,\R,\D,\I\right>$ a QML model. For every $w\in\W$ there is a DFOL model $\M$ that satisfies $\BRQML$, and $\M$ is such that
  $$
  \QMLM,w\models \phi[a] \mbox{ if and only if } \M \models i:\tauQML(\phi)[a],
  $$ 
  where $i$ is greater or equal to the number of nested modal operators of $\phi$. }
\end{thm}

\begin{pf}
Given $\QMLM=\langle\W,\R,\D,\I\rangle$, for every $w\in\W$ and for every index $i\in I$, we define a DFOL model $\M$. Let $i$ be the maximum number of nested modal operator\ftext{s} in $\phi$. 
For every $j\leq i$ we define the set $\W_j\subseteq\W$ as follows: 
\begin{itemize}\itemsep=-\parsep
\item $\W_i=\{w\}$; 
\item $\W_{j-1}=\R(\W_j)=\{w'\in\W \mid w''\R w'\mbox{ for some
    $w''\in\W_j$}\}$ for $j \leq i$.
\end{itemize}
We then define $\M$ as follows: 
\begin{itemize}\itemsep=-\parsep
\item $\dom_j=\D(w)$ for $j\geq i$;
\item $\dom_{j-1} = \dom_j\cup\bigcup_{w'\in W_{j-1}}\D(w')$ for $1\leq j\leq i$;
\item $\M_0=\{\I(w') \mid w'\in\W_0\}$;
\item \rtext{$\M_j$ contains an interpretation $m(w')$ of the language $L_j$ for every $w'\in\W_j$; $m(w')$ extends $\I(w')$ with the interpretation of the predicate $\Box\name\phi$, obtained by setting $(\Box\name\phi)^{m(w')}=\{\langle
    d_1,\dots,d_n\rangle\in\dom_j^n\mid\QMLM,w'\models\Box\phi[x_1:=d_1,\dots, x_n:=d_n]\}$ when $x_1,\dots, x_n$ are all the free variables of $\phi$;}
\item \rtext{$\dr_{i,j}=\{\langle d,d\rangle\mid d\in\dom_{\max(i,j)}\}$.}
\end{itemize}
It can be easily proved that $\M$ satisfies the bridge rules $\BRQML$. We prove the main theorem by induction on $\phi$. 
\begin{description}
\item[Base case] If $\phi$ is an atomic formula in $\L$, then $\tauQML(\phi)=\phi$.  $\QMLM,w\models\phi[a]$ iff  $\I(w)\models\phi[a]$\ftext{.} Since $\M_i$ contains only one single model $m(w)$, which coincides with $\I(w)$ on the interpretation of the symbols in $\L$, then \ftext{$\M_i\models\tauQML(\phi)[a]$}, and therefore $\M\models i:\tauQML(\phi)[a]$.
\item[Step case] The cases for connectives and quantifiers are routine; let us consider the case of $\Box\phi$. 
  Suppose that $\QMLM,w\models\Box\phi[a]$. This holds if and only if for all $w'$, with $w\R w'$, $\QMLM,w'\models\phi[a]$. 
  By construction of $m(w)$, $m(w)\models\Box\name\phi[a]$ and since \ftext{$m(w)$} is the only element of $\M_i$, we have that $M_i\models\Box\name\phi[a]$, and therefore $\M\models i:\tauQML(\Box\phi)[a]$. 
\end{description}
\end{pf}

\rtext{The DFOL encoding of QML shown above decouples the semantics of the modal operator $\Box$, captured by bridge rules \eqref{eq:box-in} and \eqref{eq:box-out}, and the assumptions on the possible worlds domains, captured by the bridge rule
\eqref{eq:increasing-domain}. Different assumptions on the possible worlds domains can be encoded by means of different bridge rules on the equality predicate, thus retaining the bridge rules for $\Box$ unchanged. 
On the contrary, in QML, the semantics of the $\Box$ operator needs to be adapted to the different assumptions made on the worlds domains. As an example, decreasing domains can be axi\ftext{o}matized using the bridge rule:
\begin{align}
  \label{eq:decreasing-domain}
\rightarrow i+n:\fromvar{x}{i}=\fromvar{x}{i};
\end{align}
and constant domains can be axiomatised by adopting both
\eqref{eq:increasing-domain} and \eqref{eq:decreasing-domain}.}

A further source of \ftext{variations} in the semantics of QML concerns the
interpretation of terms in the scope of a modal operator. For instance
what is the meaning of the formula $\Box P(a)$ when $a$ denotes, in
the current world, an object which does not exist in one of the
accessible world\ftext{s}? Or similarly, in evaluating $\Box P(a)$ should $a$ be
interpreted in the current worlds or in all the accessible worlds? 
In DFOL this ambiguity is solved by providing a syntax for both
semantics\ftext{: }the formula $\Box(\name{P(x)}(a))$ 
corresponds to the semantics of $\Box P(a)$ where $a$ is evaluated in the current
world (``de re''), \ftext{whereas} the formula $\Box\name{P(a)}$ corresponds to the 
\ftext{semantics of $\Box P(a)$} where $a$ is evaluated in each accessible world (``de dicto'').

\subsubsection{Counterpart semantics for QML} 
\label{sec:QML-counterpart}
To overcome all the difficulties introduced \ftext{in the Kripke semantics} by the interpretation of objects across different worlds, a new semantics for QML called \emph{counterpart theory}\ftext{,} has been \ftext{recently} introduced \cite{belardinelli2006counterpart, schwarz2013contingent,kracht2005semantics}. Counterpart semantics
extends Kripke semantics by adding relations between objects
in different worlds. \ftext{These relations are similar to domain relations in DFOL.}  The semantics proposed in
\cite{belardinelli2006counterpart,schwarz2013contingent} extends
standard QML models with unconstrained domains, \ftext{(i.e., for every
$w\in\W$, $\D(w)$ is an arbitrary non empty set)} with a counterpart relation
$\C$ that maps every pair $\langle w,v\rangle\in\ftext{\W\times\W}$
  to a subset of $\D(w)\times\D(v)$. Satisfiability of modal
formulas is defined as follows:
\begin{itemize}
\item \ftext{$\CPM,w \models \Box\phi(x_1,\dots,x_n)[a]$},\ftext{,} where $a$ is an
  assignment to the free variables of $\phi$ in $\D(w)$, if and only
  if $\CPM,w\models\phi[a']$, for every world $w'$ with $w\R w'$ and
  for every assignment $a'$ to the free variables of $\phi$
  into the domain $\D(w')$, such \ftext{that} $\langle
  a(x_i),a'(x_i)\rangle\in\C(w,w')$. 
\end{itemize}
With this semantics (which we call basic counterpart semantics),
however, the (K) schema no longer holds.  The approach presented in
\cite{belardinelli2006counterpart,schwarz2013contingent} overcomes this
drawback by deviating from first\ftext{-}order semantics either by adopting typed first\ftext{-}order or free logics (with partial assignments to variables). Since DFOL is based on a first\ftext{-}order semantics for local
models, we have to limit the comparison to the basic counterpart
semantics.  If we take the language of QML and the 
transformation form QML to DFOL seen in the previous section, then we can formalise the counterpart semantics in
DFOL. Let $\BRCP$ be the set of bridge rules
$\eqref{eq:box-out},\eqref{eq:box-in},\{\com_{i,j,k}\}_{i>j>k\in
  I}$.
Let $\tauCP$ be defined as $\tauQML$. The following theorem formally states the correspondence
between \ftext{c}ounterpart semantics and DFOL.

\begin{thm}
  Let $\CPM=\left<\W,\R,\D,\C,\I\right>$ a counterpart frame.
  For every $w\in\W$ there is a
  DFOL model $\M$ that satisfies $\BRCP$ such that
  $$
  \CPM,w\models \phi[a] \mbox{ if and
  only if } \M \models i:\tauCP(\phi)[a]
  $$ 
  where $i$ is greater or equal
  to the number of nested modal operators of $\phi$.
\end{thm}

\begin{pf}[outline]
The proof is the same as the one given for QML, with the only
difference that the the domain relation of $\M$ is defined as follows\ftext{:} 
\begin{itemize}
\item $\dr_{ij}$ for $i>j$ is defined in two phases. First we define 
$\dr_{i,i-1}$, and then $\dr_{i,i-k}$ as the composition of the relations 
  $\dr_{i,i-1},\dots,\dr_{i-k+1,i-k}$. 
  \begin{align}
  \dr_{j,j-1}& =\bigcup_{w'\in M_j \atop w''\in M_{j-1}}C_{w',w''} \\
  \dr_{j,j-k}& =\dr_{j,j-1}\circ\dots\circ\dr_{j-k+1,j-k}
  \end{align}
\end{itemize}
\end{pf}

\subsection{Quantified logic of contexts} 
Quantified logic of contexts (QLC) is a formalism for reasoning about propositions with context dependent truth values.  QLC was originally introduced in~\cite{buvac2} and further developed in~\cite{makarios2006,gratton2013}. In what follows we refer to the original formulation of QLC for two reasons: first, it is a formulation closer to the original logic of context introduced by John McCarthy in \cite{mccarthy23} and  to DFOL; second, the work in~\cite{makarios2006,gratton2013} extends the formalism introduced in \cite{buvac2} with the possibility of quantifying over contexts, which is not allowed in DFOL, as it would correspond to quantifying over indices in $I$.

In the formulation described in \cite{buvac2}, the language of QLC is a two sorted first\ftext{-}order language $\L$ extended with the modal operator $\ist(k,\phi)$ formalising the fact that the formula $\phi$ is true in the context $k$. 
The two sorts of $\L$ are: a sort for objects and a sort for contexts. 
The set of terms of sort context is hereafter denoted with $\K$. 
A QLC model $\QLCM$ is defined starting from two disjoint sets $\Delta^c$ and $\Delta^d$, used to interpret terms of sort context and terms of sort object, respectively. $\QLCM$ is a function that associates to each element of $\Delta^c$ a set of interpretations of $\L$ on $\Delta=\Delta^d\cup\Delta^c$ such that 
\begin{align}
\label{eq:rigid-designator-qlc}
\mbox{for all terms $t$ of $\L$, $t^{\I}=t^{\I'}$ for all $\I\in\QLCM(c)$ and $\I'\in\QLCM(c')$}
\end{align}
Restriction \eqref{eq:rigid-designator-qlc} amounts to assume that terms are rigid designators: i.e., a term denotes the same object in all contexts.

QLC can be translated in an equivalent DFOL on the set $\K$ of indices. The language of the $k$-th context, $L_k$, extends the first\ftext{-}order language $\L$ of QLC with an extra sort for interpreting well\ftext{-}formed formulas (wffs), the binary predicate $\ist(x,y)$ with $x$ of sort context and $y$ of sort wff, and a function symbol $f_\phi(x_1,\dots,x_n)$ for every formula $\phi$ with $n$ distinct free variables $x_1,\dots,x_n$ (if $\phi$ is a closed formula $f_\phi$ is a constant).  The term $f_{\phi(x_1,\dots,x_n)}(c_1,\dots,c_n)$ is used to denote the formula $\phi(x_1,\dots,x_n)$ where each $x_i$ is replaced with $c_i$.  
Formulas of the form $\ist(k,w)$ are complete. 
We define a translation $\tau$ from the language of QLC to DFOL as hinted below. Without loss of generality we assume that QLC formulas are in prenex normal form, and that existential quantifiers have been removed by introducing Skolem constants/functions.

\begin{enumerate}[$(i)$]\itemsep=-\parsep
\item $\tau(\phi)=\phi$ if $\phi$ is any expression (term or formula) of $\L$;
\item $\tau$ distributes over connectives;
\item $\tau(\ist(k,\phi(x_1,\dots,x_n)))=\ist(k,f_{\tau(\phi)}(\tau(x_1),\dots,\tau(x_n))$. 
\end{enumerate}

In QLC, the semantics of the $\ist$ operator is analogous to the
one of the $\Box$ operator in modal
logics. Namely $\I\models\ist(k,\phi)$ iff for all
$\I'\in\QLCM(k^\I)$, $\I'\models\phi$. The two semantics however are
not completely equivalent. Indeed in QLC, the schema 
$
\ist(k,\phi \vee \ist(k',\psi))\imp 
\ist(k,\phi) \vee \ist(k,\ist(k',\psi))
$
is valid, while the corresponding modal axiom 
$\Box(\phi\vee\Box\psi)\imp\Box\phi\vee\Box\Box\psi$ is not.

QLC semantics is
  axiomatised by adding specific axioms and specific inference
  rules. An example of axiom is the one corresponding to the modal
  axiom (K) for $\ist$:
  $\ist(k,\phi\imp\psi)\imp \ist(k,\phi)\imp\ist(k,\phi)$. An example
  of rules is the ``enter context'' rule:
$k:\ist(k',\phi)\Rightarrow k':\phi$.
In DFOL, the semantics of the $\ist$ predicate is axiomatised by means of bridge rules. The relation between the truth of 
the formula $\ist(k,f_\phi)$ in a context $h$ and 
the truth of $\phi$ in the context $k$ is axiomatised with the following bridge rules: 
\begin{align}
\label{eq:context-enter}
h : \ist(k,f_\phi(\tovar{x_1}{k},\dots,\tovar{x_n}{k})) 
&\rightarrow 
k : \phi(x_1,\dots,x_n) 
\\ 
\label{eq:context-exit}
k : \phi(\tovar{x_1}{h},\dots,\tovar{x_n}{h}) 
&\rightarrow 
h : \ist(k,f_\phi(x_1,\dots,x_n))
\end{align}
QLC assumes constant domains, which in DFOL corresponds to isomorphic domains. Isomorphic domains can be imposed by the restriction $\fun_{kh},\tot_{kh},\inj_{kh}$ and $\inv_{kh}$ for every pair of contextual terms $k,h\in\K$.
 Furthermore, the rigid designation assumption \eqref{eq:rigid-designator-qlc} can be axiomatised using the bridge rule:
\begin{align}
\label{eq:rigid-designator-dfol}
k : x = t \rightarrow h: \fromvar{x}{k}=t
\end{align}
Let $\BR_{QLC}$ be the set of bridge rules 
$\{\fun_{kh},\tot_{kh},\inj_{kh},\inv_{kh},\eqref{eq:rigid-designator-dfol},
\eqref{eq:context-exit},
\eqref{eq:context-enter}\}$. 
\begin{thm}
For every QLC model $\QLCM$ and any assignment $a$ to the variables of $\L$, 
there is a DFOL model $\M$ that satisfies $\BR_{QLC}$ such that 
$\QLCM \models k:\phi[a]$ if and only if $\M \models k:\tau(\phi)[\tau(a)]$, where $\tau(a)$ is the DFOL assignment 
such that $\tau(a)_i(x)=\tau(a)_k(\fromvar{x}{h})=
\tau(a)_k(\tovar{x}{h})=a(x)$%
\footnote{Variables are rigid designators}. 
\end{thm}

\begin{pf}
For every $k\in\K$\ftext{,} 
$\dom_k=\Delta^c\cup\Delta^d\cup\Delta^\text{wff}$. 
$\Delta^\text{wff}$ is a 
countable set disjoint from $\Delta^c\cup\Delta^d$ that is used 
to interpret the terms of sort wff (i.e.\ftext{,} the terms of the form $f_\phi(\dots)$). 
For any $\I\in\QLCM(c)$ with $c\in\Delta^c$,  
we define $\tau(\I,a)$, which is an interpretation of the language $L_k$ 
\ftext{obtained} extending $\I$ as follows: 
\begin{enumerate}[$(i)$]\itemsep=-\parsep
\item $f_\phi^{\tau(\I)}(t^{\tau(\I)}_1,\dots,t^{\tau(\I)}_n)=
   f_{\phi(t_1,\dots,t_n)}^{\tau(\I)}$ for every $n$-tuple of terms 
   $\langle t_1,\dots,t_n\rangle$. 
\item
$
\ist^{\tau(\I)}=\{\langle c,w\rangle\in\Delta^c\times\Delta^\text{wff} \mid 
\QLCM(c)\models\phi\mbox{ and } \phi^{\tau(\I,a)}=d\}
$
\end{enumerate}
The DFOL model $\tau(\QLCM,a)$, corresponding to the 
QLC model $\QLCM$ and the assignment $a$, is then defined as 
the pair $\langle\{M_i\},\{r_{ij}\}\rangle$ as follows: 
\begin{enumerate}[$(i)$]\itemsep=-\parsep
\item $r_{kh}$ is the identity function on $\Delta^c\cup\Delta^d\cup\Delta^\text{wff}$; 
\item $M_k=\{\tau(\I) \mid \I\in\QLCM(k^{\I,a})$, 
where $k^{\I,a}$ is the interpretation of the term $k$ w.r.t., $\I$ and the assignment $a$. 
\end{enumerate}
We show by induction that $\QLCM\models k:\phi[a]$ iff $\tau(\QLCM,a)\models k :\phi[a]$
\begin{description}
\item[Base case] if $\phi$ is a formula of $\L$ (i.e., it does not contain the $\ist$ operator), then $\QLCM \models k:\phi[a]$ iff for all $\I\in\QLCM(k^{\I,a})$, $\I\models\phi[a]$ 
iff $\M_k\models\phi[\tau(a)]$ iff $\tau(\QLCM,a)\models k:\phi[\tau(a)]$. 
\item[Step case] We show only the case of the $\ist$ operator.
The other cases are \ftext{routine}. $\QLCM \models k:\ist(h,\phi)[a]$ iff for all $\I\in\QLCM(h^{\I,a})$, $\I\models\phi[a]$ iff 
$\QLCM\models h:\phi[a]$. By induction this holds iff 
$\M_h\models\phi[\tau(a)]$ which is true iff 
$\M_k\models\ist(h,f_\phi(x_1,\dots,x_n))[\tau(a)]$. 
Notice that, the last step follows 
from the fact that $\tau(\QLCM)$ satisfies the bridge rules 
$\BR_{QLC}$
\end{description}
\end{pf}

\subsection{Ontology mapping and ontology integration formalisms}
A number of formalisms for distributed knowledge representation
have originated in the field of ontology integration and are based on
Description Logics (DL\ftext{s}) as a logic for the representation of a single
knowledge base (ontology). A comparison between DFOL and several of
these formalisms is described
in~\cite{SerafiniStuckenschmidtWache-IJCAI05}. In the following 
we recall and extend the results for \emph{Distributed Description Logics}, \emph{$\epsilon$-connection}, \emph{Package-based Description Logics} (P-DL)~\cite{Bao:2009:PDL:1560559.1560578} and peer-to-peer (P2P) data
integration~\cite{calvanese-pods-2004}.

\subsubsection{Distributed Description Logics} 
\emph{Distributed Description Logics}
(DDL)~\cite{SerafiniBorgidaTamilin-IJCAI05,
Homola-serafini-augmenting-ddl:2010,santipantakis2015distributed} and
\emph{C-OWL}~\cite{Bouquet-CWOL-ISWC03} 
are logical formalisms for ontology mapping where ontologies are
expressed using description logics. 
DDL extends description logics with a local semantics similar to \ftext{that} of DFOL and
so-called bridge rules to represent semantic relations between
different T-Boxes. A distributed interpretation for DDL on a family of
DL languages $\{L_i\}$ is a family $\{\I_i\}$ of interpretations,
one for each $L_i$, plus a family $\{\dr_{ij}\}_{i\neq j\in I}$ of
domain relations. While the original proposal only considers
subsumption between concept expressions, DDL has been extended 
in \cite{ghidini2008mapping,DBLP:conf/semweb/SerafiniT07,ghidinietal:ddl:CONTEXT07,OM-ISW2012} to support
mappings between (binary) relations, individuals, concept-to-relation (and vice-versa), and, finally, fuzzy values. 
In DDL, ontology mappings are expressions of the form 
$\co{\fls}{i}\into\co{\frs}{j}$ and 
$\co{\fls}{i}\onto\co{\frs}{j}$ where 
$\fls$ and $\frs$ are either concepts, individuals, or role
expressions of
the descriptive languages $L_i$ and $L_j$ respectively\footnote{In this
definition, to be more homogeneous, we consider interpretations of individuals to be sets containing a single object
  rather than the object itself.}. 
The satisfiability conditions of DDL ontology mappings are: 
\begin{itemize}\itemsep=-\parsep
\item $\DI\models i:\fls\into j:\frs$ if
  $\dr_{ij}(\fls^\Ii)\subseteq \frs^\Ij$;
\item $\DI\models i:\fls\onto j:\frs$ if
  $\dr_{ij}(\fls^\Ii)\supseteq \frs^\Ij$;
\end{itemize}
Since the notion of DDL model is based on the same principles as \ftext{that} of DFOL, we can
directly translate DDL bridge rules into DFOL bridge rules. In particular, there are no additional assumptions
about the nature of the domains that need to be modelled. The
translation is the following:
$$
\begin{array}{|l|l|} \hline
\mbox{\bf DDL} & \mbox{\bf DFOL} \\ \hline
\co{\fls}{i}\into \co{\frs}{j} & \co{\fls(\tovar{x}{j})}{i}\rightarrow\co{\frs(x)}{j} \\
\co{\fls}{i}\onto \co{\frs}{j} & \co{\frs(x)}{j}\rightarrow\co{\fls(\tovar{x}{j})}{j} \\
\hline
\end{array}
$$
In~\cite{Homola-serafini-augmenting-ddl:2010} additional constraints on
the domain relation are added in order to augment the information flow 
between different ontologies induced by DDL ontology mappings. Of
particular interest are the bridge rules that support the transitive propagation of
mappings:  
\begin{align}
i:A\into j:B \mbox{ and  } j:B\into k:C \Rightarrow i:A\into k:C \\ 
i:A\onto j:B \mbox{ and  } j:B\onto k:C \Rightarrow i:A\onto k:C  
\end{align}
which correspond to the entailment between the following DFOL bridge rules: 
\begin{align}
	\label{eq:DDL1}
\begin{array}{r}
i:A(\tovar{x}{j})\rightarrow j:B(x) \\
j:B(\tovar{x}{k})\rightarrow k:C(x) 
\end{array} \Longrightarrow \
i:A(\tovar{x}{k})\rightarrow k:C(x)  \\ 
\label{eq:DDL2}
\begin{array}{r}
k:C(x)\rightarrow j:B(\tovar{x}{k}) \\
j:B(x)\rightarrow i:A(\tovar{x}{j}) 
\end{array} \Longrightarrow \
k:C(x)\rightarrow i:A(\tovar{x}{k}) 
\end{align}
Since conditions \eqref{eq:DDL1} and \eqref{eq:DDL2} must hold for any interpretation of $A$, $B$, and $C$, then the entailments can
be obtained by imposing condition $\dr_{ij}\circ\dr_{jk}=\dr_{ik}$ among domain relations, which corresponds to bridge rules $\com_{ijk}$. 

\subsubsection{$\epsilon$-connections} 
A further approach for defining mappings between DL knowledge bases has emerged from the investigation of so-called \emph{$\epsilon$-connections} between abstract description systems~\cite{e-connections,cuenca1}. 
In the $\epsilon$-connections framework, for every pair of ontologies $i$ and $j$ there is a set $\epsilon_{ij}$ of \emph{links}, which represent binary relations between the domain of the $i$-th ontology and the domain of the $j$-th ontology. Links from $i$ to $j$ can be used to define $i$ concepts in a way that is analogous to how roles are used to define concepts. 
In the table below we report the syntax and the semantics of the $i$-concepts definition based on links, where $E$ denotes a link from $i$ to $j$ and $\psi$ denotes a concept in $j$. 
The only assumption about the relation between domains is global inconsistency, that is, the fact that the inconsistency of a local knowledge base makes the whole system inconsistent.
 
In comparing DFOL with $\epsilon$-connections we can notice that in DFOL there is only one  relation from $i$ to $j$, while in $\epsilon$-connection there are many possible relations. However, \cite{SerafiniStuckenschmidtWache-IJCAI05} shows how to represent each $r_{ij}$ as a relation in $\epsilon_{ij}$ and provides a detailed description of how the concept definition based on links of $\epsilon$-connections can be codified in DFOL. In a nutshell, to represent $\epsilon$-connections in DFOL it is enough to label each arrow variable with the proper link name. The arrow variable $\rtovar{x}{i}{own}$ is read as the arrow variable $\tovar{x}{i}$ where $\dr_{ij}$ is intended to be the interpretation of \relation{Own}$_{ij}$.
 With this syntactic extension of DFOL, the concept definition based on links (denoted as $E$) can be codified in DFOL as follows\footnote{A more detailed comparison between $\epsilon$-connections and DFOL is contained in~\cite{SerafiniStuckenschmidtWache-IJCAI05}.}:
$$
\begin{array}{|l|l|} \hline
\mbox{\bf $\epsilon$-conn.} & \mbox{\bf DFOL} \\ \hline
\fls\isa\exists E.\frs & \co{\fls(x)}{i}\rightarrow\co{\frs(\rfromvar{x}{i}{E})}{j} \\
\fls\isa\forall E.\frs & \co{\fls(\rtovar{x}{j}{E})}{i}\rightarrow\co{\frs(x)}{j} \\
\fls\isa\geq n E.\frs & \co{\bigwedge_{k=1}^n\fls(x_1)}{i} \rightarrow  \co{\bigwedge_{k\neq h=1}^n \frs(\rfromvar{x_k}{i}{E})\wedge x_k\neq x_h}{j} \\
\fls\isa\leq n E.\frs & \co{\fls(x)\wedge\bigwedge_{k=1}^{n+1}x=\rtovar{x}{j}{E}_k}{i} \rightarrow \co{\bigvee_{k=1}^{n+1}\left(\frs(x_k)\supset\bigvee_{h\neq
             k}x_h=x_k\right)}{j} \\ \hline
\end{array}
$$

\subsubsection{Package-based Description Logics} (P-DL)~\cite{Bao:2009:PDL:1560559.1560578} is a formalism focused on ontology import, that is, it allows a subset of concepts, relations, and individuals
defined in one ontology to be imported into another ontology where they are then reused. These ontologies are called, in P-DL terms, packages. In a nutshell, a package\ftext{-}based ontology is a \shoiq ontology ${\cal P}$ which is partitioned into a finite set of packages $\{P\}_{i\in I}$, using an index set $I$. Each $P_i$ uses its own alphabet of terms. The
alphabets are not mutually disjoint, but for any term $t$ there is a unique home package of $t$, denoted by home($t$). The importing of a term of $P_i$ in $P_j$ is denoted with the expression $P_i \pdl{t} P_j$, while $P_i \pdl{*} P_j$ is used to denote the transitive closure of $\rightarrow$. A distributed interpretation of ${\cal P}$ is composed of a set of interpretations, one for each package, plus a set of domain relations similar to the ones of DFOL. The main difference with DFOL is that each $r_{ij}$ is an injective partial function, and that if $i \pdl{*} j$ and $j \pdl{*} k$ then $r_{ik}$ is defined as the composition $r_{ij}\circ r_{jk}$.  As shown in Figure~\ref{fig:dr-bridge-rules}, these restrictions can be formalised using the bridge rules $\inj_{ij}$ and $\com_{ijk}$. The semantics of the import of a term $t$, expressed by $P_i \pdl{t} P_j$, is defined as $r_{ij}(t^{I_i}) = t_{I_j}$. If $t$ is a description logic concept (unary predicate) $\phi(x)$, then the import can be represented in DFOL by the pair of bridge rules: 
\begin{align}
	\co{\phi(\tovar{x}{j})}{i} & \rightarrow \co{\phi(x)}{j}\\
	\co{\phi(x)}{j} & \rightarrow \co{\phi(\fromvar{x}{i})}{i}
\end{align}
while if $t$ is an individual $a$, then the import can be represented in DFOL by the bridge rule:   
\begin{align}
	\co{x = a}{i} & \rightarrow \co{\tovar{x}{i} = a}{j}
\end{align}
If $t$ is a role (binary predicate) $\phi(x,y)$ then we can define analogous bridge rules, and also impose that $r_{ij}$ is role preserving, that is, if $(x,y) \in \phi^{I_i}$, then $r_{ij}(x) \neq \emptyset$ iff $r_{ij}(y) \neq \emptyset$.

\subsubsection{Logical foundation of peer-to-peer (P2P) data integration} 
The work in~\cite{calvanese-pods-2004} defines an epistemic semantics for P2P systems and applies it to different architectures of P2P systems. The epistemic semantics is based on the introduction of a modal operator $K$ used to express what is known by peers. Mapping assertions  of the form $cq_S \leadsto cq_G$ represent the fact that all the data satisfying the (conjunctive) query $cq_S$ over the sources also satisfy the concept in the peer schema represented by $cq_G$. This mapping assertion is captured, in the epistemic semantics, by an axiom of the form\ftext{:}
$$
\forall \mathbf{x} (K(\exists \mathbf{y}\ body_{cq_S}(\mathbf{x},\mathbf{y})) \imp \exists \mathbf{z}\ body_{cq_G}(\mathbf{x},\mathbf{z}))
$$ 
which states that only what is \emph{known} in $S$ is transferred to $G$. Differently from DFOL, the epistemic semantics presented in \cite{calvanese-pods-2004} provides a unique model of the P2P system, based on a single domain of interpretation. Similarly to DFOL, this semantics addresses the problem of the representation of complete and incomplete information. In fact, the modal operator $K$ can be considered a way of dealing with non-complete formulas and to consider, in the mappings, only their ``complete part'' (that is, the tuples that belong to the interpretation of that formula in all possible models).   

\subsection{Annotated Logic}
Annotated logics
\cite{subrahmanian1} is a formalism that has been applied to a variety of aspects in knowledge representation, expert systems, quantitative reasoning, and hybrid databases.  In annotated logics it is possible to integrate a set of logical theories in an unique amalgamated theory. The amalgamated theory is the disjoint union of the original theories plus a set of clauses (called amalgamated clauses) which resolve conflicts due to inconsistent facts and compose uncertain information of different theories. One of the main similarities with our approach is the capability to cope with inconsistent knowledge bases. Annotated logics provide an explicit way to solve conflicts. The main difference between annotated logics and DFOL concerns the ability to represent different interpretation domains. Annotated logics have a single logical language, and the same symbol in different knowledge bases is interpreted in the same object.  This of course might be solved by indexing the constant with the name of the knowledge base. In this case explicit relational symbols between objects of different knowledge bases should be introduced.  

\def\dfolo{DFOL$_0$\xspace}
\subsection{Original DFOL}
\label{sec:originalDFOL}
We conclude this section by illustrating the difference between the version of DFOL presented in this paper and the original version introduced in \cite{ghidini5}, here denoted with \dfolo. The first difference concerns \emph{arrow variables}. 
In the current version of DFOL arrow variables are part of the syntax of the local languages and of the bridge rules. In \dfolo\ arrow variables are a meta-notation of the calculus, which is used to keep track of the dependencies between variables in different modules. Arrow variables, therefore, are not part of the logical language of \dfolo\ and no semantics is provided for them. Also, the bridge rules of \dfolo (called interpretation constraints) do not contain arrow variables. Thus the \dfolo\ bridge rule 
$
i:\phi(x) \rightarrow j:\psi(x)
$
corresponds to the DFOL bridge rule 
$
i:\phi(\tovar{x}{j}) \rightarrow j:\psi(x)
$\ftext{.}
The introduction of bridge rules in the logical language of DFOL is not only a matter of syntax: arrow variables extend the expressivity of the language. Indeed, in \dfolo there is no way of expressing a constraint represented by a DFOL bridge rules with arrow variables in the conclusion. For instance, the fact 
``{\it for all the objects of type $A$ in $i$ there is a corresponding object of type $B$ in $j$}'' is represented by means of the DFOL bridge rule $i:A(x)\rightarrow j:B(\fromvar{x}{i})$ and is not expressible in \dfolo.

A second important difference is the fact that arrow variables allow to \emph{unify} the two types of constraints introduced in \dfolo: domain constraints and interpretation constraints. Domain constraints between $i$ and $j$ are constraints on the domain of $i$ and $j$, while interpretation constraints are constraints between the interpretations of the symbols in $i$ and $j$ modulo the transformation via domain relation $\dr_{ij}$. As shown in section 
\ref{sec:interpretation-cosntraints} bridge rules with arrow variables enable the formalisation of a wide set of relations between (two or more) domains together with relations between predicates.  
On the contrary, the only domain constraints allowed in \dfolo\ are $r_{ij}$ being total or  surjective, which correspond to the bridge rules $\mathsf{T}_{ij}$ and $\mathsf{S}_{ij}$ in Figure~\ref{fig:dr-bridge-rules}. 

Third, in \dfolo\ bridge rules (interpretation constraints) connect only two KBs, that is, they are of the form $i:\phi(\bx) \rightarrow j:\psi(\bx')$, where $\bx'\subseteq \bx$ are two sets of variables. In DFOL we \emph{generali\ftext{s}e bridge rules} by allowing more than one index in the premise. 

Finally, in this paper we have defined a notion of \emph{logical consequence between bridge rules}. This notion is very important when formalising reasoning about ontology mapping. In fact it makes possible to check the consistency, the redundancy, and the inter-dependency of sets of ontology mappings (see 
\cite{meilicke2006improving,meilicke2007repairing} for example). Given a set of DFOL bridge rules $\BR$, the problem of checking if the bridge rule $br = i_1:\phi_1,\dots,i_n\phi_n\rightarrow i:\phi$ is a logical consequence of $\BR$ can be formulated as the problem of proving $i_1:\phi_1,\dots,i_n:\phi_n\models_{\BR} i:\phi$. 

\def\Not{\mathbf{not}}
\def\N{\mathcal{N}}
\def\DFOLMCS{\mathsf{DFOL_{MCS}}}

\subsection{Equilibria based Multi-context systems} 

The last decade has seen a number non-monotonic extensions of the
original multi-context systems (MCS) introduced in the 90's
\cite{giunchiglia38,ghidini10}.  The work in
\cite{roelofsen-and-serafini-2005} extends MCS with minimal beliefs,
while \cite{roelofsen-brewka-and-serafini-2007} introduces default
reasoning with contexts. A notable generalisation of MCS was proposed
in~\cite{brewka-and-eiter-2007}, where the focus of the work is the
ability to deal with distributed heterogeneous reasoning systems, that
is, systems that adopt different logics in the different contexts
(knowledge modules).  The semantics of this version of MCS is
called equilibria based semantics. Given the semantics of a set of
local logics (that can be either monotonic or non monotonic logics)
the equilibria based semantics is obtained by composing the local
semantics with a methodology inspired to the answer set programming
paradigm. 

When comparing DFOL with equilibria based MCS, we need to take into account some important aspects. 
DFOL and equilibria based MCS have been developed to tackle
different forms of heterogeneity. From the one hand, DFOL focuses on
capturing the heterogeneity that arises in integrating knowledge bases
that describe different but overlapping or interconnected domains expressed in a
set of first\ftext{-}order languages. On the other hand, equilibria based MCS
focus on capturing the heterogeneity that arises in integrating
knowledge bases expressed using different logics.
From this perspective\ftext{,} DFOL can be seen as a special case of equilibria 
based MCS. 
However, due to their generality, equilibria based MCS do not provide a 
specific investigation on specific relations between heterogeneous
domains, semantic shifting of symbols across different contexts, 
rigid and non-rigid semantics of constants, and so on. These aspects are the ones that DFOL
analyses in terms of specific bridge rules. 
A second difference concerns the different reasoning tasks the two
systems are focused on. Equilibria based MCS is a formalism developed
with the aim of supporting query answering. Thus, the emphasis is on
the computation of equilibria KBs which can then be queried. DFOL is
instead focused on the notion of logical consequence and bridge rules
entailment. Thus, the emphasis is on the definition of a semantics and
a calculus that axiomatise mapping entailment.

In the following comparison, therefore, we will concentrate only on the common
aspects of DFOL and MCS. In particular we restrict to a
specific version of DFOL, where bridge rules involve only closed
formulas, and to a specific version of equilibria based MCS, where the local logics are
propositional or first\ftext{-order} classical logics. In other words we omit arrow variables and the contribution of the domain relation in the semantics of DFOL and the ability to deal with different logics in equilibria based MCS. Note that a complete formal comparison of equilibria based
MCS and DFOL is out of the scope \ftext{of} this paper. In the following we
provide some insights and an example about this correspondence. \ftext{A thorough comparison is left for future work}.

As already said we focus on MCS where each KB is formalised by means of a propositional or first\ftext{-order} classical logic. This does not make  the resulting system monotonic. In fact, an important characteristic of equilibria based MCS are bridge rules that can introduce a form of non monotonicity. Let us go a bit more into details considering a correspondence between DFOL and equilibria based MCS when the local KBs are modelled using propositional logic. 

Let $MCS$ be an MCS defined on a set $I$ of classical propositional
logics with languages $\{L_i\}_{i\in I}$.  
The corresponding DFOL logic $\DFOLMCS$ is obtained by extending 
the set of contexts (indexes) with a meta context $mc$ that contains the propositional letter 
$\Not(j:p)$ for every propositional
letter $p\in L_j$, $j \in I$.  We also assume that $\Not(j:p)$ in $mc$ is complete. Since propositions of the form $\Not(i:p)$ occur only in $mc$, we simplify the notation by omitting the index $mc$. 

The semantics of $\Not(j:p)$ is fixed by the bridge rules \eqref{eq:rup} and \eqref{eq:rdw}\footnote{Similar bridge rules have been introduced and studied in \cite{criscuolo-giunchiglia-serafini-I-02,criscuolo-giunchiglia-serafini-II-02} under the name of reflection rules. \ftext{They have been widely applied to the modular representation of beliefs in multi-agent systems (see, e.g.,~\cite{DBLP:journals/ijis/BenerecettiGSV99,ghidini9, fisher1}).}}, while \eqref{eq:botup} enables to export an inconsistency to the meta context $mc$:
\begin{align}
\label{eq:rup} j:p & \rightarrow \neg\Not(j:p) \\
\label{eq:rdw}  \neg\Not(j:p) & \rightarrow  j:p  \\
\label{eq:botup}  i:\bot & \rightarrow mc:\bot 
\end{align}
As a consequence of these rules 
\begin{equation}
	\label{eq:notprop}
	\text{$\M\models \Not(i:p)$ if and only if 
there is a model $m\in\M_i$ such that $m\not\models p$.}
\end{equation}
Furthermore, $M_{mc}$ is completely defined by the set
$\{M_{i}\}_{i\in I}$. In other words, $M_{mc}\models\Not(i:p)$ if and only if 
$\M_i\not\models p$. 

For each bridge rule of $MCS$ which is of the form: 
$$
i:p \leftarrow i_1:p_1,\dots,i_n:p_n,\Not(j_i:q_1),\dots,\Not(j_m:q_m)
$$
we add the following bridge rule $br$ into $\DFOLMCS$: 
$$
i_1:p_1,\dots,i_n:p_n,\Not(j_1:q_1), \dots,\Not(j_m:q_m)\rightarrow i:p
$$
Following the logic programming notation, we use $head(br)$ to denote 
$i:p$ and $body(br)$ to denote the set
$\{i_1:p_1,\dots,i_n:p_n,\Not(j_1:q_1), \dots,\Not(j_m:q_m)\}$ 

Given a model $\M$ for $\DFOLMCS$, its \emph{local reduction} $LR(\M)$ is
the $\DFOLMCS$ model obtained by removing from each $M_i$ any model
$m$ such that there exist\ftext{s} a model $m'\in M_i$, with $m \neq m'$, such that
$m'\models p$ implies $m\models p$ for all propositional letters $p \in L_i$. 
Notice that if $\M$ is a $\DFOLMCS$ model then $LR(\M)$ is also a $\DFOLMCS$ model. 

We use local reductions to compute minimal models of $\DFOLMCS$ as follows:
\begin{itemize}
\item $\M^{(0)}$ is the $\DFOLMCS$ model such that
  $M^{(0)}_i$ contains all the models of $L_i$ that satisfy the local
  axioms.
\item $\M^{(k+1)}$ is obtained by deleting from $M^{(k)}_i$ all the
  models that do not satisfy the consequence of some bridge rule 
  if all its premises are satisfied by $\M^{(k)}$. 
  Formally $M^{(k+1)}_i=\{m\in M^{(k)}_i \mid \M^{(k)}\models body(br)
  \Rightarrow m\models head(br) \mbox{ for all $br\in
    \BR_{\DFOLMCS}$}\}$ and $\M^{k+1}_{mc}$ is updated according to condition \eqref{eq:notprop}.
\item $\M^*$ is the fix-point of this operator. The fix\ftext{-}point exists since the
  bridge rules have only a finite number of premises. 
  \item The minimal model of $\DFOLMCS$ is equal to $LR(\M^*)$.
\end{itemize}
Such a minimal model is the analogous of the grounded equilibrium as defined
in \cite{brewka-and-eiter-2007}. To show how this construction works we consider the example similar to
Example 3 in \cite{brewka-and-eiter-2007}. 

\begin{exmp}
  Consider the $\DFOLMCS$ consisting of two contexts $1$ and $2$.
  Suppose that there are no local axioms and that the set of bridge
  rules $\BR$ contains the two standard bridge rules
  $2:q\rightarrow 1:p$ and $1:p\rightarrow 2:q$ and the non monotonic
  bridge rule 
  \ftext{\begin{equation}
  	\label{eq:nonmonotonic}
	\Not(1:p)\rightarrow 2:r.
  \end{equation}
  }
    In the following table we show step by step the construction of
  $RL(M^*)$. 
  $$\footnotesize
  \begin{array}{llll} 
   \M & M_1 & M_2 & M_{mc} \\ \hline 
   \M^{(0)} & \{\{\},\{p\}\} & \{\{\},\{q\},\{r\},\{q,r\}\} & \{\{\Not(1:p),\Not(2:q), \Not(2:r)\}\} \\ 
   \M^{(1)} & \{\{\},\{p\}\} & \{\{r\},\{q,r\}\} & \{\{\Not(1:p),\Not(2:q)\}\} \\ 
   \M^{(2)} & \{\{\},\{p\}\} & \{\{r\},\{q,r\}\} & \{\{\Not(1:p),\Not(2:q)\}\} \\ 
   RL(\M^{(2)})& \{\{\}\} & \{\{r\}\} & \{\{\Not(1:p),\Not(2:q)\}\} \\ 
  \end{array}
  $$
  \ftext{$\M^{(0)}$ is the $\DFOLMCS$ model such that each $M^0_i$ satisfies the local axioms. Since there are no local axioms $M^0_1$ and $M^0_2$ contain all the possible local models. The only applicable rule to compute $M^1$ is bridge rule \eqref{eq:nonmonotonic}, which can be applied because $M^0_{mc}$ satisfies $\Not(2:r)$. Thus, $M^1$ is obtained by removing all the models that do not satisfy $r$ from $M^0_2$ and by removing $\Not(2:r)$ from $M^1_{mc}$ to comply with condition \eqref{eq:notprop}. After that no more rules are applicable. Thus $\M^{(2)}$ provides the fix-point $\M^*$. The last step of our computation concerns the computation of the local reduction $RL(\M^{(2)}$ of $\M^{(2)}$ which terminates the computation.
As we can see from its construction, the resulting model $RL(\M^{(2)})$ satisfies only $2:r$, $1:\neg p$ and $2:\neg q$. }
\end{exmp} 


%% file: conclusions.tex

\section{Conclusions}
\label{sec:conclusions}

In this paper we have presented a systematic account of Distributed First Order Logic (DFOL) and we have shown how the notions of \emph{domain relation}, \emph{arrow variable} and \emph{bridge rule} enable the characterisation of a wide range of semantic relationships between different KBs belonging to a distributed knowledge base systems modelled by means of (subsets of) first\ftext{-}order logics, each KB having its own domain of interpretation. Moreover, we have defined a sound and complete calculus which characterises the notion of DFOL logical consequence, and we have illustrated how to use it to infer logical relations between distributed knowledge.

\section*{Acknowledgments}
This paper has benefitted from many discussions with Massimo Benerecetti, Paolo Bouquet, Loris Bozzato and Holger Wache.
The influence of the work on multi-context logics by Fausto Giunchiglia and his co-authors has been significant throughout the paper. \ftext{We thank Chiara Di Francescomarino, Ivan Donadello, and Riccardo De Masellis for their help in polishing the paper, and} the anonymous reviewers for their constructive comments, which helped us to improve the manuscript.

%% file: appendixB.tex

\section{Examples of DFOL Deductions} 
\label{sec:appendix-deductions}

In this section we provide examples of deductions in DFOL by proving the bridge rule entailments described in Proposition \ref{prp:entailment-of-br}. As usual, we write $\IC_{(n)}$ to denote the inference rule obtained from the bridge rule $(n)$. 

\begin{paragraph}{Conjunction}
$$
\resizebox{.75\textwidth}{!}{$\displaystyle 
\left.\begin{array}{r@{\quad}l}
	(\fun_{ij}) & \co{\tovar{x}{j}=\tovar{y}{j}}{i}\rightarrow\co{x=y}{j}\\
    (a) & \co{\phi(x)}{i}\rightarrow\co{\psi(\fromvar{x}{i})}{j} \\ 
    (b)& \co{\phi'(x)}{i}\rightarrow\co{\psi'(\fromvar{x}{i})}{j} 
	\end{array}\right\}
	\models
	\co{\phi\con\phi'(x)}{i}\rightarrow\co{\psi\con\psi'(\fromvar{x}{i})}{j} 
$}
$$
$$
\resizebox{.8\textwidth}{!}{$\displaystyle 
   \begin{ndproof}
			    (1) & i:\phi(x)\wedge\phi'(x) & (1) & & \reason{Assumption} \\ 
			    (2) & i:\phi(x) & (1) & & \reason{From (1) by $\ande{i}$} \\ 
			    (3) & j:\psi(\fromvar{x}{i}) & & (1) & \reason{From (2) by $\BR_{\eqref{eq:br1}}$} \\ 
			    (4) & i:\phi'(x) & (1) & & \reason{From (1) by $\ande{i}$} \\
		(5) & i:x=z & (5) & & \reason{Assumption} \\
		(6) & i:\phi'(z) & (1)(5)& & \reason{From (4), (5) by $\eqe{i}$} \\		
		(7) & j:\psi'(\fromvar{z}{i}) & & (1)(5) & \reason{From (6) by $\BR_{\eqref{eq:br2}}$} \\		
		(8) & j:\psi(\fromvar{x}{i}) & (8) & & \reason{Assumption} \\
		(9) & j:\psi'(\fromvar{z}{i}) & (9) & & \reason{Assumption} \\
		(10) & 	j:\psi(\fromvar{x}{i}) \con \psi'(\fromvar{z}{i})& (8)(9) & & \reason{From (8) and (9) by $\andi{j}$} \\	
		(11) & 	j:\psi(\fromvar{x}{i}) \con \psi'(\fromvar{z}{i})& (9) & (1) & \reason{From (3) and (10) by $\cut{ji}$ discharging (8)} \\	
		(11) & 	j:\psi(\fromvar{x}{i}) \con \psi'(\fromvar{z}{i})& & (1) & \reason{From (7) and (11) by $\cut{ji}$ discharging (9)} \\	
			 	(12) & j:\fromvar{x}{i}=y & (12) & & \reason{Assumption} \\ 
			    (13) & i:x=\tovar{y}{j}& & (12) & \reason{From (12) by $\toI{j}I_{ji}$} \\
			    (14) & j:\fromvar{z}{i}=w & (14) & & \reason{Assumption} \\ 
			    (15) & i:z=\tovar{w}{j} & & (14) & \reason{From (14) by $\toI{j}I_{ji}$} \\ 
			    (16) & i:\tovar{y}{i}=\tovar{w}{i} & (5) & (12)(14) & \reason{from (5), (13) and (15) by $\eqe{i}$ and applications of $\cut{}$ to handle existential arrow variables} \\ 
			    (17) & j:y=w & & (5)(12)(14) & \reason{From (16) by $\IC_{\fun_{ij}}$}  \\ 
		(18) & j:\fromvar{x}{i}=\fromvar{z}{i} & &(5)(12)(14) &  \reason{From (12), (14) and (18) by $\eqe{j}$}\\
		(19) & j:\psi\con \psi'(\fromvar{x}{i})& & (1)(5)(12)(14) & \reason{From (11) and (18) by $\eqe{j}$} \\ 	
			    (20) & j:\fromvar{x}{i}=\fromvar{x}{i} & & (1) & \reason{From (3) by $\eqi{j}$} \\ 
			    (21) & j:\exists y \ \fromvar{x}{i}=y & & (1) & \reason{From (20) by $\exi{j}$} \\ 
			    (22) & j:\fromvar{z}{i}=\fromvar{z}{i} & & (1)(5) & \reason{From (7) by $\eqi{j}$} \\ 
			    (23) & j:\exists w \ \fromvar{z}{i}=w & & (1) & \reason{From (22) by $\exi{j}$} \\ 
		(24) & j:\psi\con \psi'(\fromvar{x}{i}) & &(1)(5)(14) &  \reason{From (19) and (21) by $\exe{j}$ discharging (12)}\\		
		(25) & j:\psi\con \psi'(\fromvar{x}{i}) & &(1)(5) &  \reason{From (23) and (24) by $\exe{j}$ discharging (14)}\\ 
		(26) & i:z=z & & & \reason{By $\eqi{i}$}\\
		(27) & i:\exists x \ x=z  & & & \reason{From (26) by $\exi{i}$}\\
		(27) & j:\psi\con \psi'(\fromvar{x}{i}) & & (1) &  \reason{From (25) and (26) by $\exe{ji}$ discharging (5)}\\ \hline
	\end{ndproof}
	$}
$$
\end{paragraph}

\begin{paragraph}{Composition}\ \\
	
$$\left.\begin{array}{r@{\ \ }l}
	(\com_{ijk}) & \co{\fromvar{x}{i}=z\to{k}}{j}\rightarrow\co{\fromvar{x}{i}=z}{k} \\ 
    (a) & \co{\phi(x)}{i}\rightarrow\co{\psi(\fromvar{x}{i})}{j} \\ 
	(b) & \co{\psi(x)}{j}\rightarrow\co{\theta(\fromvar{x}{j})}{k} 
    \end{array}\right\}
    \models
	\co{\phi(x)}{i}\rightarrow\co{\theta(\fromvar{x}{i})}{k} 
$$

$$
\begin{ndproof}
    (1) & i:\phi(x)
        & (1) & & \reason{Assumption} \\ 
    (2) & j:\psi(\fromvar{x}{i})
        & & (1) & \reason{From (1) by $\BR_{(a)}$} \\ 
    (3) & j:\fromvar{x}{i}=y 
        & (3) & & \reason{Assumption} \\ 
    (4) & j:\psi(y) 
        & (3) & (1) & \reason{From (2) and (3) by $\eqe{j}$} \\ 
    (5) & k:\theta(\fromvar{y}{j}) 
        & & (3) (1) & \reason{From (4) by $\BR_{(b)}$} \\ 
    (7) & k:\fromvar{y}{j}=z 
        & (7) & & \reason{Assumption}  \\ 
    (8) & j:y=z\to{k} 
        & & (7) & \reason{From (7) by $\toI{k}I_{kj}$}  \\ 
    (9) & j:\fromvar{x}{i}=z\to{k} 
        & & (3)(7) & \reason{From (3) and (8) by $\eqe{j}$} \\ 
    (10) & k:\fromvar{x}{i}=z 
        & & (3)(7) & \reason{From (9) by $BR_{\com_{ijk}}$} \\ 
    (11) & k:\fromvar{x}{i}=\fromvar{y}{j} 
        & & (3)(7) & \reason{From (7) and (10) by $\eqe{k}$} \\ 
    (12) & k:\theta(\fromvar{x}{i}) 
        & & (1)(3)(7) & \reason{From (5) and (11) by  $\eqe{k}$} \\ 
    (13) & k:\fromvar{y}{j}=\fromvar{y}{j} 
        & &(1)(3) & \reason{From (5) by $\eqi{k}$} \\ 
    (14) & k:\exists z.y\fromvar{i}=z 
        & &(1)(3) & \reason{From (13) by $\exi{k}$} \\ 
    (15) & k:\theta(\fromvar{x}{i}) 
        & & (1)(3) & \reason{From (14) and (12) by  $\exe{k}$ discharging (7)} \\ 
    (16) & j:\fromvar{x}{i}=\fromvar{x}{i}
        & &(1) & \reason{From (2) by $\eqi{j}$} \\ 
    (17) & j:\exists y.\fromvar{x}{i}=y 
        & & (1) & \reason{From (16) by $\exi{j}$} \\ 
    (18) & k:\theta(\fromvar{x}{i})
        & & (1) & \reason{From (17) and (15) by  $\exe{jk}$ discharging (3)} \\ \hline
\end{ndproof}
$$
\end{paragraph}

\newpage
\begin{paragraph}{Existential quantification} \ 
$$ 
	(a) \  \ \co{\phi(x)}{i}\rightarrow\co{\psi(\fromvar{x}{i})}{j}
	\models
	\co{\exists x\phi(x)}{i}\rightarrow\co{\exists x \psi(x)}{j}
$$
under the assumption that $\phi$ is a complete formula.
$$
\begin{ndproof}
    (1) & i:\phi(x) 
        & (1) & & \reason{Assumption} \\ 
    (2) & j:\psi(\fromvar{x}{i})
        & & (1) & \reason{From (1) by $\BR_{(a)}$} \\ 
	(3) & j:\psi(\fromvar{x}{i})
		& (3) & & \reason{Assumption} \\ 	
    (4) & j:\exists x.\psi(x) 
        & (3)& & \reason{From (3) by $\exi{j}$} \\ 
    (5) & j:\exists x.\psi(x) 
        & & (1)& \reason{From (2) and (4) by $\cut{ji}$ discharging (3)} \\ 
    (6) & i:\exists x.\phi(x) 
        & (6) & &  \reason{Assumption} \\ 
    (7) & j:\exists x.\psi(x) 
        & & (7) & \reason{From (5) and (6) by $\exe{ji}$ discharging (1)}\\ \hline
\end{ndproof}
$$
	Notice that the application $\exe{ji}$ in step (7) satisfies restriction R\ref{item:restr-disch-complete-global-assumptions} only if $\phi$ is a complete formula.
\end{paragraph}

\begin{paragraph}{Universal quantification}\ 
	$$\left.\begin{array}{r@{\ \ }l}
	(\sur_{ij}) & \co{x=x}{j}\rightarrow\co{\exists y \ y=\tovar{x}{j}}{i} \\ 
    (a) & \co{\phi(\tovar{x}{j})}{i}\rightarrow\co{\psi(x)}{j} \\ 
    \end{array}\right\}
    \models
	\co{\forall x \ \phi(x)}{i} \rightarrow \co{\forall x \ \psi(x)}{j}
$$
$$
\begin{ndproof}
   (1) & i:\forall x.\phi(x) 
       & (1) & & \reason{Assumption} \\ 
   (2) & i:\phi(y) 
       & (1) & & \reason{From (1) by $\alle{i}$}  \\
   (3) & i:y=\tovar{x}{j} 
       & (3) & & \reason{Assumption} \\
   (4) & i:\phi(\tovar{x}{j})
       & (1)(3) &  & \reason{From (2) and (3) by $\eqe{i}$} \\
   (5) & j:\psi(x)
       & & (1)(3) & \reason{From (4) by $\BR_{(a)}$} \\ 
   (6) & j : x=x
       & & & \reason{By $\eqi{j}$}\\ 
   (7) & i:\exists y.y=\tovar{x}{j} 
       & & & \reason{From (6) by $\BR_{\sur_{ij}}$} \\
   (8) & j:\psi(x) 
       & & (1) & \reason{From (5) and (7) by $\exe{i}$ discharging (3)}\\
   (9) & \co{\forall x \ \psi(x)}{j}
       & & (1) & \reason{From (8) $\alli{i}$}\\ \hline
\end{ndproof}
$$
\end{paragraph}

\begin{paragraph}{Disjunction}\ 
$$
\left.\begin{array}{r@{\ \ }l}
   (a) & \co{\phi(x)}{i}\rightarrow\co{\psi(\fromvar{x}{i})}{j} \\ 
   (b) & \co{\phi'(x)}{i}\rightarrow\co{\psi'(\fromvar{x}{i})}{j}
 \end{array}\right\}   
\models 
\co{\phi\vee\phi'(x)}{i}\rightarrow\co{\psi\vee\psi'(\fromvar{x}{i})}{j}$$
under the assumption that at least one among $\phi(x)$ and $\phi'(x)$ is a complete formula. 

$$
\begin{ndproof}
   (1) & i:\phi(x) 
       & (1) & & \reason{Assumption} \\ 
   (2) & j:\psi(\fromvar{x}{i}) 
       & & (1) & \reason{From (1) by \BR} \\ 
   (3) & j:\psi(\fromvar{x}{i})  
       & (3) & & \reason{Assumption} \\ 
   (4) & j:\psi\vee\psi'(\fromvar{x}{i}) 
       & (3) & & \reason{From (3) by $\ori{j}$} \\ 
   (5) & j:\psi\vee\psi'(\fromvar{x}{i}) 
       & & (1) & \reason{From (2) and (4) by $\cut{ji}$} \\ 
   (6) & i:\phi'(x) 
       & (6) & & \reason{Assumption} \\ 
   (7) & j:\psi'(\fromvar{x}{i}) 
       & & (6) & \reason{From (6) by \BR} \\
   (8) & j:\psi'(\fromvar{x}{i})  
       & (8) & & \reason{Assumption} \\ 
   (9) & j:\psi\vee\psi'(\fromvar{x}{i}) 
	   & (8) & & \reason{From (8) by $\ori{j}$} \\ 
  (10) & j:\psi\vee\psi'(\fromvar{x}{i}) 
	   & & (6) & \reason{From (7) and (9) by $\cut{ji}$} \\
  (11) & i:\phi\vee\phi'(x) 
       & (11) & & \reason{Assumption} \\ 
  (12) & j:\psi\vee\psi'(\fromvar{x}{i}) 
       & & (11) & \reason {from (5), (10), and (11) by $\ore{ji}$ discharging (1) and (6)}\\ \hline
  \end{ndproof}
$$
Note that $\ore{ji}$ can be applied at step (12) only if at least eone among $\phi$ and $\phi'$ is complete. 
\end{paragraph}

\newpage
\begin{paragraph}{Instantiation}\ 
$$\left.
  \begin{array}{r@{\ \ }l}
     (a) & \co{x=t}{i}\rightarrow\co{\fromvar{x}{i}=s}{j} \\ 
     (b) & \co{\phi(\tovar{x}{j})}{i}\rightarrow\co{\psi(x)}{j}
   \end{array}\right\}
   \models
       \co{\phi(t)}{i}\rightarrow\co{\psi(s)}{j}$$
under the assumption that $t$ is a complete ground term. 

$$
   \begin{ndproof}
   (1) & i:x=t 
       & (1) & & \reason{Assumption} \\ 
   (2) & j:\fromvar{x}{i}=s 
       & & (1) & \reason{From (1) by $\BR_{(a)}$} \\ 
   (3) & j:y = \fromvar{x}{i} 
       & (3) & & \reason{Assumption} \\ 
   (4) & i:\tovar{y}{j}=x
       & & (3) & \reason{From (3) by $\toI{j}I$} \\ 
   (5) & i:\tovar{y}{j}=x
	   & (5)&  & \reason{Assumption} \\   
   (6) & i:\tovar{y}{j}=t 
       &  (1) (5) & & \reason{From (1) and (5) by $\eqe{i}$} \\ 
   (7) & i:\tovar{y}{j}=t 
	   &  (1) & (3)& \reason{From (4) and (6) by $\cut{i}$ discharging (5)} \\ 
   (8) & i:\phi(t) 
       & (8) & & \reason{Assumption} \\ 
   (9) & i:\phi(\tovar{y}{j}) 
       & (1)(8) &  (3) & \reason{From (7) and (8) by $\eqe{i}$} \\ 
   (10) & j:\psi(y)
       & & (1) (3) (8)& \reason{From (9) by $\BR_{(b)}$} \\ 
   (11) & j:\psi(s)
       & & (6) (1) (3) & \reason{From (10),(3) and (2) by $\eqe{j}$} \\ 
   (12) & j:\fromvar{x}{i}=\fromvar{x}{i}
        & & (1) & \reason{From (2) by $\eqi{j}$} \\ 
   (13) & j:\exists y \ y=\fromvar{x}{i}
        & & (1) & \reason{From (12) by $\exi{j}$} \\ 
   (12) & j:\psi(s)
        & & (6) (1) & \reason{From (11) and (13) by $\exe{j}$ discharging (3)} \\ 
   (13) & i:t=t 
        & & & \reason{By $\eqi{i}$} \\ 
   (14) & i:\exists x \ x=t 
        & & & \reason{From (13) by $\exi{i}$} \\ 
   (15) & j:\psi(s) 
        & & (6) & \reason{From (14) and (12) by $\exe{ji}$ discharging (1)}\\ \hline
   \end{ndproof}
$$
\end{paragraph}

\newpage
\begin{paragraph}{Inversion}
$$\left.
  \begin{array}{r@{\ \ }l}
	(\fun_{ij}) & \co{\tovar{x}{j}=\tovar{y}{j}}{i}\rightarrow\co{x=y}{j}\\ 
    (\incp_{ji}) & \co{\bot}{j}\rightarrow\co{\bot}{i  } \\
	(a) & \co{\phi(x)}{i}\rightarrow\co{\psi(\fromvar{x}{i})}{j}
   \end{array}\right\}	
	\models
    \co{\neg\psi(\fromvar{x}{i})}{j}\rightarrow\co{\neg\phi(x)}{i}$$
under the assumption that $\phi(x)$ is a complete formula. 
  $$
  \begin{ndproof}
   (1) & i:\phi(x) 
       & (1) & & \reason{Assumption} \\ 
   (2) & j:\psi(\fromvar{x}{i}) 
       & & (1) & \reason{From (1) by $\BR_{(a)}$} \\ 
   (3) & j:\neg\psi(\fromvar{x}{i})
       & (3) & & \reason{Assumption} \\ 
   (4) & j:\bot
       & (3) & (1) & \reason{From (2) and (3) by $\impe{j}$} \\ 
   (5) & i:\bot 
       & & (1) (3) & \reason{From (4) by $\BR_{\incp_{ji}}$} \\
   (6) & i:\neg\phi(x)
       & & (3) & \reason{From (5) by $\bot_{i}$ discharging (1)} \\ \hline
  \end{ndproof}
 $$
\end{paragraph}

%% file: soundness.tex
\section{Proof of the Soundness Theorem}
\label{sec:soundness}

\begin{thm}[Soundness]
\label{theo:soundness}
$
(\Gamma,\Sigma)\vdash_{\IC}\co{\phi}{i} \Longrightarrow
\Gamma,e(\Sigma)\models_{\IC} \co{\bigwedge_{i:\sigma \in \Sigma}\sigma\imp\phi}{i}
$
\end{thm}
The proof of the Soundness theorem makes use of the following lemma and notation. 

\begin{lem}
  \label{lem:extension-of-a}
  Let $a$ and $a'$ be two assignments that agree on the values assigned to the arrow and free variables of $\phi$. Then:
  \begin{itemize}
  	\item $\M\models\co{\phi}{i}[a]$ if and only if  $\M\models\co{\phi}{i}[a']$. 
	\item $ m\models \phi[a_i]$ if and only if  $m\models \phi[a'_i]$.
  \end{itemize}   
\end{lem}
The proof of Lemma~\ref{lem:extension-of-a} follows easily from the definition of satisfiability and from the fact that $a$ and $a'$ agree on the interpretation of all the variables of $\phi$.

We write $a(\co{x}{i}=d)$ to denote the assignment obtained from $a$ by setting $a_i(x)=d$ and by letting both $a_j(\tovar{x}{i})$ and $a_j(\fromvar{x}{i})$
undefined. Let $t$ be an $i$-term, $m$ a local model of $L_i$, and $a$ an assignment admissible for $t$, we write $m(t)[a]$ to denote the interpretation
of $t$ in the local model (first\ftext{-}order interpretation) $m$ under the assignment $a$. Let $a$ be an assignment strictly admissible for $\Gamma$ and $\Delta \subseteq \Gamma$, we define $a|_{\Delta}$ to be the reduction of $a$ strictly admissible for $\Delta$:
$$
(a|_{\Delta})_i(x)  = \begin{cases}
					a_i(x)& \text{if $x$ is a regular variable,}\\
					a_i(x)& \text{if $x$ is an arrow variable occurring in $\Delta$,}\\
					\text{undefined} & \text{otherwise.} 
				\end{cases}
$$
Given two assignments $a$ and $a'$ defined over two sets of (regular and arrow) variables $X$ and $X'$, such that they agree on the set of variables $X \cap X'$ they have in common\footnote{Remember that two assignments can differ on the sets of arrow variables they provide an assignment for.}, we define the assignment $a + a'$ as follows:
$$
(a+a')_i(x)  = \begin{cases}
					a_i(x)=a'_i(x) & \text{if both $a_i(x)$ and $a'_i(x)$ are defined,}\\
					a_i(x) & \text{if only $a_i(x)$ is defined,} \\
                    a'_i(x) & \text{if only $a'_i(x)$ is defined.} 
				\end{cases}
$$
Finally, for the sake of readability we write $\bigwedge \Sigma$ as a shorthand for $\bigwedge_{i:\sigma \in \Sigma}\sigma$. 
                    
\begin{pot}{\ref{theo:soundness}} 
  	The proof is by induction on the structure of the derivation of $\co{\phi}{i}$ from $(\Gamma,\Sigma)$. We first prove the theorem for  $(\Gamma,\Sigma)\vdash_{\IC}\co{\phi}{i}$ with a one step derivation (base case). Then we prove the theorem for $(\Gamma,\Sigma)\vdash_{\IC}\co{\phi}{i}$ with a deduction $\Pi$ of length $n+1$, by assuming that the theorem holds for all deductions $\Pi'$ of length $\leq n$ and proving that it holds also for $\Pi$ (inductive step). 
\onlyintechrep{We prove the inductive step by examining all the inference rules $\rho$ used in the final step of $\Pi$.}
\onlyinpaper{For the sake of space we include here the proof of the inductive step for a subset of paradigmatic inference rules $\rho$ among the ones presented at page \pageref{def:mlsystem}. The complete proof for all inference rules can be found in \cite{DFOL-techrep}.} 

\begin{description}
	\item{\textbf{Base Case}:} If $(\Gamma,\Sigma)\vdash_{\IC}\co{\phi}{i}$ with a one step derivation, then either $\co{\phi}{i}$ is an assumption or a direct application of $\eqi{i}$ with $n=0$. If $\co{\phi}{i}$ is an assumption, then we have to prove that $e(\co{\phi}{i})\models \co{\phi\imp\phi}{i}$. 
  Let us assume that $\M\models e(\co{\phi}{i})[a]$. This means that $a_i$ is defined on all the 
  arrow variables of $\phi$. Therefore, for every $m\in M_i$, 
  $m\models\phi\imp\phi[a]$ and the proof is done. If $\co{\phi}{i}$ is the consequence of an application of $\eqi{i}$ with $n=0$, then it is of the form $\co{t=t}{i}$ and $t$ does not contain any arrow variable. This implies that $\M\models i:t=t[a]$ for all models $\M$ and assignments $a$, which also concludes the proof.  
\item[$\impi{i}$] If $(\Gamma,\Sigma)\vdash_{\IC}\co{\phi\imp\psi}{i}$, and the last rule used is $\impi{i}$, then $(\Gamma, \Sigma\cup\{\co{\phi}{i}\})\vdash_{\IC}\co{\phi \imp\psi}{i}$ also holds. 

To prove that $\Gamma,e(\Sigma)\brmodels i:\bigwedge\Sigma\imp(\phi\imp\psi)$, let $\M$ be a \IC-model and $a$ an assignment strictly admissible for $\Gamma, e(\Sigma)$ such that $\M\models\Gamma, e(\Sigma)[a]$. From the restriction R\ref{item:no-existential-arrow-variables}, $\phi$ and $\psi$ cannot contain existential arrow variables (that is, arrow variables not contained in the premises). Therefore what we have to prove is that $\M\models \co{\bigwedge\Sigma\imp(\phi\imp\psi)}{i}[a]$. We do this by distinguishing two cases: 
\begin{description}
\item[$\co{\phi}{i}$ is a local assumption.] 
  From the fact that $(\Gamma, \Sigma\cup\{\co{\phi}{i}\})\vdash_{\IC}\co{\phi \imp\psi}{i}$ is a deduction  $\Pi'$ of length $\leq n$ we can apply the inductive hypothesis and obtain that $\Gamma, e(\Sigma \cup \{\co{\phi}{i}\})\brmodels\co{(\bigwedge\Sigma\wedge\phi)\imp\psi}{i}$.
  Since $\co{\phi}{i}$ does not contain any existential arrow variable (restriction R\ref{item:no-existential-arrow-variables}), then $\M\models\Gamma, e(\Sigma \cup \{\co{\phi}{i}\})[a]$. Thus, from the inductive hypothesis we have that $\M \models \co{(\bigwedge\Sigma \wedge\phi)\imp\psi}{i}[a]$. 
  Since $(\bigwedge\Sigma \wedge\phi)\imp\psi$ is equivalent to
  $\bigwedge\Sigma \imp(\phi\imp\psi)$, and they contain
  the same arrow variables, we can conclude that $\M\models \co{\bigwedge\Sigma \imp(\phi\imp\psi)}{i}$, and this ends the proof.  
\item[$\co{\phi}{i}$ is a global assumption.] 
	In this case from the inductive hypothesis we obtain that $\Gamma \cup \{\co{\phi}{i}\}, e(\Sigma)\brmodels\co{\bigwedge\Sigma\imp\psi}{i}$.
	From the restriction R\ref{item:restr-disch-complete-global-assumptions} we have that $\co{\phi}{i}$ is a complete formula, and either $\M\models \co{\phi}{i}[a]$ or $\M\models \co{\neg\phi}{i}[a]$. Let us consider the two cases separately.
  If $\M\models \co{\phi}{i}[a]$ we can use the inductive hypothesis to prove that $\M\models \co{\bigwedge\Sigma\imp\psi}{i}[a]$. This, in turn, implies that $\M\models \co{\bigwedge\Sigma\imp(\phi\imp\psi)}{i}[a]$; if $\M\models \co{\neg\phi}{i}[a]$ then $\M\models \co{\bigwedge\Sigma\imp(\phi\imp\psi)}{i}[a]$ from the definition of first\ftext{-}order satisfiability, and this ends the proof.  
\end{description}
\onlyintechrep{
\item[$\impe{i}$] If $(\Gamma,\Sigma)\vdash_{\IC}\co{\psi}{i}$ and the last rule used is $\impe{i}$, then from the inductive hypothesis there are two formulae $\co{\phi}{i}$ and $\co{\phi \imp \psi}{i}$ such that $\Gamma_1, e(\Sigma_1)\brmodels \co{\bigwedge\Sigma_1\imp\phi}{i}$ and $\Gamma_2,e(\Sigma_2)\brmodels \co{\bigwedge\Sigma_2\imp(\phi\imp\psi)}{i}$, with $\Gamma = \Gamma_1 \cup \Gamma_2$ and $\Sigma = \Sigma_1 \cup \Sigma_2$. We have to prove 
$\Gamma_1\cup\Gamma_1,e(\Sigma_1\cup\Sigma_2)\brmodels \co{\bigwedge(\Sigma_1\cup\Sigma_2)\imp\psi}{i}$.
 
Let $a$ be an assignment strictly admissible for $\Gamma,e(\Sigma)$ such that $\M\models\Gamma,e(\Sigma)[a]$. Since neither $\phi$ nor $\psi$ contain existential arrow variables (restriction R\ref{item:no-existential-arrow-variables}) we have to prove that $\M\models \co{\bigwedge\Sigma \imp\psi}{i}[a]$ holds. 
Let $a|_{\Gamma_1,e(\Sigma_1)}$ and $a|_{\Gamma_2,e(\Sigma_2)}$ be the restrictions of $a$ strictly admissible for $\Gamma_1,e(\Sigma_1)$ and $\Gamma_2,e(\Sigma_2)$ respectively. We have that  $\M\models\Gamma_1,e(\Sigma_1)[a|_{\Gamma_1,e(\Sigma_1)}]$ and $\M\models\Gamma_2,e(\Sigma_2)[a|_{\Gamma_2,e(\Sigma_2)}]$ hold. From the inductive hypothesis and the fact that neither $\phi$ nor $\psi$ contain existential arrow variables (restriction R\ref{item:no-existential-arrow-variables}) we have that 
$\M\models \co{\bigwedge\Sigma_1 \imp\phi}{i}[a|_{\Gamma_1,e(\Sigma_1)}]$ and 
$\M\models \co{\bigwedge\Sigma_2\imp(\phi\imp\psi)}{i}[a|_{\Gamma_2,e(\Sigma_2)}]$.
Let $m$ be a local model in $\M_i$ such that $m \models \sigma[a]$ for all formulae $\sigma \in \Sigma_1 \cup \Sigma_2$. Since $a = a|_{\Gamma_1,e(\Sigma_1)}+a|_{\Gamma_2,e(\Sigma_2)}$, then $m\models \phi[a]$ and $m \models \phi \imp \psi[a]$. Thus $m \models \psi[a]$ and  
from this we can conclude that $\M\models \co{\bigwedge\Sigma \imp\psi}{i}[a]$ holds.
\item[$\andi{i}$] (similar to $\impe{i}$). 
If $(\Gamma,\Sigma)\vdash_{\IC}\co{\phi \con \psi}{i}$ and the last rule used is $\andi{i}$, then from the inductive hypothesis we know that 
$\Gamma_1,e(\Sigma_1)\brmodels \co{\bigwedge\Sigma\imp\phi}{i}$ and $\Gamma_2,e(\Sigma_2)\brmodels \co{\bigwedge\Sigma_2\imp\psi}{i}$, with $\Gamma = \Gamma_1 \cup \Gamma_2$ and $\Sigma = \Sigma_1 \cup \Sigma_2$. We have to prove that $\Gamma_1\cup\Gamma_2,e(\Sigma_1\cup\Sigma_2)\brmodels \co{\bigwedge(\Sigma_1\cup\Sigma_2)\imp(\phi\wedge\psi)}{i}$.

Let $a$ be an assignment strictly admissible for $\Gamma,e(\Sigma)$ such that $\M\models\Gamma,e(\Sigma)[a]$. Since neither $\phi$ nor $\psi$ contain existential arrow variables (restriction R\ref{item:no-existential-arrow-variables}) we have to prove that $\M\models \co{\bigwedge\Sigma \imp(\phi \con \psi)}{i}[a]$ holds.
 
Let $a|_{\Gamma_1,e(\Sigma_1)}$ and $a|_{\Gamma_2,e(\Sigma_2)}$ be two restrictions of $a$ strictly admissible for $\Gamma_1,e(\Sigma_1)$ and $\Gamma_2,e(\Sigma_2)$, respectively. We have that  $\M\models\Gamma_1,e(\Sigma_1)[a|_{\Gamma_1,e(\Sigma_1)}]$ and $\M\models\Gamma_2,e(\Sigma_2)[a|_{\Gamma_2,e(\Sigma_2)}]$ hold. From the inductive hypothesis and the fact that neither $\phi$ nor $\psi$ contain existential arrow variables (restriction R\ref{item:no-existential-arrow-variables}) we have that 
$\M\models \co{\bigwedge\Sigma_1\imp\phi}{i}[a|_{\Gamma_1,e(\Sigma_1)}]$ and 
$\M\models \co{\bigwedge\Sigma_2\imp\psi}{i}[a|_{\Gamma_2,e(\Sigma_2)}]$. 
Let $m$ be a local model in $\M_i$ such that $m \models \sigma[a]$ for all formulae $\sigma \in \Sigma_1 \cup \Sigma_2$. Since $a = a|_{\Gamma_1,e(\Sigma_1)}+a|_{\Gamma_2,e(\Sigma_2)}$, then $m\models \phi[a]$ and $m \models \psi[a]$. Thus $m \models \phi \con \psi[a]$ and  
from this we can conclude that $\M\models \co{\bigwedge\Sigma \imp(\phi \con \psi)}{i}[a]$ holds.
\item[$\ande{i}$]
If $(\Gamma,\Sigma)\vdash_{\IC}\co{\phi}{i}$ and the last rule used is $\ande{i}$, then from the inductive hypothesis there is a formula $\co{\psi}{i}$ such that  
$\Gamma,e(\Sigma)\brmodels \co{\bigwedge\Sigma\imp(\phi \con \psi)}{i}$ and we have to prove that $\Gamma,e(\Sigma)\brmodels \co{\bigwedge\Sigma\imp\phi}{i}$. 

Let $a$ be an assignment strictly admissible for $\Gamma,e(\Sigma)$ such that $\M \models \Gamma,e(\Sigma)[a]$. From the inductive hypothesis and and the fact that $\phi$ and $\psi$ do not contain existential arrow variables we can infer $\M\models \co{\bigwedge\Sigma_i\imp(\phi\wedge\psi)}{i}[a]$, and from this and the notion of satisfiability in $\M$ we can infer $\M\models \co{\bigwedge\Sigma\imp\phi}{i}[a]$. 

The proof for the elimination of the lefthand side conjunct is analogous. 
\item[$\ori{i}$]
If $(\Gamma,\Sigma)\vdash_{\IC}\co{\phi \dis \psi}{i}$ and the last rule used is $\ori{i}$, then from the inductive hypothesis we know that   
$\Gamma,e(\Sigma)\brmodels \co{\bigwedge\Sigma\imp\phi}{i}$ holds and we have to prove that $\Gamma,e(\Sigma)\brmodels \co{\bigwedge\Sigma\imp(\phi \dis \psi)}{i}$. 

Let $a$ be an assignment strictly admissible for $\Gamma,e(\Sigma)$ such that $\M\models\Gamma,e(\Sigma)[a]$. 
From the inductive hypothesis, the fact that $\phi$ does not contain existential arrow variables (restriction R\ref{item:no-existential-arrow-variables}) we have that $\M\models \co{\bigwedge\Sigma\imp\phi}{i}[a]$. From the fact that $\ori{i}$ cannot introduce existential arrow variables (restriction R\ref{item:restr-introduce-arrow-variable}) we know that $a$ is also admissible for $\co{\psi}{i}$. Thus it is easy to show that all local models in $\M_i$ satisfy $\phi\dis\psi$ under the assignment $a$ and therefore that $\M\models \co{\bigwedge\Sigma\imp(\phi\vee\psi)}{i}[a]$ holds.  

The proof for the introduction of the lefthand side disjunct is analogous. 
\item[$\ore{ji}$] If $(\Gamma,\Sigma)\vdash_{\IC}\co{\theta}{j}$ and the last rule used is $\ore{ji}$, then there is a formula $\co{\phi \dis \psi}{i}$, and three pairs $(\Gamma_i, \Sigma_i)$, $i=1,\ldots,3$ with $\Gamma_i \subseteq \Gamma \cup \Sigma$  and $\Sigma_i \subseteq \Gamma \cup \Sigma$ such that 
\begin{align*}
	(\Gamma_1, \Sigma_1) & \vdash_{\IC} \co{\phi \dis \psi}{i},\\
	(\Gamma_2,\co{\phi}{i}, \Sigma_2) & \vdash_{\IC} \co{\theta}{j}, \\
	(\Gamma_3,\co{\psi}{i}, \Sigma_3) & \vdash_{\IC} \co{\theta}{j}.
\end{align*}
We prove that $\Gamma,e(\Sigma)\brmodels\co{\bigwedge\Sigma \imp \theta}{j}$ by considering three different cases: 
\begin{description}
\item[$\co{\phi}{i}$ and $\co{\psi}{i}$ are both local assumptions.]
In this case $i=j$, $\Gamma = \Gamma_1 \cup \Gamma_2 \cup \Gamma_3$, $\Sigma =  \Sigma_1 \cup \Sigma_2 \cup \Sigma_3$, and from the inductive hypothesis we have that:
\begin{align}
	\label{eq:ore-1.1}\Gamma_1, e(\Sigma_1) & \brmodels \co{\bigwedge \Sigma_1 \imp (\phi \dis \psi)}{i}, \\
	\label{eq:ore-1.2}\Gamma_2, e(\Sigma_2 \cup \{\co{\phi}{i}\}) & \brmodels \co{(\bigwedge\Sigma_2 \con \phi) \imp \theta}{i}, \\
	\label{eq:ore-1.3}\Gamma_3, e(\Sigma_3 \cup \{\co{\psi}{i}\}) & \brmodels \co{(\bigwedge\Sigma_3 \con \psi) \imp \theta}{i}.
\end{align}

Let $a$ be an assignment strictly admissible for $\Gamma,e(\Sigma)$ such that $\M\models\Gamma,e(\Sigma)[a]$, and let $m$ be an arbitrary local model in $\M_i$. We have to prove that there is an extension $a'$ of $a$ for $\co{\theta}{i}$ such that if $m \models \sigma[a']$ for all $\sigma \in \Sigma$ then $m \models \theta[a']$. 

Let $a|_{\Gamma_1,e(\Sigma_1)}$ be the restriction of $a$ strictly admissible for $\Gamma_1,e(\Sigma_1)$. 
Then $\M  \models\Gamma_1,e(\Sigma_1)[a|_{\Gamma_1,e(\Sigma_1)}]$, and from the inductive hypothesis \eqref{eq:ore-1.1} $a|_{\Gamma_1,e(\Sigma_1)}$ can be extended to an assignment $\bar{a}|_{\Gamma_1,e(\Sigma_1)}$ admissible for $\co{\phi \dis \psi}{i}$ such that 
\begin{equation}
	\label{eq:ore-1.4}
	\M  \models \co{\Sigma_1 \imp (\phi \dis \psi)}{i}[\bar{a}|_{\Gamma_1,e(\Sigma_1)}].
\end{equation}
Let $a|_{\Gamma_2,e(\Sigma_2)}$ be the restriction of $a$ strictly admissible for $\Gamma_2,e(\Sigma_2)$. Then $\M  \models\Gamma_2,e(\Sigma_2)[a|_{\Gamma_2,e(\Sigma_2)}]$. Restriction R\ref{item:restriction-cut} imposes that the existential arrow variables occurring in $\co{\phi \dis \psi}{i}$ do not occur in any of the assumptions in $\Gamma_2$ and $\Sigma_2$. Thus we can extend $a|_{\Gamma_2,e(\Sigma_2)}$ to an assignment $\bar{a}|_{\Gamma_2,e(\Sigma_2)}$ admissible for $\co{\phi}{i}$ by adding to $a|_{\Gamma_2,e(\Sigma_2)}$ the assignment of the existential arrow variables of $\co{\phi}{i}$ according to $\bar{a}|_{\Gamma_1,e(\Sigma_1)}$.
Then it is easy to show that $\M \models\Gamma_2,e(\Sigma_2 \cup \{\co{\phi}{i}\})[\bar{a}|_{\Gamma_2,e(\Sigma_2)}]$, and from the inductive hypothesis \eqref{eq:ore-1.2} we can infer that $\bar{a}|_{\Gamma_2,e(\Sigma_2)}$ can be extended to an assignment $\doublebar{a}|_{\Gamma_2,e(\Sigma_2)}$ for $\co{\theta}{i}$ such that 
\begin{equation}
	\label{eq:ore-1.5}
	\M \models \co{(\bigwedge \Sigma_2 \con \phi) \imp \theta}{i}[\doublebar{a}|_{\Gamma_2,e(\Sigma_2)}].
\end{equation}
With a similar construction we can use \eqref{eq:ore-1.3} to extend $a|_{\Gamma_3,e(\Sigma_3)}$ to an assignment $\doublebar{a}|_{\Gamma_3,e(\Sigma_3)}$ for $\co{\theta}{i}$ such that 
\begin{equation}
	\label{eq:ore-1.6}
	\M \models \co{(\bigwedge \Sigma_3 \con \psi) \imp \theta}{i}[\doublebar{a}|_{\Gamma_3,e(\Sigma_3)}].
\end{equation}
Let $m$ be a local model in $\M_i$ such that $m \models \sigma[a]$ for all $\sigma \in \Sigma$. Since $m$ satisfies all the formulae in $\Sigma_1$, then we can use \eqref{eq:ore-1.4} to obtain $m \models \phi \dis \psi[\bar{a}|_{\Gamma_1,e(\Sigma_1)}]$. Let us assume that $m\models \phi[\bar{a}|_{\Gamma_1,e(\Sigma_1)}]$, then we can use \eqref{eq:ore-1.5} to infer $m\models \theta[\doublebar{a}|_{\Gamma_2,e(\Sigma_2)}]$. Analogously if $m\models \psi[\bar{a}|_{\Gamma_1,e(\Sigma_1)}]$, then  we can use \eqref{eq:ore-1.6} to infer $m\models \theta[\doublebar{a}|_{\Gamma_3,e(\Sigma_3)}]$, and this ends the proof.

\item [$\co{\phi}{i}$ is a local assumption and $\co{\psi}{i}$ is a global assumption.]
If this is the case\footnote{The case $\co{\phi}{i}$ global assumption and $\co{\psi}{i}$ local assumption is analogous.}, then $i=j$, $\Gamma = \Gamma_1 \cup \Gamma_2 \cup \Gamma_3$, $\Sigma =  \Sigma_1 \cup \Sigma_2 \cup \Sigma_3$, $\co{\psi}{i}$ is a complete formula (restriction R\ref{item:restr-disch-complete-global-assumptions}) and from the inductive hypothesis we have that:
\begin{align}
	\label{eq:ore-2.1}\Gamma_1, e(\Sigma_1) & \brmodels \co{\bigwedge \Sigma_1 \imp (\phi \dis \psi)}{i}, \\
	\label{eq:ore-2.2}\Gamma_2, e(\Sigma_2 \cup \{\co{\phi}{i}\}) & \brmodels \co{(\bigwedge\Sigma_2 \con \phi) \imp \theta}{i}, \\
	\label{eq:ore-2.3}\Gamma_3 \cup \{\co{\psi}{i}\}, e(\Sigma_3) & \brmodels \co{\bigwedge\Sigma_3 \imp \theta}{i}.
\end{align}

We proceed as in the previous case by assuming that $a$ is an assignment strictly admissible for $\Gamma,e(\Sigma)$ such that $\M\models\Gamma,e(\Sigma)[a]$, and let $m$ be an arbitrary local model in $\M_i$. We have to prove that there is an extension $a'$ of $a$ for $\co{\theta}{i}$ such that if $m \models \sigma[a']$ for all $\sigma \in \Sigma$ then $m \models \theta[a']$. 

Let $a|_{\Gamma_1,e(\Sigma_1)}$ be the restriction of $a$ strictly admissible for $\Gamma_1,e(\Sigma_1)$. 
Then $\M  \models\Gamma_1,e(\Sigma_1)[a|_{\Gamma_1,e(\Sigma_1)}]$, and from the inductive hypothesis \eqref{eq:ore-2.1} $a|_{\Gamma_1,e(\Sigma_1)}$ can be extended to an assignment $\bar{a}|_{\Gamma_1,e(\Sigma_1)}$ admissible for $\co{\phi \dis \psi}{i}$ such that 
\begin{equation}
	\label{eq:ore-2.4}
	\M  \models \co{\Sigma_1 \imp (\phi \dis \psi)}{i}[\bar{a}|_{\Gamma_1,e(\Sigma_1)}].
\end{equation}
Let $a|_{\Gamma_2,e(\Sigma_2)}$ be the restriction of $a$ strictly admissible for $\Gamma_2,e(\Sigma_2)$. With a proof equal to the one for the previous case we can extend $a|_{\Gamma_2,e(\Sigma_2)}$ to an assignment $\bar{a}|_{\Gamma_2,e(\Sigma_2)}$ admissible for $\co{\phi}{i}$ by adding to $a|_{\Gamma_2,e(\Sigma_2)}$ the assignment of the existential arrow variables of $\co{\phi}{i}$ according to $\bar{a}|_{\Gamma_1,e(\Sigma_1)}$.
Then it is easy to show that $\M \models\Gamma_2,e(\Sigma_2 \cup \{\co{\phi}{i}\})[\bar{a}|_{\Gamma_2,e(\Sigma_2)}]$, and from the inductive hypothesis \eqref{eq:ore-2.2} we can infer that $\bar{a}|_{\Gamma_2,e(\Sigma_2)}$ can be extended to an assignment $\doublebar{a}|_{\Gamma_2,e(\Sigma_2)}$ for $\co{\theta}{i}$ such that 
\begin{equation}
	\label{eq:ore-2.5}
	\M \models \co{(\bigwedge \Sigma_2\con\phi) \imp \theta}{i}[\doublebar{a}|_{\Gamma_2,e(\Sigma_2)}].
\end{equation}

With a similar construction we can use \eqref{eq:ore-2.3} to extend $a|_{\Gamma_3,e(\Sigma_3)}$ to an assignment $\bar{a}|_{\Gamma_3,e(\Sigma_3)}$ admissible also for $\co{\psi}{i}$ such that if $\M \models \co{\psi}{i}[\bar{a}|_{\Gamma_3,e(\Sigma_3)}]$, then $\bar{a}|_{\Gamma_3,e(\Sigma_3)}$ can be extended to an assignment $\doublebar{a}|_{\Gamma_3,e(\Sigma_3)}$ for $\co{\theta}{i}$ such that 
\begin{equation}
	\label{eq:ore-2.6}
	\M \models \co{\Sigma_3 \imp \theta}{i}[\doublebar{a}|_{\Gamma_3,e(\Sigma_3)}].
\end{equation}
Let $m$ be a local model in $\M_i$ such that $m \models \sigma[a]$ for all $\sigma \in \Sigma$. Since $m$ satisfies all the formulae in $\Sigma_1$, then we can use \eqref{eq:ore-2.4} to obtain $m \models \phi \dis \psi[\bar{a}|_{\Gamma_1,e(\Sigma_1)}]$. Let us assume that $m\models \phi[\bar{a}|_{\Gamma_1,e(\Sigma_1)}]$, then we can use \eqref{eq:ore-2.5} to infer $m\models \theta[\doublebar{a}|_{\Gamma_2,e(\Sigma_2)}]$. If $m\models \psi[\bar{a}|_{\Gamma_1,e(\Sigma_1)}]$, we can use the fact that $\co{\psi}{i}$ is a complete formula to infer $\M \models \co{\psi}{i}[\bar{a}|_{\Gamma_3,e(\Sigma_3)}]$. Therefore, we can use \eqref{eq:ore-2.6} to infer $m\models \theta[\doublebar{a}|_{\Gamma_3,e(\Sigma_3)}]$, and this ends the proof.

\item [$\co{\phi}{i}$ and $\co{\psi}{i}$ are both global assumptions.] 
We consider the two cases $i=j$ and $i\neq j$ separately.

If $i=j$, then $\Gamma = \Gamma_1 \cup \Gamma_2 \cup \Gamma_3$, $\Sigma =  \Sigma_1 \cup \Sigma_2 \cup \Sigma_3$, $\co{\phi}{i}$ and $\co{\psi}{i}$ are complete formulae (restriction R\ref{item:restr-disch-complete-global-assumptions}) and from the inductive hypothesis we have that:
\begin{align}
	\label{eq:ore-3.1}\Gamma_1, e(\Sigma_1) & \brmodels \co{\bigwedge \Sigma_1 \imp (\phi \dis \psi)}{i}, \\
	\label{eq:ore-3.2}\Gamma_2 \cup\{\co{\phi}{i}\}, e(\Sigma_2) & \brmodels \co{\bigwedge\Sigma_2 \imp \theta}{i}, \\
	\label{eq:ore-3.3}\Gamma_3 \cup\{\co{\psi}{i}\}, e(\Sigma_3) & \brmodels \co{\bigwedge\Sigma_3 \imp \theta}{i}.
\end{align}

We proceed as in the previous cases by assuming that $a$ is an assignment strictly admissible for $\Gamma,e(\Sigma)$ such that $\M\models\Gamma,e(\Sigma)[a]$, and  $m$ is an arbitrary local model in $\M_i$. We have to prove that there is an extension $a'$ of $a$ for $\co{\theta}{i}$ such that if $m \models \sigma[a']$ for all $\sigma \in \Sigma$ then $m \models \theta[a']$. 

Let $a|_{\Gamma_1,e(\Sigma_1)}$ be the restriction of $a$ strictly admissible for $\{\Gamma_1,e(\Sigma_1)\}$. 
Then $\M  \models\Gamma_1,e(\Sigma_1)[a|_{\Gamma_1,e(\Sigma_1)}]$, and from the inductive hypothesis \eqref{eq:ore-3.1} $a|_{\Gamma_1,e(\Sigma_1)}$ can be extended to an assignment $\bar{a}|_{\Gamma_1,e(\Sigma_1)}$ admissible for $\co{\phi \dis \psi}{i}$ such that 
\begin{equation}
	\label{eq:ore-3.4}
	\M  \models \co{\Sigma_1 \imp (\phi \dis \psi)}{i}[\bar{a}|_{\Gamma_1,e(\Sigma_1)}].
\end{equation}
Let $a|_{\Gamma_2,e(\Sigma_2)}$ be the restriction of $a$ strictly admissible for $\{\Gamma_2,e(\Sigma_2)\}$. With a proof equal to the one for the previous cases we can extend $a|_{\Gamma_2,e(\Sigma_2)}$ to an assignment $\bar{a}|_{\Gamma_2,e(\Sigma_2)}$ admissible also for $\co{\phi}{i}$ by adding to $a|_{\Gamma_2,e(\Sigma_2)}$ the assignment of the existential arrow variables of $\co{phi}{i}$ according to $\bar{a}|_{\Gamma_1,e(\Sigma_1)}$.
Then it is easy to show that if $\M \models \co{\phi}{i}[\bar{a}|_{\Gamma_2,e(\Sigma_2)}]$ we can use \eqref{eq:ore-2.2} to infer that $\bar{a}|_{\Gamma_2,e(\Sigma_2)}$ can be extended to an assignment $\doublebar{a}|_{\Gamma_2,e(\Sigma_2)}$ for $\co{\theta}{i}$ such that 
\begin{equation}
	\label{eq:ore-3.5}
	\M \models \co{\bigwedge \Sigma_2 \imp \theta}{i}[\doublebar{a}|_{\Gamma_2,e(\Sigma_2)}].
\end{equation}
With a similar construction we can use \eqref{eq:ore-2.3} to extend $a|_{\Gamma_3,e(\Sigma_3)}$ to an assignment $\bar{a}|_{\Gamma_3,e(\Sigma_3)}$ admissible also for $\co{\psi}{i}$ such that if $\M \models \co{\psi}{i}[\bar{a}|_{\Gamma_3,e(\Sigma_3)}]$, then $\bar{a}|_{\Gamma_3,e(\Sigma_3)}$ can be extended to an assignment $\doublebar{a}|_{\Gamma_3,e(\Sigma_3)}$ for $\co{\theta}{i}$ such that 
\begin{equation}
	\label{eq:ore-3.6}
	\M \models \co{\Sigma_3 \imp \theta}{i}[\doublebar{a}|_{\Gamma_3,e(\Sigma_3)}].
\end{equation}
Let $m$ be a local model in $\M_i$ such that $m \models \sigma[a]$ for all $\sigma \in \Sigma$. Since $m$ satisfies all the formulae in $\Sigma_1$, then we can use \eqref{eq:ore-3.4} to obtain $m \models \phi \dis \psi[\bar{a}|_{\Gamma_1,e(\Sigma_1)}]$. Let us assume that $m\models \phi[\bar{a}|_{\Gamma_1,e(\Sigma_1)}]$. Since $\co{\phi}{i}$ is a complete formula then $\M \models \co{\phi}{i}[\bar{a}|_{\Gamma_2,e(\Sigma_2)}]$. Therefore we can use \eqref{eq:ore-3.5} to infer $m\models \theta[\doublebar{a}|_{\Gamma_2,e(\Sigma_2)}]$. If $m\models \psi[\bar{a}|_{\Gamma_1,e(\Sigma_1)}]$ with $\co{\psi}{i}$ complete formula, then $\M \models \co{\psi}{i}[\bar{a}|_{\Gamma_3,e(\Sigma_3)}]$. Therefore we can use \eqref{eq:ore-3.6} to infer $m\models \theta[\doublebar{a}|_{\Gamma_3,e(\Sigma_3)}]$, and this ends the proof.

If $i\neq j$, the proof can be obtained as in the previous case, just taking into account that $\Gamma = \Gamma_1 \cup \Sigma_1 \cup \Gamma_2 \cup \Gamma_3$, and $\Sigma = \Sigma_2 \cup \Sigma_3$ (that is, the local assumptions of $\Sigma_1$ become global due to the change of index from $i$ to $j$ triggered by $\ore{ji}$). 

\end{description}
\item[$\bot_i$]
If $(\Gamma,\Sigma)\vdash_{\IC}\co{\neg \phi}{i}$ and the last rule used is $\bot_{i}$, then we know that $\Gamma,\co{\phi}{i},\Sigma\vdash_{\IC} \co{\bot)}{i}$.

To prove that $\Gamma,e(\Sigma)\brmodels \co{\bigwedge\Sigma\imp\phi}{i}$, let $\M$ be a \IC-model and $a$ an assignment strictly admissible for $\Gamma,  e(\Sigma)$ such that $\M\models\Gamma, e(\Sigma)[a]$. From the restriction R\ref{item:restr-introduce-arrow-variable}, the arrow variables in $\co{\phi}{i}$ must be contained in some of the other assumptions used to infer $\co{\bot}{i}$. If not, the rule $\bot_{i}$ would introduce new existential arrow variables by discharging $\co{\neg \phi}{i}$. Thus we have to prove that $\M\models\co{\bigwedge\Sigma \imp \phi}{i}[a]$, and we do it by distinguishing two cases: 

\begin{description}
\item[$\co{\neg\phi}{i}$ is a local assumption.] In this case the inductive hypothesis enables us to infer $\Gamma,e(\Sigma \cup \{\co{\neg \phi}{i}\})\brmodels \co{(\bigwedge\Sigma \con \neg \phi)\imp\bot}{i}$. 
Since all the arrow variables in $\co{\neg \phi}{i}$ are also contained in $\Gamma \cup \Sigma$, then $\M\models\Gamma, e(\Sigma \cup \{\co{\neg \phi}{i}\})[a]$. Thus all the local models $m \in M_i$ satisfy $\co{(\bigwedge\Sigma \con \neg \phi)\imp\bot}{i}$ which is classically equivalent to $\co{\bigwedge\Sigma \imp\phi}{i}$. Thus $\M\models\co{\bigwedge\Sigma \imp \phi}{i}[a]$.  
\item[$\co{\neg\phi}{i}$ is a global assumption.] In this case $\co{\neg \phi}{i}$, and therefore $\co{\phi}{i}$, are complete formulae. Moreover the inductive hypothesis enables us to infer that $\Gamma \cup \{\co{\neg \phi}{i}\},e(\Sigma)\brmodels \co{(\bigwedge\Sigma \con \neg \phi)\imp\bot}{i}$ holds.

Since $\co{\phi}{i}$ is a complete formula, either $\M\models \co{\phi}{i}[a]$ or $\M\models \co{\neg\phi}{i}[a]$. In the first case $\M\models\co{\bigwedge\Sigma \imp \phi}{i}[a]$ is trivially satisfied. In the second case, we can use the inductive hypothesis and obtain that $\M \models \co{(\bigwedge\Sigma \con \neg \phi)\imp\bot}{i}$. Again, this is equivalent to say that $\M \models \co{\bigwedge\Sigma \imp\phi}{i}$ and this ends the proof. 
\end{description}
\item[$\alli{i}$]   If $(\Gamma,\Sigma)\vdash_{\IC}\co{\forall x\ \phi}{i}$ and the last rule used is $\alli{i}$ then, from the inductive hypothesis we know that  $\Gamma, e(\Sigma)\brmodels \co{\bigwedge\Sigma\imp\phi}{i}$ holds and we have to prove that $\Gamma, e(\Sigma)\brmodels \co{\bigwedge\Sigma\imp\forall x\ \phi}{i}$.

Let $a$ be an assignment strictly admissible for $\Gamma,e(\Sigma)$ such that $\M \models \Gamma,e(\Sigma)[a]$. For every $d\in\dom_i$ we have that 
  $\M\models\Gamma,e(\Sigma)[a(\co{x}{i}=d)]$. This is guaranteed by the fact that $x$ does not occur free in the formulae in $\Gamma$ and $\Sigma$ with index $i$ 
  and that $\tovar{x}{i}$ and $\fromvar{x}{i}$ do not occur in any formula in $\Gamma$ (restriction R\ref{item:restriction-alli}).
From the inductive hypothesis and and the fact that $\phi$ does not contain existential arrow variables (restriction R\ref{item:no-existential-arrow-variables}) we can infer $\M\models \co{\bigwedge\Sigma\imp\phi}{i}[a(\co{x}{i}=d)]$. Since $x$ does not occur in $\Sigma$ we can infer (via first\ftext{-}order satisfiability) that 
$\M\models \co{\bigwedge\Sigma\imp\forall x \ \phi}{i}[a]$. 
\item[$\alle{i}$] If $(\Gamma,\Sigma)\vdash_{\IC}\co{\phi}{i}$ and the last rule used is $\alle{i}$, then from the inductive hypothesis we have that $\Gamma, e(\Sigma)\brmodels \co{\bigwedge\Sigma\imp\forall x\ \phi}{i}$ holds and we have to prove that $\Gamma, e(\Sigma)\brmodels \co{\bigwedge\Sigma\imp\phi^t_x}{i}$.

Let $a$ be an assignment strictly admissible for $\Gamma,e(\Sigma)$ such that $\M \models \Gamma,e(\Sigma)[a]$. Then from the inductive hypothesis and the fact that $\alle{i}$ cannot be applied to formulae containing existential arrow variables (restriction R\ref{item:no-existential-arrow-variables}) we have that $\M\models \co{\bigwedge \Sigma\imp\forall x\ \phi}{i}[a]$. Since $\alle{i}$ cannot introduce new existential variables (restriction R\ref{item:restr-introduce-arrow-variable}) $a$ is admissible also for $t$. Let $m$ be a local model in $M_i$ such that $m \models \bigwedge \Sigma[a]$ (if not $m\models\Sigma\imp\phi^t_x[a]$ trivially holds), then $m\models \forall x\ \phi[a]$. Let $d=m(t)[a]$ be the object in $\dom_i$ assigned to the interpretation of $t$ in model $m$ by $a$, then $m\models\Sigma\imp\phi^t_x[a]$, which implies that $\M\models \co{\Sigma\imp\phi^t_x}{i}[a]$.
\item[$\exi{i}$] If $(\Gamma,\Sigma)\vdash_{\BR} \co{\exists x\ \phi}{i}$ and the last rule used is $\exi{i}$, then from the inductive hypothesis we have that $\Gamma, e(\Sigma)\brmodels \co{\bigwedge\Sigma\imp\phi^t_x}{i}$ holds and we have to prove that $\Gamma, e(\Sigma)\brmodels \co{\bigwedge\Sigma\imp\exists x\ \phi}{i}$.

Let $a$ be an assignment strictly admissible for $\Gamma,e(\Sigma)$ such that $\M \models \Gamma,e(\Sigma)[a]$. Then from the inductive hypothesis and the fact that $\exi{i}$ cannot be applied to formulae containing existential arrow variables (restriction R\ref{item:no-existential-arrow-variables}) we have that $\M\models \co{\bigwedge\Sigma\imp\phi^t_x}{i}[a]$, that is, $m \models \bigwedge\Sigma\imp\phi^t_x[a]$ for all $m \in \M_i$. But this easily imply $\M\models \co{\bigwedge\Sigma\imp\exists x \phi}{i}[a]$ from the definition of first\ftext{-}order satisfiability. 
\item[$\exe{ji}$] If $(\Gamma,\Sigma)\vdash_{\BR} \co{\psi}{j}$ and the last rule used is $\exe{ji}$, then there exist a formula $\co{\exists x\ phi}{i}$ and two pairs pairs $(\Gamma_i, \Sigma_i)$, $i=1,2$ with $\Gamma_i \subseteq \Gamma \cup \Sigma$  and $\Sigma_i \subseteq \Gamma \cup \Sigma$ such that 
\begin{align*}
	(\Gamma_1, \Sigma_1) & \vdash_{\IC} \co{\exists x\ \phi}{i},\\
	(\Gamma_2,\co{\phi}{i},\Sigma_2) & \vdash_{\IC} \co{\psi}{j}. 
\end{align*}
We prove that $\Gamma,e(\Sigma)\brmodels\co{\bigwedge\Sigma \imp \psi}{j}$ by considering two different cases:

\begin{description}
\item[$\co{\phi}{i}$ is a local assumption.] 
In this case $i=j$, $\Gamma = \Gamma_1 \cup \Gamma_2$, $\Sigma =  \Sigma_1 \cup \Sigma_2$, and from the inductive hypothesis we have:
\begin{align}
	\label{eq:exe-1.1}\Gamma_1, e(\Sigma_1) & \brmodels \co{\bigwedge \Sigma_1 \imp \exists x \ \phi)}{i}, \\
	\label{eq:exe-1.2}\Gamma_2, e(\Sigma_2 \cup \{\co{\phi}{i}\}) & \brmodels \co{(\bigwedge\Sigma_2 \con \phi) \imp \psi}{i}.
\end{align}

Let $a$ be an assignment strictly admissible for $\Gamma,e(\Sigma)$ such that $\M\models\Gamma,e(\Sigma)[a]$, and let $m$ be an arbitrary local model in $\M_i$. We have to prove that there is an extension $a'$ of $a$ for $\co{\psi}{i}$ such that if $m \models \sigma[a']$ for all $\sigma \in \Sigma$ then $m \models \psi[a']$. 

Let $a|_{\Gamma_1,e(\Sigma_1)}$ be the restriction of $a$ strictly admissible for $\Gamma_1,e(\Sigma_1)$. 
Then $\M  \models\Gamma_1,e(\Sigma_1)[a|_{\Gamma_1,e(\Sigma_1)}]$, and from the inductive hypothesis \eqref{eq:exe-1.1} $a|_{\Gamma_1,e(\Sigma_1)}$ can be extended to an assignment $\bar{a}|_{\Gamma_1,e(\Sigma_1)}$ admissible for $\co{\exists x\ \phi}{i}$ such that 
\begin{equation}
	\label{eq:exe-1.4}
	\M  \models \co{\Sigma_1 \imp \exists x\ \phi}{i}[\bar{a}|_{\Gamma_1,e(\Sigma_1)}].
\end{equation}
Let $a|_{\Gamma_2,e(\Sigma_2)}$ be the restriction of $a$ strictly admissible for $\Gamma_2,e(\Sigma_2)$. Then $\M  \models\Gamma_2,e(\Sigma_2)[a|_{\Gamma_2,e(\Sigma_2)}]$. Restriction R\ref{item:restriction-cut} imposes that the existential arrow variables occurring in $\co{\exists x\ \phi}{i}$ do not occur in any of the assumptions in $\Gamma_2$ and $\Sigma_2$. Thus we can extend $a|_{\Gamma_2,e(\Sigma_2)}$ to an assignment $\bar{a}|_{\Gamma_2,e(\Sigma_2)}$ admissible for $\co{\phi}{i}$ by adding to $a|_{\Gamma_2,e(\Sigma_2)}$ the assignment of the existential arrow variables of $\co{\exists x\ \phi}{i}$ according to $\bar{a}|_{\Gamma_1,e(\Sigma_1)}$.
Then it is easy to show that $\M \models\Gamma_2,e(\Sigma_2 \cup \{\co{\phi}{i}\})[\bar{a}|_{\Gamma_2,e(\Sigma_2)}]$, and from the inductive hypothesis \eqref{eq:exe-1.2} we can infer that $\bar{a}|_{\Gamma_2,e(\Sigma_2)}$ can be extended to an assignment $\doublebar{a}|_{\Gamma_2,e(\Sigma_2)}$ for $\co{\psi}{i}$ such that 
\begin{equation}
	\label{eq:exe-1.5}
	\M \models \co{(\bigwedge \Sigma_2 \con \phi) \imp \psi}{i}[\doublebar{a}|_{\Gamma_2,e(\Sigma_2)}].
\end{equation}
Let $m$ be a local model in $\M_i$ such that $m \models \sigma[a]$ for all $\sigma \in \Sigma$. Since $m$ satisfies all the formulae in $\Sigma_1$, then we can use \eqref{eq:exe-1.4} to obtain $m \models \exists x \phi[\bar{a}|_{\Gamma_1,e(\Sigma_1)}]$. Let $d=m(t)[a]$. Thus, $m \models \phi[\bar{a}|_{\Gamma_1,e(\Sigma_1)}(\co{x}{i}=d)]$. From the restriction R\ref{item:restriction-exe} we know that $x$ does not occur free in any of the assumptions of $\Gamma$ and $\Sigma$ and in $\co{\psi}{i}$. Therefore $\M \models\Gamma_2,e(\Sigma_2 \cup \{\co{\phi}{i}\})[\bar{a}|_{\Gamma_2,e(\Sigma_2)}(\co{x}{i}=d)]$, and we can use the inductive hypothesis as in \eqref{eq:exe-1.5} to obtain 
$\M \models \co{(\bigwedge \Sigma_2 \con \phi) \imp \psi}{i}[\doublebar{a}|_{\Gamma_2,e(\Sigma_2)}(\co{x}{i}=d)]$. $m\models \Sigma_2[\doublebar{a}|_{\Gamma_2,e(\Sigma_2)}(\co{x}{i}=d)]$, from the fact that it satisfies all the formulae in $\Sigma$ under the assignment $a$, and that $[\doublebar{a}|_{\Gamma_2,e(\Sigma_2)}(\co{x}{i}=d)]$ is built from a restriction of $a$. Analogously $m\models \phi[\doublebar{a}|_{\Gamma_2,e(\Sigma_2)}(\co{x}{i}=d)]$, by construction of $\doublebar{a}|_{\Gamma_2,e(\Sigma_2)}(\co{x}{i}=d)$. Thus $m\models \psi[\doublebar{a}|_{\Gamma_2,e(\Sigma_2)}(\co{x}{i}=d)]$, and the proof is done.
\end{description}

\item[$i:\phi$ is a global assumption.]
Here we distinguish two cases: $i=j$ and $i\neq j$.

If $i=j$, then $\Gamma = \Gamma_1 \cup \Gamma_2$, $\Sigma =  \Sigma_1 \cup \Sigma_2$, $\co{\phi}{i}$ is a complete formula (restriction R\ref{item:restr-disch-complete-global-assumptions}) and from the inductive hypothesis we have:
\begin{align}
	\label{eq:exe-2.1}\Gamma_1, e(\Sigma_1) & \brmodels \co{\bigwedge \Sigma_1 \imp \exists x \ \phi)}{i}, \\
	\label{eq:exe-2.2}\Gamma_2\cup \{\co{\phi}{i}\}, e(\Sigma_2) & \brmodels \co{\bigwedge\Sigma_2 \imp \psi}{i}.
\end{align}

Let $a$ be an assignment strictly admissible for $\Gamma,e(\Sigma)$ such that $\M\models\Gamma,e(\Sigma)[a]$, and let $m$ be an arbitrary local model in $\M_i$. We have to prove that there is an extension $a'$ of $a$ for $\co{\psi}{i}$ such that if $m \models \sigma[a']$ for all $\sigma \in \Sigma$ then $m \models \psi[a']$. 

Let $a|_{\Gamma_1,e(\Sigma_1)}$ be the restriction of $a$ strictly admissible for $\Gamma_1,e(\Sigma_1)$. 
Then $\M  \models\Gamma_1,e(\Sigma_1)[a|_{\Gamma_1,e(\Sigma_1)}]$, and from the inductive hypothesis \eqref{eq:exe-2.1} $a|_{\Gamma_1,e(\Sigma_1)}$ can be extended to an assignment $\bar{a}|_{\Gamma_1,e(\Sigma_1)}$ admissible for $\co{\exists x\ \phi}{i}$ such that 
\begin{equation}
	\label{eq:exe-2.4}
	\M  \models \co{\Sigma_1 \imp \exists x\ \phi}{i}[\bar{a}|_{\Gamma_1,e(\Sigma_1)}].
\end{equation}
Let $a|_{\Gamma_2,e(\Sigma_2)}$ be the restriction of $a$ strictly admissible for $\Gamma_2,e(\Sigma_2)$. Then $\M  \models\Gamma_2,e(\Sigma_2)[a|_{\Gamma_2,e(\Sigma_2)}]$. Restriction R\ref{item:restriction-cut} imposes that the existential arrow variables occurring in $\co{\exists x\ \phi}{i}$ do not occur in any of the assumptions in $\Gamma_2$ and $\Sigma_2$. Thus we can extend $a|_{\Gamma_2,e(\Sigma_2)}$ to an assignment $\bar{a}|_{\Gamma_2,e(\Sigma_2)}$ admissible for $\co{\phi}{i}$ by adding to $a|_{\Gamma_2,e(\Sigma_2)}$ the assignment of the existential arrow variables of $\co{\exists x\ \phi}{i}$ according to $\bar{a}|_{\Gamma_1,e(\Sigma_1)}$.
Then it is easy to show that if $\M \models \co{\phi}{i}[\bar{a}|_{\Gamma_2,e(\Sigma_2)}]$, then we can use the inductive hypothesis \eqref{eq:exe-2.2} to infer that $\bar{a}|_{\Gamma_2,e(\Sigma_2)}$ can be extended to an assignment $\doublebar{a}|_{\Gamma_2,e(\Sigma_2)}$ for $\co{\psi}{i}$ such that 
\begin{equation}
	\label{eq:exe-2.5}
	\M \models \co{\bigwedge \Sigma_2 \imp \psi}{i}[\doublebar{a}|_{\Gamma_2,e(\Sigma_2)}].
\end{equation}
Let $m$ be a local model in $\M_i$ such that $m \models \sigma[a]$ for all $\sigma \in \Sigma$. Since $m$ satisfies all the formulae in $\Sigma_1$, then we can use \eqref{eq:exe-2.4} to obtain $m \models \exists x \phi[\bar{a}|_{\Gamma_1,e(\Sigma_1)}]$. Let $d=m(t)[a]$. Thus, $m \models \phi[\bar{a}|_{\Gamma_1,e(\Sigma_1)}(\co{x}{i}=d)]$. From the restriction R\ref{item:restriction-exe} we know that $x$ does not occur free in any of the assumptions of $\Gamma$ and $\Sigma$ and in $\co{\psi}{i}$. Thus $\M \models\Gamma_2,e(\Sigma_2)[\bar{a}|_{\Gamma_2,e(\Sigma_2)}(\co{x}{i}=d)]$. 
From the way $\bar{a}|_{\Gamma_2,e(\Sigma_2)}(\co{x}{i}=d)$ is built, we know that $m \models \phi[\bar{a}|_{\Gamma_2,e(\Sigma_2)}(\co{x}{i}=d)]$, and from the fact that $\co{\phi}{i}$ is a complete formula we can obtain $\M \models \phi[\bar{a}|_{\Gamma_2,e(\Sigma_2)}(\co{x}{i}=d)]$. Thus we can repeat the \ftext{reasoning} steps to infer \eqref{eq:exe-2.5} to obtain $\M \models \co{\bigwedge \Sigma_2 \imp \psi}{i}[\doublebar{a}|_{\Gamma_2,e(\Sigma_2)}(\co{x}{i}=d)].$

$m\models \Sigma_2[\doublebar{a}|_{\Gamma_2,e(\Sigma_2)}(\co{x}{i}=d)]$, from the fact that it satisfies all the formulae in $\Sigma$ under the assignment $a$, and that $[\doublebar{a}|_{\Gamma_2,e(\Sigma_2)}(\co{x}{i}=d)]$ is built from a restriction of $a$. Thus $m\models \psi[\doublebar{a}|_{\Gamma_2,e(\Sigma_2)}(\co{x}{i}=d)]$, and this ends the proof.

If $i\neq j$, the proof can be obtained as in the previous case, just taking into account that $\Gamma = \Gamma_1 \cup \Sigma_1 \cup \Gamma_2$, and $\Sigma = \Sigma_2$ (that is, the local assumptions of $\Sigma_1$ become global due to the change of index from $i$ to $j$ triggered by $\exe{ji}$). 

\item[$\eqi{i}$] If $(\Gamma,\Sigma)\vdash_{\IC}\co{t=t}{i}$ and the last rule used is $\eqi{i}$, then from the inductive hypothesis there are $n$ formulae $\co{\phi_1}{i}, \ldots,\co{\phi_n}{i}$ such that $\Gamma_1, e(\Sigma_1)\brmodels \co{\bigwedge\Sigma_1\imp \phi_1}{i}, \ldots, \Gamma_n, e(\Sigma_n)\brmodels \co{\bigwedge\Sigma_2\imp \phi_n}{i}$, with $\Gamma =\bigcup_{1\leq k\leq n} \Gamma_k$ and $\Sigma =\bigcup_{1\leq k\leq n} \Sigma_k$. 

Let $a$ be an assignment strictly admissible for $\Gamma,e(\Sigma)$ such that $\M\models\Gamma,e(\Sigma)[a]$. 
Let $a|_{\Gamma_k,e(\Sigma_k)}$ be the restrictions of $a$ strictly admissible for $\Gamma_k,e(\Sigma_k)$. Since $\co{\phi_k}{i}$ cannot contain existential arrow variables, then $\M\models \co{\bigwedge\Sigma_k \imp\phi_k}{i}[a|_{\Gamma_k,e(\Sigma_k)}]$. Thus each $a|_{\Gamma_k,e(\Sigma_k)}$ is admissible for $\co{\phi_k}{i}$ for all $1\leq k \leq n$, and since $a=(a|_{\Gamma_1,e(\Sigma_1)}+,\ldots,+a|_{\Gamma_n,e(\Sigma_n)}$, we also have that $a|_{\Gamma_k,e(\Sigma_k)}$ is admissible for $\co{\phi_k}{i}$ for all $1\leq k \leq n$. 
Since $\eqi{i}$ cannot introduce new existential arrow variables (restriction R\ref{item:restr-introduce-arrow-variable}) $a$ is admissible also for $\co{t=t}{i}$. Thus $\M\models \co{t=t}{i}[a]$ from first\ftext{-}order satisfiability of $=$.
\item[$\eqe{i}$] If $(\Gamma,\Sigma)\vdash_{\IC}\co{\phi^u_x}{i}$ and the last rule used is $\eqi{i}$, then from the inductive hypothesis there are two formulae $\co{\phi^t_x}{i}$ and $\co{t=u}{i}$ such that $\Gamma_1, e(\Sigma_1)\brmodels \co{\bigwedge\Sigma_1\imp \phi^t_x}{i}$, $\Gamma_2, e(\Sigma_2)\brmodels \co{\bigwedge\Sigma_2\imp t=u}{i}$, $\Gamma =\Gamma_1 \cup \Gamma_2$ and $\Sigma =\Sigma_1 \cup \Sigma_2$. 

Let $a$ be an assignment strictly admissible for $\Gamma,e(\Sigma)$ such that $\M\models\Gamma,e(\Sigma)[a]$. Since $\eqe{i}$ cannot introduce new arrow variables (restriction R\ref{item:restr-introduce-arrow-variable}), then we have to prove that $\M\models \co{\bigwedge\Sigma\imp \phi^u_x}{i}[a]$.
Let $a|_{\Gamma_k,e(\Sigma_k)}$ be the restrictions of $a$ strictly admissible for $\Gamma_k,e(\Sigma_k)$, $k=1,2$. Then $\M\models\Gamma_1,e(\Sigma_1)[a|_{\Gamma_1,e(\Sigma_1)}]$ and $\M\models\Gamma_2,e(\Sigma_2)[a|_{\Gamma_2,e(\Sigma_2)}]$.
From the inductive hypothesis and restriction R\ref{item:no-existential-arrow-variables} we can obtain that $\M \models \co{\bigwedge\Sigma_1\imp \phi^t_x}{i}[a|_{\Gamma_1,e(\Sigma_1)}]$ and $\M \models \co{\bigwedge\Sigma_2\imp t=u}{i}[a|_{\Gamma_2,e(\Sigma_2)}]$.
Let $m \in M_i$ be a local model which satisfies $\Sigma$ under the assignment $a$. Using the inductive hypothesis and the fact that $a= a|_{\Gamma_1,e(\Sigma_1)}+a|_{\Gamma_2,e(\Sigma_2)}$ we can obtain that $m \models \phi^t_x[a]$ and $m\models t=u[a]$. Then $m \models \phi^u_x[a]$ from the definition of first\ftext{-}order satisfiability.   
}
\item[$\fromI{i}I_{ij}$] If $(\Gamma,\Sigma)\vdash_{\IC}\co{\fromvar{x}{i}=y}{j}$ and the last rule used is $\fromI{i}I_{ij}$, then $i \neq j$, and  
$\Sigma = \emptyset$\footnote{Remember that the application of a $b$-rule makes all the local assumption become global.}. Thus, we have to show that $\Gamma \brmodels \co{\fromvar{x}{i}=y}{j}$.

From the shape of the $\fromI{i}I_{ij}$ rule we know that there is a formula $\co{x=\tovar{y}{j}}{i}$ and two sets $\Gamma_1, \Sigma_1$ with $\Gamma = \Gamma_1 \cup \Sigma_1$ such that $\Gamma_1, \Sigma_1 \vdash_{\IC}\co{x=\tovar{y}{j}}{i}$. Thus from the inductive hypothesis we know that $\Gamma_1, e(\Sigma_1)\brmodels \co{\bigwedge\Sigma_1\imp (x=\tovar{y}{j})}{i}$. 

Let $a$ be an assignment strictly admissible for $\Gamma$ such that $\M\models\Gamma[a]$. We have to prove that there is an extension $a'$ of $a$ for $\co{\fromvar{x}{i}=y}{j}$ such that $\M \models \co{\fromvar{x}{i}=y}{j}$. From $\M\models\Gamma[a]$ we have that $\M\models\Gamma_1, e(\Sigma_1)[a]$, and from the inductive hypothesis and the fact that $\fromI{i}I_{ij}$ cannot be applied to formulae containing existential arrow variables (restriction R\ref{item:no-existential-arrow-variables}) we also have that $\M\co{\bigwedge\Sigma_1\imp (x=\tovar{y}{j})}\models[a]$. Again from $\M\models\Gamma[a]$ we have that $\M\models\co{x=\tovar{y}{j}}{i}[a]$. This implies that $a_i(x)=a_i(\tovar{y}{j})$. From restriction R\ref{item:restr-introduce-arrow-variable} we know that 
$\fromvar{x}{i}$ is an existential variable in $\co{\fromvar{x}{i}=y}{j}$. Thus $a_j(\fromvar{x}{i})$ is undefined, and we can obtain a new assignment $a'$ by adding to $a$ the value $a'_j(\fromvar{x}{i}) = a_j(y)$. We have to show that $(a'_i(x),a'_j(\fromvar{x}{i})) \in \dr_{ij}$. 
This follows from the fact that $(a'_i(x),a'_j(\fromvar{x}{i})) = (a_i(x)=a_j(y)) = (a_i(\tovar{y}{j}),a_j(y))$, and from the fact that $(a_i(\tovar{y}{j}),a_j(y)) \in \dr_{ij}$ because of the fact that $a$ is an assignment (see Definition~\ref{def:assignment}). Since $a'_j(\fromvar{x}{i}) = a'_j(y)$ we have that $m\models\co{\fromvar{x}{i}=y}{j}[a']$ and this ends the proof. 
\onlyintechrep{\item[$\toI{i}I_{ij}$] The proof is analogous to the one of $\fromI{i}I_{ij}$.}
\item[\BR:] If $(\Gamma,\Sigma)\vdash_{\IC}\co{\phi}{i}$ and the last rule used is $\BR$, then $\Sigma = \emptyset$ because the application of a $b$-rule makes all the local assumptions become global. 
Thus we have to prove that $\Gamma \brmodels\co{\phi}{i}$. From the inductive hypothesis we know that there are $n$ formulae $\co{\phi_1}{i_1}, \ldots,\co{\phi_n}{i_n}$ such that $\Gamma_1, e(\Sigma_1)\brmodels \co{\bigwedge\Sigma_1\imp \phi_1}{i_1}, \ldots, \Gamma_n, e(\Sigma_n)\brmodels \co{\bigwedge\Sigma_n\imp \phi_n}{i_n}$ with $\Gamma_k \subseteq \Gamma$ and $\Sigma_k \subseteq \Gamma$ for all $1\leq k\leq n$. 

Let $a$ be an assignment strictly admissible for  $\Gamma$ such that $\M\models\Gamma[a]$. 
Since $\Sigma_k \subseteq \Gamma$, then $a$ is admissible for all the variables in $\Sigma_k$, for all $1\leq k\leq n$.  
Let $a|_{\Gamma_k,e(\Sigma_k)}$ be the restrictions of $a$ strictly admissible for $\Gamma_k,e(\Sigma_k)$. Then,  $\M\models\Gamma_k,e(\Sigma_k)[a|_{\Gamma_k,e(\Sigma_k)}]$ holds. From the restriction R\ref{item:no-existential-arrow-variables} which states that each $\co{\phi_k}{i_k}$ cannot contain existential arrow variables, and the inductive hypothesis, we infer that $\M\models\co{\bigwedge\Sigma_k\imp \phi_k}{i_k}[a|_{\Gamma_k,e(\Sigma_k)}]$. 
Since $\M\models \Gamma$ and $\Sigma_k \subseteq \Gamma$ we have that $\M\models \co{\phi_k}{i_k}$ for all $1\leq k\leq n$.
From the definition of satisfiability of a bridge rule we know that $a$ can be extended to an assignment $a'$ such that $\M \models\co{\phi}{i}[a']$, and this ends the proof. 
\item[$\cut{ji}$] If $(\Gamma,\Sigma)\vdash_{\BR} \co{\psi}{i}$ and the last rule used is $\cut{ji}$, then there exists a formula $\co{\phi}{j}$ and two pairs $(\Gamma_k, \Sigma_k)$, $k=1,2$ with $\Gamma_k\subseteq \Gamma \cup \Sigma$  and $\Sigma_k \subseteq \Gamma \cup \Sigma$ such that 
\begin{align*}
	(\Gamma_1, \Sigma_1) & \vdash_{\IC} \co{\phi}{j},\\
	(\Gamma_2,\co{\phi}{j},\Sigma_2) & \vdash_{\IC} \co{\psi}{i}. 
\end{align*}
We prove that $\Gamma,e(\Sigma)\brmodels\co{\bigwedge\Sigma \imp \psi}{i}$ by considering two different cases:

\begin{description}
\item[$\co{\phi}{j}$ is a local assumption.] 
In this case $i=j$, $\Gamma = \Gamma_1 \cup \Gamma_2$, $\Sigma =  \Sigma_1 \cup \Sigma_2$, and from the inductive hypothesis we have:
\begin{align}
	\label{eq:cut-1.1}\Gamma_1, e(\Sigma_1) & \brmodels \co{\bigwedge \Sigma_1 \imp \phi}{i}, \\
	\label{eq:cut-1.2}\Gamma_2, e(\Sigma_2 \cup \{\co{\phi}{i}\}) & \brmodels \co{(\bigwedge\Sigma_2 \con \phi) \imp \psi}{i}.
\end{align}

Let $a$ be an assignment strictly admissible for $\Gamma,e(\Sigma)$ such that $\M\models\Gamma,e(\Sigma)[a]$, and let $m$ be an arbitrary local model in $\M_i$. We have to prove that there is an extension $a'$ of $a$ for $\co{\psi}{i}$ such that if $m \models \sigma[a']$ for all $\sigma \in \Sigma$ then $m \models \psi[a']$. 

Let $a|_{\Gamma_1,e(\Sigma_1)}$ be the restriction of $a$ strictly admissible for $\Gamma_1,e(\Sigma_1)$. 
Then $\M  \models\Gamma_1,e(\Sigma_1)[a|_{\Gamma_1,e(\Sigma_1)}]$, and from the inductive hypothesis \eqref{eq:cut-1.1} $a|_{\Gamma_1,e(\Sigma_1)}$ can be extended to an assignment $\bar{a}|_{\Gamma_1,e(\Sigma_1)}$ admissible for $\co{\phi}{i}$ such that 
\begin{equation}
	\label{eq:cut-1.4}
	\M  \models \co{\Sigma_1 \imp \phi}{i}[\bar{a}|_{\Gamma_1,e(\Sigma_1)}].
\end{equation}
Let $a|_{\Gamma_2,e(\Sigma_2)}$ be the restriction of $a$ strictly admissible for $\Gamma_2,e(\Sigma_2)$. Then $\M  \models\Gamma_2,e(\Sigma_2)[a|_{\Gamma_2,e(\Sigma_2)}]$. Restriction R\ref{item:restriction-cut} imposes that the existential arrow variables occurring in $\co{\phi}{i}$ do not occur in any of the assumptions in $\Gamma_2$ and $\Sigma_2$. Thus we can extend $a|_{\Gamma_2,e(\Sigma_2)}$ to an assignment $\bar{a}|_{\Gamma_2,e(\Sigma_2)}$ admissible for $\co{\phi}{i}$ by adding to $a|_{\Gamma_2,e(\Sigma_2)}$ the assignment of the existential arrow variables of $\co{\phi}{i}$ according to $\bar{a}|_{\Gamma_1,e(\Sigma_1)}$.
Then it is easy to show that $\M \models\Gamma_2,e(\Sigma_2 \cup \{\co{\phi}{i}\})[\bar{a}|_{\Gamma_2,e(\Sigma_2)}]$, and from the inductive hypothesis \eqref{eq:cut-1.2} we can infer that $\bar{a}|_{\Gamma_2,e(\Sigma_2)}$ can be extended to an assignment $\doublebar{a}|_{\Gamma_2,e(\Sigma_2)}$ for $\co{\psi}{i}$ such that 
\begin{equation}
	\label{eq:cut-1.5}
	\M \models \co{(\bigwedge \Sigma_2 \con \phi) \imp \psi}{i}[\doublebar{a}|_{\Gamma_2,e(\Sigma_2)}].
\end{equation}
Let $m$ be a local model in $\M_i$ such that $m \models \sigma[a]$ for all $\sigma \in \Sigma$. Since $m$ satisfies all the formulae in $\Sigma_1$, then we can use \eqref{eq:cut-1.4} to obtain $m \models \phi[\bar{a}|_{\Gamma_1,e(\Sigma_1)}]$. By construction we also have that $m \models \phi[\doublebar{a}|_{\Gamma_2,e(\Sigma_2)}]$ and $m \models \Sigma_2[\doublebar{a}|_{\Gamma_2,e(\Sigma_2)}]$. Thus $m \models \psi[\doublebar{a}|_{\Gamma_2,e(\Sigma_2)}]$ and this ends the proof.

\item[$\co{\phi}{j}$ is a global assumption.]
Here we distinguish two cases: $i=j$ and $i\neq j$.

If $i=j$, then $\Gamma = \Gamma_1 \cup \Gamma_2$, $\Sigma = \Sigma_1 \cup \Sigma_2$, $\co{\phi}{i}$ is a complete formula (restriction R\ref{item:restr-disch-complete-global-assumptions}) and from the inductive hypothesis we have:
\begin{align}
	\label{eq:cut-2.1}\Gamma_1, e(\Sigma_1) & \brmodels \co{\bigwedge \Sigma_1 \imp \phi}{i}, \\
	\label{eq:cut-2.2}\Gamma_2\cup \{\co{\phi}{i}\}, e(\Sigma_2) & \brmodels \co{\bigwedge\Sigma_2 \imp \psi}{i}.
\end{align}

Let $a$ be an assignment strictly admissible for $\Gamma,e(\Sigma)$ such that $\M\models\Gamma,e(\Sigma)[a]$, and let $m$ be an arbitrary local model in $M_i$. We have to prove that there is an extension $a'$ of $a$ for $\co{\psi}{i}$ such that if $m \models \sigma[a']$ for all $\sigma \in \Sigma$ then $m \models \psi[a']$. 

Let $a|_{\Gamma_1,e(\Sigma_1)}$ be the restriction of $a$ strictly admissible for $\Gamma_1,e(\Sigma_1)$. 
Then $\M  \models\Gamma_1,e(\Sigma_1)[a|_{\Gamma_1,e(\Sigma_1)}]$, and from the inductive hypothesis \eqref{eq:cut-2.1} $a|_{\Gamma_1,e(\Sigma_1)}$ can be extended to an assignment $\bar{a}|_{\Gamma_1,e(\Sigma_1)}$ admissible for $\co{\phi}{i}$ such that 
\begin{equation}
	\label{eq:cut-2.4}
	\M  \models \co{\Sigma_1 \imp \phi}{i}[\bar{a}|_{\Gamma_1,e(\Sigma_1)}].
\end{equation}
Let $a|_{\Gamma_2,e(\Sigma_2)}$ be the restriction of $a$ strictly admissible for $\Gamma_2,e(\Sigma_2)$. Then $\M  \models\Gamma_2,e(\Sigma_2)[a|_{\Gamma_2,e(\Sigma_2)}]$. Restriction R\ref{item:restriction-cut} imposes that the existential arrow variables occurring in $\co{\exists x\ \phi}{i}$ do not occur in any of the assumptions in $\Gamma_2$ and $\Sigma_2$. Thus we can extend $a|_{\Gamma_2,e(\Sigma_2)}$ to an assignment $\bar{a}|_{\Gamma_2,e(\Sigma_2)}$ admissible for $\co{\phi}{i}$ by adding to $a|_{\Gamma_2,e(\Sigma_2)}$ the assignment of the existential arrow variables of $\co{\phi}{i}$ according to $\bar{a}|_{\Gamma_1,e(\Sigma_1)}$.
Then it is easy to show that if $\M \models \co{\phi}{i}[\bar{a}|_{\Gamma_2,e(\Sigma_2)}]$, then we can use the inductive hypothesis \eqref{eq:cut-2.2} to infer that $\bar{a}|_{\Gamma_2,e(\Sigma_2)}$ can be extended to an assignment $\doublebar{a}|_{\Gamma_2,e(\Sigma_2)}$ for $\co{\psi}{i}$ such that 
\begin{equation}
	\label{eq:cut-2.5}
	\M \models \co{\bigwedge \Sigma_2 \imp \psi}{i}[\doublebar{a}|_{\Gamma_2,e(\Sigma_2)}].
\end{equation}
Let $m$ be a local model in $\M_i$ such that $m \models \sigma[a]$ for all $\sigma \in \Sigma$. Since $m$ satisfies all the formulae in $\Sigma_1$, then we can use \eqref{eq:cut-2.4} to obtain $m \models \phi[\bar{a}|_{\Gamma_1,e(\Sigma_1)}]$. Since $\phi$ is a complete formula, we have that $\M \models\co{\phi}{i}[\bar{a}|_{\Gamma_1,e(\Sigma_1)}]$. 
From the way $\bar{a}|_{\Gamma_2,e(\Sigma_2)}(\co{x}{i}=d)$ is built, we know that $\M \models\co{\phi}{i}[\bar{a}|_{\Gamma_1,e(\Sigma_1)}]$. Thus we can apply the inductive hypothesis to infer \eqref{eq:cut-2.5} to obtain $\M \models \co{\bigwedge \Sigma_2 \imp \psi}{i}[\doublebar{a}|_{\Gamma_2,e(\Sigma_2)}(\co{x}{i}=d)].$

$m\models \Sigma_2[\doublebar{a}|_{\Gamma_2,e(\Sigma_2)}(\co{x}{i}=d)]$, from the fact that it satisfies all the formulae in $\Sigma$ under the assignment $a$, and that $[\doublebar{a}|_{\Gamma_2,e(\Sigma_2)}(\co{x}{i}=d)]$ is built from a restriction of $a$. Thus $m\models \psi[\doublebar{a}|_{\Gamma_2,e(\Sigma_2)}(\co{x}{i}=d)]$, and this ends the proof.

If $i\neq j$, the proof can be obtained as in the previous case, just taking into account that $\Gamma = \Gamma_1 \cup \Sigma_1 \cup \Gamma_2$, and $\Sigma = \Sigma_2$ (that is, the local assumptions of $\Sigma_1$ become global due to the change of index from $j$ to $i$ triggered by $\cut{ji}$).
\end{description}
\end{description}

\end{pot}

%% file: completeness.tex

\section{Proof of the Completeness Theorem}
\label{sec:completeness}

\begin{thm}[Completeness]
\label{thm-completeness}
$
\Gamma,e(\Sigma)\models_{\IC} \co{\bigwedge_{i:\sigma \in \Sigma}\sigma\imp\phi}{i} 
\Longrightarrow
(\Gamma,\Sigma)\vdash_{\IC}\co{\phi}{i} 
$
\end{thm}

The contrapositive will be proved: it will be shown that if $(\Gamma,\Sigma)\not \vdash_{\IC}\co{\phi}{i}$ then there exists a \IC-model $\M^c$ and an assignment $a$ such that $\M \models \Gamma[a]$, $\M \models e(\Sigma)[a]$ but $\M \not \models \co{\bigwedge_{i:\sigma \in \Sigma}\sigma\imp\phi}{i}[a]$. 
The technique we use is based on the construction of canonical model $\Mc$ using the method of models constructed from constants originally due to Henkin~\cite{Henkin:completeness:1949} (see also \cite{chang5}). This method is based on the ability of constructing a canonical models $\M^c$ starting from maximal consistent sets of formulae and an appropriate set of constants (or existential witnesses). The situation in DFOL is more complex than the one of FOL due to the following three reasons: first, the presence of sets of formulas belonging to different languages; second, the presence of partial knowledge; and finally, the presence of arrow variables. 

The generalisation of the Henkin technique to the case of DFOL is composed by the following steps:

\begin{enumerate}
	\item We \ftext{generalise} the notion of consistency to $k$-consistency (Definitions \ref{def:k-consistecy});
	\item We introduce the operators $cl$ and $\otimes_i$ (Definitions \ref{def:closure} and \ref{def:product}) to be able to deal with sets of sets of formulas and we show some relevant properties of these operators (Lemma~\ref{lem:prop-oplus});
	\item We modify the Henkin technique to extend a consistent set $\Gamma$ of DFOL formulas to a set with existential witnesses (Lemma~\ref{lem:consistency-of-Gamma}); 
	\item We introduce the notion of $k$-saturated set of formulas (Definition \ref{def:saturated}) and show how to saturate a $k$-consistent set of formulas (Lemma \ref{lem:saturation-of-Gamma});
	\item We define the canonical model $\Mc$ as a compatibility relation over sets of (local) models satisfying maximal-k-consistent sets of formulas (Definition \ref{def:canonical-model});
	\item We show that $\Mc$ is a $\IC$-model (Lemma \ref{lem:model-construction}).
\end{enumerate}

As already observed in \cite{ghidini10}, the first step in proving completeness for logical systems whose formulas are scattered among different languages is the introduction of   specific notions of consistency and maximal consistency, which \ftext{generalise} the analogous concepts given in \cite{chang5}.	
\begin{defn}[$k$-consistency]
	\label{def:k-consistecy}
$\Gamma$ is \emph{$k$-consistent} if\/ $\Gamma
\not\vdash_{\IC}\co{\bot}{k}$. 
\end{defn}

This generalisation is needed as a set of DFOL formulas $\Gamma$ can be locally inconsistent with respect to an index $i$ without being inconsistent with respect to some different index $j$, that is, $\Gamma\vdash_{\IC}\co{\bot}{i}$ but $\Gamma\not\vdash_{\IC}\co{\bot}{j}$. Thus, we have to prove that every $k$-consistent set of formulas $\Gamma$ of DFOL has a (canonical) model, which associates a non empty set $M_k$ of local models to the language $L_k$.  


The second step in proving completeness for DFOL is to be able to work with sets of labelled formulas. In fact the scattering of the system in different languages implies that we have to build a canonical model structured as a family $\M_0, \M_1, \ldots$ of sets of interrelated (local) models spanning over different languages.   
From now on we use $\Gamma_i$ to denote a set of formulas in $L_i$, and $\GGamma_i$ to denote a (finite or infinite) set \{$\Gamma^0_i, \Gamma^1_i, \ldots, \Gamma^n_i, \ldots$\} of sets of formulas in $L_i$. We instead use $\GGamma$ to denote a family $\{\GGamma_0, \GGamma_1, \ldots\}$ of sets $\GGamma_i$ of sets of formulas, one for each $i$ in $I$. 
The \emph{closure} of a set $\Gamma_i$ of formulas in $L_i$, denoted as $cl(\Gamma_i)$, is defined as the set of formulas derivable from $\Gamma_i$ using the  deduction rules of first\ftext{-}order logic in the context $i$, and considering the arrow variables occurring in $\Gamma_i$ as constants. The \emph{closure} of a set $\GGamma$ is instead defined as the set containing all (and only) the formulas $\co{\phi}{i}$ that belong to the deductive closure $cl(\Gamma_i)$ of all $\Gamma_i \in \GGamma_i$, for all $i \in I$. Formally,

\begin{defn}[Closure of $\GGamma$]
	\label{def:closure}
Let $\GGamma = \{\GGamma_0, \GGamma_1, \ldots \}$ be a family of sets of sets of formulas in $\{L_0, L_1, \ldots\}$. The \emph{closure} of $\GGamma$, in symbols $cl(\GGamma)$, is defined as:
$$
cl(\GGamma)=\bigcup_{i\in I}
\left\{\co{\phi}{i} \; \left| \; \phi\in
    \left(\bigcap_{\Gamma\in\GGamma_i}cl(\Gamma)\right)\right.\right\} 
$$
\end{defn}
We say that $\co{\phi}{i}$ is \emph{derivable} from $\GGamma$ iff $\co{\phi}{i}$ is derivable from $cl(\GGamma)$. Analogously, we say that $\GGamma$ is \emph{$k$-consistent}, iff
$cl(\GGamma)$ is $k$-consistent. We also write $\GGamma\vdash_{\IC}\co{\phi}{i}$ as a shorthand for $cl(\GGamma)\vdash_{\IC}\co{\phi}{i}$. 

	Note that the computation of $cl(\GGamma)$ does not take into account any $b$-rule in $\MC{\IC}$ but only the (local) $i$-rules for first\ftext{-}order logic with equality of the
	different knowledge sources. Instead the deduction of formulas from $cl(\GGamma)$ is performed taking into account the entire deductive calculus of $\MC{\IC}$.

Given a set of arrow variables $AV = \bigcup_{i \in I}AV_i$, 
we define the \emph{closure of $\GGamma$ w.r.t. AV} as before but considering the arrow variables in $AV$ like regular first\ftext{-}order terms. We define the notions of derivability from $\GGamma$ w.r.t. $AV$ and of $k$-consistency of $\GGamma$ w.r.t. $AV$, accordingly. 

We illustrate the difference between the closure of $\GGamma$ and the closure of $\GGamma$ w.r.t. $AV$ by means of an example. 
\begin{exmp}
	Let $\GGamma$ be a set such that each $\Gamma_i \in \GGamma_i$ satisfies the fact that a deduction $\Gamma_i, x = \bothvar{z} \der{} \psi$ involving only formulas in $i$ exists. If there is no other way of proving $\psi$ from all $\Gamma_i$, then $\psi \not \in cl(\GGamma)$. This holds because an additional formula $x = \bothvar{z}$ outside $\Gamma_i$ is used to infer it. 
	If $\bothvar{z} \in AV$ and we treat $\bothvar{z}$ as a first\ftext{-}order regular term, then $\exists x. x=\bothvar{z}$ becomes a regular first\ftext{-}order (valid) formula and therefore we can apply the $\exists$ elimination rule to obtain a deduction of $\psi$ from $\Gamma_i$ (for all $i$) as follows:
	$$
	\infer[\exe{ii}]
	{\psi}
	{\deduce[\Pi]
	{\psi}
	{\Gamma_i, [x = \bothvar{z}]} & \exists x. x=\bothvar{z}}
	$$
	Thus, while $\psi$ does not belong to the closure of $\GGamma$ it belongs to the closure of $\GGamma$ w.r.t. $AV$.
\end{exmp}

To be able to generalise the Henkin technique to DFOL, where we need to manipulate sets of sets of formulas $\GGamma$, we introduce an operator $\otimes_i$, that we use to perform a kind of cross product between sets of formulas.  

\begin{defn}[$\otimes_i$]
	\label{def:product}
Let $\Sigma_i$ be a set of formulas in $L_i$. $\GGamma' =
\GGamma\otimes_i\Sigma_i$ is defined as:   
$$
\GGamma' = \left \{
\begin{array}{l}
\GGamma'_{j}=\GGamma_{j} \mbox{ for all } j\neq i\\[.3cm]
\GGamma'_i= \{\Gamma\cup\{\sigma\} \; | \; \Gamma\in\GGamma_i,
\sigma\in\Sigma_i\} 
\end{array}
\right.
$$
\end{defn}

\begin{exmp}
	Let $\GGamma = \{\GGamma_1, \GGamma_2\}$ where:
	\begin{itemize}
	\item $\GGamma_1 = \{\{\co{A}{1}\}\}$
	\item $\GGamma_2 = \{\{\co{B}{2}, \co{C}{2}\}, \{\co{\neg C}{2}\}\}$
	\end{itemize}
	and let $\Sigma_2 = \{\co{D}{2}, \co{E}{2}\}$. The set $\GGamma' =
	\GGamma\otimes_2\Sigma_2$ is built as follows:
	\begin{itemize}
	\item $\GGamma'_1 = \GGamma_1 = \{\{\co{A}{1}\}\}$
	\item $\GGamma'_2 = \{\{\co{B}{2}, \co{C}{2}, \co{D}{2}\}, \{\co{\neg
	    C}{2}, \co{D}{2}\}, \{\co{B}{2}, \co{C}{2}, \co{E}{2}\},
	  \{\co{\neg C}{2}, \co{E}{2}\}\}$
	\end{itemize}
\end{exmp}


\begin{lem}
\label{lem:prop-oplus}
Let $\{\phi_1,\ldots,\phi_n\}$ be a set of formulas of $L_i$. 
$\GGamma\otimes_i\{\phi_1,\ldots,\phi_n\}\vdash_{\IC}\co{\psi}{j}$ if and only if
$cl(\GGamma),\co{\phi_1\vee\ldots\vee\phi_n}{i}\vdash_{\IC}\co{\psi}{j}$. 
\end{lem}

\begin{pf}
	From the notion of derivability from $\GGamma$ the statement of the theorem can be reformulated in $cl(\GGamma\otimes_i\{\phi_1,\ldots, \phi_n\})\vdash_{\IC}\co{\psi}{j}$ if and only if $cl(\GGamma),\co{\phi_1\vee\ldots\vee\phi_n}{i}\vdash_{\IC}\co{\psi}{j}$. In the proof we use $\Phi$ to denote the set of $L_i$-formulas $\{\phi_1,\ldots, \phi_n\}$.

\gap
\noindent
\textsc{``If'' direction}. Assume that $cl(\GGamma),\co{\phi_1\vee\ldots\vee\phi_n}{i}\vdash_{\IC}\co{\psi}{j}$ with a proof
$\Pi$, and let $\co{\psi_1}{j_1},\ldots,\co{\psi_m}{j_m}$ be the undischarged assumptions of $\Pi$. We show that all the undischarged assumptions of $\Pi$ can be derived from $cl(\GGamma\otimes_i\{\phi_1,\ldots, \phi_n\})$. 

For each undischarged assumption  $\co{\psi_k}{j_k}$, $k=1,\ldots, m$, we have to consider two cases: (a) $\co{\psi_k}{i_k} \in cl(\GGamma)$, and (b) $\co{\psi_k}{i_k} = \co{\phi_1\vee\ldots\vee\phi_n}{i}$: 
\begin{enumerate}[(a)]
\item If $\co{\psi_k}{i_k} \in cl(\GGamma)$ then we have to consider two separate cases: if $i_k \neq i$ then $\co{\psi_k}{i_k} \in cl(\GGamma)\otimes_i \Phi$ since $cl(\GGamma)$ and $cl(\GGamma\otimes_i\Phi)$ contain the same set of formulas with index $j \neq i$; if $i_k = i$ it is easy to see
  from the definition of the operator $cl(.)$ that 
  \begin{equation*}
    \text{if } \psi_k\in\bigcap_{\Gamma_{i}\in\GGamma_{i}}cl(\Gamma_{i}), \text{ then }
        \psi_k\in\bigcap_{\Gamma_{i}\in\GGamma_{i}} \bigcap_{\phi_l \in \Phi}
    cl(\Gamma_{i}\cup\phi_l)
   \end{equation*}
	This is true because derivability in FOL (and therefore the FOL deductive closure of a set of $L_i$-formulas) is monotone. 
\item If $\co{\psi_k}{i_k} = \co{\phi_1\vee\ldots\vee\phi_n}{i}$, then we can easily see that $\co{\phi_1\vee\ldots\vee\phi_n}{i}$ is
  derivable, via $\ori{i}$, from all the sets $\Gamma_{i}\cup\phi_l$ (with $\Gamma_{i}\in\GGamma_{i}$ and $\phi_l \in \Phi$). Therefore, from the definition of $cl(\GGamma)$,  $\co{\phi_1\vee\ldots\vee\phi_n}{i} \in cl(\GGamma\otimes_i \Phi)$.
\end{enumerate}
Since all the undischarged assumptions $\co{\psi_1}{j_1},\ldots,\co{\psi_m}{j_m}$ of $\Pi$ are derivable from the set $cl(\GGamma \otimes_i \Phi)$ we can easily extend $\Pi$ to build a deduction of $\co{\psi}{i}$ from $\GGamma \otimes_i \Phi$.

\gap
\noindent
\textsc{``Only if'' direction}. Assume that $\GGamma\otimes_i\Phi\vdash_{\IC}\co{\psi}{j}$ with a proof $\Pi$, and let $\co{\psi_1}{j_1},\ldots,\co{\psi_m}{j_m}$ be the undischarged assumptions of $\Pi$. Then all $\co{\psi_1}{j_1},\ldots,\co{\psi_m}{j_m}$ belong to $cl(\GGamma\otimes_i \Phi)$ and we show that they can be derived from $cl(\GGamma),\co{\phi_1\vee\ldots\vee\phi_n}{i}$.

For each undischarged assumption  $\co{\psi_k}{j_k}$, $k=1,\ldots, m$, we have to consider two cases: (a) $i_k\neq i$ and (b) $i_k=i$. 
\begin{enumerate}[(a)]
\item Assume that $i_k\neq i$. From the definition of $\otimes_i$ and the fact that it only modifies the set $\GGamma_i$,
  and from the definition of closure ($cl(.)$) over a set $\Gamma_i$ of FOL formulas we can easily see that 
\begin{equation*}
	\co{\psi_k}{i_k} \in cl(\GGamma\otimes_i\Phi) \text{ if and only if } \psi_k\in\bigcap_{\Gamma_{i_k}\in\GGamma_{i_k}}cl(\Gamma_{i_k})
\end{equation*}	
Thus, since $\co{\psi_k}{i_k} \in cl(\GGamma\otimes_i\Phi)$, we can prove that $\psi_k\in cl(\GGamma)$. 

\item Assume that $i_k=i$. From the definition of $\otimes_i$ and $cl(.)$ we have that 
\begin{equation*}
	\co{\psi_k}{i_k} \in cl(\GGamma\otimes_i\Phi) \text{ if and only if }
	  \psi_k\in\bigcap_{\Gamma_{i}\in\GGamma_{i}} \bigcap_{\phi_l \in \Phi}cl(\Gamma_{i}\cup\phi_l)
\end{equation*}

This implies that for all $\Gamma_i\in\GGamma_i$, $\psi_k$ is derivable from all $\Gamma_i\cup\phi_l$ for all $\phi_l \in \Phi$ with a \ftext{f}irst\ftext{-}order deduction. We can therefore apply $\ore{ii}$ to obtain that $\co{\psi_k}{i_k}$ is derivable from $\Gamma_i,\co{\phi_1\vee\ldots\vee\phi_n}{i}$, for all $\Gamma_i\in\GGamma_i$. Note that Restriction R\ref{item:restriction-cut} is satisfied as $\co{\phi_1\vee\ldots\vee\phi_n}{i}$ does not contain existential arrow variables (as it does not depend upon any assumption), and Restriction R\ref{item:restr-disch-complete-global-assumptions} is satisfied as the derivation of $\psi_k$ from each of the $\Gamma_i\cup\phi_l$ is local (and therefore we can apply $\ore{ii}$ and discharge the different $\phi_l$). Therefore we can easily build a deduction $\Pi_k$ of $\psi_k$ from $cl(\GGamma),\co{\phi_1\vee\ldots\vee\phi_n}{i}$.
\end{enumerate}

Since all the undischarged assumptions $\co{\psi_1}{j_1},\ldots,\co{\psi_m}{j_m}$ of $\Pi$ are derivable from
the set $cl(\GGamma),\co{\phi_1\vee\ldots\vee\phi_n}{i}$ we can build a deduction of $\co{\psi}{i}$ from $cl(\GGamma),\co{\phi_1\vee\ldots\vee\phi_n}{i}$ and this ends the proof.
\end{pf}

The proof of completeness proceeds with the construction of the canonical model $\Mc$ using the method of models constructed from constants (see \cite{Henkin:completeness:1949, chang5}). Roughly speaking the Henkin approach is based on two \ftext{fundamental} ideas: first, for each existential sentence $\exists x \ \phi(x) $ in the language, one adds a new constant $c$, the so-called \emph{existential witness}, to the language and inserts an axiom $\exists x \ \phi(x) \imp \phi(c)$ to the theory; second, the canonical model is built using equivalence classes of existential witnesses as domain elements.  
Differently from the classical proof for \ftext{f}irst\ftext{-}\ftext{o}rder \ftext{l}ogic, the presence of partial knowledge implies that a DFOL formula $\co{\exists x \phi(x)}{i}$ does not necessarily entail a formula $\co{\phi(c)}{i}$ for some $c$ (unless $\phi$ is a complete formula). This fact forces us to modify the original technique of existential witnesses as follows: we start from a 
$k$-consistent set of formulas $\Gamma$ and examine all the formulas $\co{\phi(x)}{j}$ in $\{L_i\}$ with one free variable; if $\co{\exists x \phi(x)}{j}$ is a complete formula then we add the existential witness $\co{\phi(c)}{j}$ as usual, while if it is a ``regular'' (that is, non-complete) formula we add an infinite set of witnesses $\co{\phi(c_1)}{j}, \ldots, \co{\phi(c_n)}{j}, \ldots$, one for each possible interpretation in $M_i$ (Lemma \ref{lem:consistency-of-Gamma}). 

We first extend the definition of existential witness to the case of multiple languages. 
\begin{defn}[Existential Witness]
	\label{def:existetial-witnesses}
	Let $\GGamma$ be a set $\{\GGamma_0, \GGamma_1, \ldots\}$ of formulas in $\{L_i\} = \{L_0, L_1, \ldots\}$ and let ${\cal C} = \{{\cal C}_0, {\cal C}_1, \ldots\}$ be a set of constant symbols such that each ${\cal C}_i$ is a set of constants of $L_i$. We say that ${\cal C}$ is a set of \emph{witnesses} for $\GGamma$ in $\{L_i\}$ iff for every formula $\co{\phi(x)}{i}$ with exactly one free variable $x$ in $L_i$ the following holds: 
	\begin{itemize}
		\item if $\co{\phi(x)}{i}$ is a complete formula, then there is a constant $c \in {\cal C}_i$ such that $\GGamma \der{\IC} \co{\exists \phi(x) \imp \phi(c)}{i}$;
		\item if $\co{\phi(x)}{i}$ is a non complete formula, then there is a set of constants $c_1, \ldots, c_k, \ldots \in {\cal C}_i$ such that $\Gamma_i \der{\IC}\co{\exists \phi(x) \imp \phi(c_i)}{i}$ for each $\Gamma_i \in \GGamma_i$.
	\end{itemize}
\end{defn}

\begin{lem}
	\label{lem:consistency-of-Gamma}
Let $\Gamma$ be a $k$-consistent set of sentences of $\{L_i\}$. Let ${\cal C}$ be a set $\{{\cal C}_i\}$ of new constant symbols of 
power $|{\cal C}| = \parallel \{L_i\}\parallel$, and let $\{L_i \cup {\cal C}_i\}$ be the family of languages defined as the simple 
expansion of each $L_i$ formed by adding $\{{\cal C}_i\}$. Then $\Gamma$ can be extended to a $k$-consistent set of set of sentences $\Gamma^*$ in 
$\{L_i \cup {\cal C}_i\}$ which has ${\cal C}$ as a set of witnesses in $\{L_i \cup {\cal C}_i\}$. 
\end{lem}

\begin{pf}
	Let 
	$$\co{\phi_1(x)}{i_1},\co{\phi_2(x)}{i_2},\co{\phi_3(x)}{i_3},\dots$$
	be an enumeration of all the formulas in $\{L_i\}$ that contain exactly one free variable $x$.  
	Let ${\cal C}$ be an infinitely enumerable set of variables. We ``split'' ${\cal C}$ into an enumerable sequence of disjoint sets $W$, $U_1$, $U_2$, $U_3$, \ldots as follows 
	\begin{center}
	  $
	  \begin{array}{lll}
	    W  & = & \{w_1, w_2, w_3, \ldots\}\\
	    U_1 & = & \{u^1_1, u^1_2, u^1_3, \ldots\}\\
	    U_2 & = & \{u^2_1, u^2_2, u^2_3, \ldots\}\\
	    U_3 & = & \{u^3_1, u^3_2, u^3_3, \ldots\}\\
	    & & \ldots
	  \end{array}
	  $
	\end{center}
	such that $V=W\cup U_1\cup U_2\cup U_3\cup\dots$. 
	We define an infinite sequence $\GGamma^0,\GGamma^1,\ldots$ as follows:   

	\begin{enumerate}
	\item $\GGamma^0$ is such that $\GGamma^0_i=\{\Gamma_i\}$ for all $i \in I$. 
	\item $\GGamma^{n}$ with $n=m+1$ is built according to the following rules:
	  \begin{itemize}
	  \item if $\GGamma^m\not\vdash_{\IC}\co{\exists y. y=\bothvar{z}}{i_n}$ for some arrow variable $\bothvar{z}$ that occurs in $\phi_n(x)$, then
	    $\GGamma^n=\GGamma^{m}$; 
	  \item otherwise $\GGamma^n$ is defined as follows:
	    \begin{enumerate}
	    \item if $\co{\phi_n(x)}{i_n}$ is a complete formula then 
			$$\GGamma^{n}=\GGamma^m\otimes_{i_n}\{\exists x\phi_n(x)\imp\phi_n(w_n)\}$$
	    \item if $\co{\phi_n(x)}{i_n}$ is not a complete formula then
	      $$\GGamma^{n}=\GGamma^m\otimes_{i_n}\{\exists x\phi_n(x)\imp\phi_n(u) \; |\; u\in U_n\}$$
	    \end{enumerate}
	  \end{itemize}
	\end{enumerate}

	Each $\GGamma^n$ has a set of witnesses in a subset of $V$ by construction. 
	We have to show that each $\GGamma_n$ is $k$-consistent. 
	We prove this by induction on $n$.  
	
	\begin{itemize}
		\item Base Case ($n=0$).
		
		From the definition of $\GGamma^0$ we can immediately see that $cl(\GGamma^0)\vdash\co{\phi}{i}$ if and only if $\Gamma\vdash\co{\phi}{i}$, for all formulas $\co{\phi}{i}$. Therefore the $k$-consistency of $\GGamma^0$ follows immediately from the
		$k$-consistency of $\Gamma$. 
		
		\item Inductive step. 
		
		Let us assume that $\GGamma^{n-1}$ is $k$-consistent. We have to prove that $\GGamma^{n}$ is $k$-consistent as well. If $\GGamma^{n} = \GGamma^{n-1}$, then the theorem is trivially true. Let us examine the case $\GGamma^{n} \neq \GGamma^{n-1}$. In this case, the definition of $\GGamma^{n}$ depends upon whether the formula $\co{\phi_n}{i_n}$ in the enumeration is complete or not. We assume, by contradiction, that $\GGamma^{n}$ is not $k$-consistent and we split the proof in two different cases depending on whether $\co{\phi_n}{i_n}$ is a complete formula or not.
				
				\begin{enumerate}
				\item If $\co{\phi_n}{i_n}$ is complete, then we can use Lemma~\ref{lem:prop-oplus} and say that there is a deduction $\Pi$ of $\co{\bot}{k}$ from $cl(\GGamma^{n-1}),\co{\exists x\phi_n(x)\imp\phi_n(w_n)}{i_n}$.
				
			  	$\Pi$ must contain an un-discharged assumption of the form $\co{\exists x\phi_n(x)\imp\phi_n(w_n)}{i_n}$. Otherwise the same proof $\Pi$ is obtainable from $cl(\GGamma^{n-1})$, which violates the assumption of $k$-consistency of $cl(\GGamma^{n-1})$. Therefore we can build the following deduction from $cl(\GGamma^{n-1})$:
				
				$$
				\infer[{\exe{=}}]
				 {\co{\bot}{k}}
				 {\deduce[\Pi_1]
				     {\co{\exists z.(\exists x\phi_n(x)\imp\phi_n(z))}{i_n}}
				     {cl(\GGamma^{n-1})}
				     & 
				    \deduce[\Pi]
				      {\co{\bot}{k}}
				      {cl(\GGamma^{n-1}) & 
				       [\co{\exists x\phi_n(x)\imp\phi_n(w_n)}{i_n}]}}
				$$
				Remember that $cl(\GGamma^{n-1}) \der{\IC} \co{\exists y. y=\bothvar{z}}{i_n}$ for all the arrow variables occurring in $\phi_n(x)$. Under this hypothesis the existence of the deduction $\Pi_1$ above is guaranteed. Therefore we have shown a deduction of $\co{\bot}{k}$ from $cl(\GGamma^{n-1})$ which contradicts the inductive hypothesis.
				
				\item If $\co{\phi_n}{i_n}$ is not complete we can prove that if $cl(\GGamma^{n}) \der{\IC} \co{\phi}{i}$, then $cl(\GGamma^{n-1}) \der{\IC}\co{\phi}{i}$. Let us assume that $cl(\GGamma^{n}) \der{\IC} \co{\phi}{i}$ with a proof $\Pi$ and that $\co{\psi}{j}$ is one of the undischarged assumptions of $\Pi$. We will prove that we can infer all the undischarged assumptions $\co{\psi}{j}$ of $\Pi$ from $cl(\GGamma^{n-1})$. This will immediately prove that $\Pi$ can be extended to a new proof $\Pi'$ of $\co{\phi}{i}$ from
				  $cl(\GGamma^{n-1})$.
				
				  Let $\co{\psi}{j} \in cl(\GGamma^{n})$. If $j \neq i_n$, then $\co{\psi}{j} \in cl(\GGamma^{n-1})$ and the proof is done.  
				  If $j = i_n$, then the definition of cl(.) says that for all $\Gamma^{n}_{i_n} \in \GGamma^{n}_{i_n}$,  $\Gamma^{n}_{i_n}$ locally entails $\psi$. From the definition of $\GGamma^{n}$ we know that all $\Gamma^{n}_{i_n}$ are of the form $\Gamma^{n-1}_{i_n} \cup \{\exists x\phi_n(x)\imp\phi_n(u)\}$. Therefore we have that for all $\Gamma^{n-1}_{i_n} \in \GGamma^{n-1}$ there are an infinite number of proofs $\Gamma^{n-1}_{i_n},\exists x\phi_n(x)\imp\phi_n(u) \der{} \psi$. 
				
			      Since all the variables $u$ are new, and $cl(\GGamma^{n-1}) \der{\IC} \co{\exists y. y=\bothvar{z}}{i_n}$ for all the arrow variables possibly occurring in $\phi_n(x)$, then, in defining $cl(\GGamma^{n-1})$ we can treat all arrow variables $\bothvar{z}$ in $\phi_n$ as terms. Therefore for all $\Gamma^{n-1}_{i_n} \in \GGamma^{n-1}_{i_n}$ the following proof of $\co{\psi}{i_n}$ holds: 
			  	  $$
				  \infer[\exe{}]
				   {\co{\psi}{i_n}}
				   {\co{\exists z.(\exists x\phi_n(x)\imp\phi_n(z))}{i_n} & 
				    \deduce[\Pi_u]
				      {\co{\psi}{i_n}}
				      {\Gamma^{n-1}_{i_n} & [\exists x.\phi_n(x)\imp\phi_n(u)]}}
				  $$
				  and $\co{\psi}{i_n} \in cl({\GGamma^{n-1}})$. 
			\end{enumerate}
		Let $\GGamma^*$ be the upper-bound of the sequence $\GGamma^0,\GGamma^1,\ldots$. From the proof above we can conclude that $\GGamma^*$ is $k$-consistent. This terminates the proof. 
	\end{itemize}
\end{pf}

\begin{lem}
For all $n\geq 0$, consider the closure of $\GGamma^{n}$ w.r.t. the set of variables $\bothvar{v}$ that occur in $\phi_{n+1}(x)$ such that
$\GGamma^n\vdash_{\IC}\co{\exists y. y=\bothvar{z}}{i_{n+1}}$. 
$\GGamma^n$ is such that if $z\not\in {\cal C}$, then 
  $\GGamma^n\vdash_{\IC}\co{\exists y.\bothvar{z}=y}{i}$ if and only if  
  $\Gamma\vdash_{\IC}\co{\exists y.\bothvar{z}=y}{i}$; 
\end{lem}

\begin{pf}
The proof is by induction on $n$ and is similar to the one for Lemma~\ref{lem:consistency-of-Gamma}.
\end{pf}

The third step in our proof is the construction of saturated sets of formulas that will determine the local models $m$ that belong to the canonical model. 

\begin{defn}[$k$-saturated]
	\label{def:saturated}
	Given a set of formulas $\Gamma$ we say that $\Gamma$ is $k$-saturated if for all formulas $\co{\phi}{k}$ in $\{L_k\}$ at least one between $\co{\phi}{k}$ and $\co{\neg \phi}{k}$ belongs to $\Gamma$. Given a set of set of formulas $\GGamma$ we say that $\GGamma$ is $k$-saturated if for all formulas $\co{\phi}{k}$ in $\{L_k\}$ at least one between $\co{\phi}{k}$ and $\co{\neg \phi}{k}$ belongs to each $\Gamma_k \in \GGamma$ in $L_k$.
\end{defn}

\begin{lem}
	\label{lem:saturation-of-Gamma}
$\GGamma^*$ can be extended to a $k$-saturated set $\SSigma^*$. 
\end{lem}

\begin{pf}
Let 
$$\co{\phi_1}{i_1},\co{\phi_2}{i_2},\co{\phi_3}{i_3},\dots$$ 
be a new enumeration of the formula in the original languages in $\{L_i\}$ extended with the variables in ${\cal C}$ and the corresponding extended variables (that is, if $c \in {\cal C}$ we consider here also formulas containing arrow variables of the form $\tovar{c}{j}$ and $\fromvar{c}{j}$). 
We define an infinite sequence of sets of sets of formulas $\SSigma^0,\SSigma^1,\dots$ as follows:  
\begin{enumerate}
\item $\SSigma^0=\GGamma^*$;
\item $\SSigma^{n}$, with $n=m+1$ is defined as follows: 
\begin{enumerate}[(a)]
  \item if $\SSigma^m\not\vdash_{\IC}\co{\exists y. y=\bothvar{z}}{i_n}$ for
    some arrow variable $\bothvar{z}$ that occurs in $\phi_n(x)$, then
    $\SSigma^n=\SSigma^{m}$; otherwise, 
  \item if $\co{\phi_n}{i_n}$ is a complete formula then:
   $$
   \SSigma^{n}=
    \left\{
      \begin{array}{ll}
        \SSigma^m\otimes_{i_n}\{\phi_n\} & \mbox{if }
          \SSigma^m\otimes_{i_n}\{\phi_n\} \mbox{ is $k$-consistent} \\  
        \SSigma^m\otimes_{i_n}\{\neg\phi_n\} & \mbox{otherwise }
      \end{array}
    \right.
   $$
  \item if $\co{\phi_n}{i_n}$ is not a complete formula then
    $$\SSigma^{n}=\SSigma^m\otimes_{i_n}\{\phi_n,\neg\phi_n\}$$
\end{enumerate}
\end{enumerate}

Let $\SSigma^*$ be the upper-bound of the sequence $\SSigma^0,\SSigma^1,\ldots$. $\SSigma^*$ is $k$-saturated by construction.
\end{pf}

\begin{lem}
$\SSigma^*$ is $k$-consistent. 
\end{lem}

\begin{pf}
We prove that for each $n\geq 0$, $\SSigma^n$ is $k$-consistent by induction on $n$.

\begin{itemize}
	\item {\bf Base Case ($n=0$).} 
	
	$\SSigma^0$ is $k$-consistent because of the $k$-consistency of $\GGamma^*$.
	
	\item {\bf Inductive Step.}
	
	Suppose that $\SSigma^{n-1}$ is $k$-consistent, and let us prove that  $\SSigma^{n}$ is $k$-consistent too. If $\SSigma^{n} = \SSigma^{n-1}$,
	then the proof is done. As usual, we split the proof in two parts, depending on whether the $n$-th formula $\co{\phi_n}{i_n}$ in the enumeration is complete or not.  

	\begin{enumerate}
	\item If $\co{\phi_n}{i_n}$ is complete then the only possibility for having $\SSigma^{n}$ $k$-inconsistent is that both $\SSigma^{n-1}\otimes_{i_n}\phi_n$ and $\SSigma^{n-1}\otimes_{i_n}\neg\phi_n$ are $k$-inconsistent. From the Lemma~\ref{lem:prop-oplus} we can deduce that $\co{\bot}{k}$ is derivable both from $cl(\SSigma^{n-1}),\co{\phi_n}{i_n}$ and from $cl(\SSigma^{n-1}),\co{\neg\phi_n}{i_n}$. Then, with an application of $\ore{=}$ we have that $\co{\bot}{k}$ is derivable from $cl(\SSigma^{n-1})$. This contradicts the fact that $\SSigma^{n-1}$ is $k$-consistent. 

	\item If $\co{\phi_n}{i_n}$ is not  complete then, $\SSigma^n=\SSigma^{n-1}\otimes_{i_n}\{\phi_n,\neg\phi_n\}$. Using Lemma~\ref{lem:prop-oplus}, $\SSigma^n\vdash_{\IC}\co{\bot}{k}$ iff $cl(\SSigma^{n-1}),\co{\phi_n\vee\neg\phi_n}{i_n}\vdash_{\IC}\co{\bot}{k}$. Since $cl(\SSigma^{n-1}) \vdash_{\IC}\co{\exists y. y=\bothvar{z}}{i_n}$ for all the arrow variables $\bothvar{z}$ that occur in $\phi_n$, then $cl(\SSigma^{n-1}) \vdash_{\IC} \co{\phi_n\vee\neg\phi_n}{i_n}$, and $cl(\SSigma^{n-1})\vdash_{\IC}\co{\bot}{k}$ holds which contradicts the fact that $\SSigma^{n-1}$ is $k$-consistent. 
	\end{enumerate}
\end{itemize}
\end{pf}

The final step in the proof is the definition of the canonical model, and the proof that this canonical model is a $\IC$-model.

\begin{defn}[Canonical Model]
	\label{def:canonical-model}
  The {\em canonical model} $\M^c=\npla{\{S_i^c\},\{\dr^c_{ij}\}}$ is defined as follows.
  \begin{description}
  \item[Domains:] 
    Let ${\cal C}$ be the set of existential witnesses introduced in the construction of $\GGamma^*$. Let $\bothvar{{\cal C}_i}$ be the set of the additional arrow variables $\bothvar{c}$
    such that $\exists x.x=\bothvar{c}$ belongs to the intersection $\bigcap \SSigma^*_i$ of all $\Sigma_i$ in $\SSigma^*_i$. 
    For two variables $c_1, c_2 \in {\cal C} \cup \bothvar{{\cal C}_i}$ we define:
    \begin{center}
      $c_1 \sim_i c_2$ if and only if $c_1=c_2 \in \bigcap \SSigma^*_i$. 
    \end{center}
    Since $\SSigma^*$ is saturated w.r.t. complete formulas, we have that each $\sim_i$ is an equivalence relation on ${\cal C} \cup \bothvar{{\cal C}_i}$. For each 
    $c\in {\cal C} \cup \bothvar{{\cal C}_i}$, let 
    $$
    [c]_i = \{c \in {\cal C}\cup \bothvar{{\cal C}_i} \ | \ c \sim_i c'\} 
    $$ 
    be the equivalence class of $c$. Similarly to the usual proof of completeness for first\ftext{-}order logic, we propose to construct a model $\M^c$ that associates to each language $L_i$ the domain $\bdom^c_i$ of all the equivalence classes $[c]_i$. Formally,
    $$
    \bdom^c_i = \{[c]_i \ | \ c \in {\cal C} \cup \bothvar{{\cal C}_i}\}
    $$  

  \item[Local models:]
    For each $i\in I$ each element $\Sigma\in\SSigma^*_i$ is saturated. That is, for each $i$-formula with arrow variables in $\bothvar{{\cal C}}$, at least one between $\phi$ and $\neg\phi$ is in $\Sigma$. In the general case $\Sigma$ may be inconsistent (this happens when both $\phi$ and $\neg\phi$ belong to $\Sigma$), but if this is not the case, then $\Sigma$ automatically determines a local interpretation of $L_i$ over $\bdom^c_i$. Therefore we define $S^c_i=\{\Sigma\in\SSigma^*_i | \Sigma \text{ is $i$-consistent}\}$.    

  \item[Domain relations:]
    For each pair $i\neq j \in I$, the domain relation $\dr^c_{ij}\subseteq\bdom^c_i\times\bdom^c_j$ is defined as 
    \begin{equation*}
      \dr_{ij} = \{\npla{[c][\fromvar{c}{i}]\ |\ \text{ if } \fromvar{c}{i}\in \bothvar{{\cal C}_j}}\} \cup \{\npla{[\tovar{c}{j}][c]\ |\ \text{ if } \tovar{c}{j}\in \bothvar{{\cal C}_i}}\}
    \end{equation*}
  \end{description}
\end{defn}

Let us prove that $\M^c$ is a model which satisfies the bridge rules $\IC$.

\begin{lem}
\label{lem:model-construction} 
$\M^c$ is a $\IC$-model. 
\end{lem}

\begin{pf} We have to prove that $\M^c$ is not empty, that is, that at least one of the $S^c_i$ is not empty, and that is satisfies the bridge rules $\IC$. 
  \begin{description}
  \item[Not emptiness:] Since $\SSigma^*$ is $k$-consistent there is an element $\Sigma\in\SSigma^*_k$ which is consistent. Therefore $S^c_k$ is not empty.

  \item[Satisfiability of \IC:] We consider the simple case of $\co{\phi(x)}{i}\rightarrow\co{\psi(\fromvar{x}{i})}{j}$. The proof for
    more complex interpretation constraints is analogous. 

    Suppose that $\Mc\models\co{\phi(x)}{i}[a]$ for an assignment $a$ with $a_i(x)=[c]$. This implies that $\phi(c)\in\bigcap\SSigma^*_i$. Because of the interpretation constraint we have that $\psi(\fromvar{c}{i})\in\bigcap\SSigma^*_j$ and also $\exists x.x=\fromvar{c}{i}\in\SSigma^*_j$. This means that $a$ can be extended with $a_j(\fromvar{x}{i})=[\fromvar{c}{i}]$ and $\Mc\models\phi(\fromvar{x}{i})[a']$. 
  \end{description}
\end{pf}
  
Let $\M^c$ be the canonical model built for $(\Gamma,\Sigma),\co{\neg \phi}{i}$ and let $a$ be an assignment which assigns all variables $x$ to $[x]$. It is easy to see that $\M \models \Gamma[a]$, and $\M \models e(\Sigma)[a]$, but $\M \not \models \co{\bigwedge_{i:\sigma \in \Sigma}\sigma\imp\phi}{i}[a]$. This concludes the completeness proof.
